\documentclass[english, LaM, oneside]{sapthesis}
\usepackage[utf8]{inputenx}
\usepackage{indentfirst}
\usepackage{microtype}
\usepackage[style=numeric,sorting=none]{biblatex}
\usepackage{physics}
\usepackage{csquotes}
\usepackage{cancel}
\usepackage{lettrine}
\usepackage{float}
\usepackage{comment}
\usepackage{makecell}
\usepackage[nottoc, notlof, notlot]{tocbibind}
\usepackage{hyperref}
\hypersetup{
			hyperfootnotes=true,			
			bookmarks=true,			
			colorlinks=true,
			linkcolor=red,
                        linktoc=page,
			anchorcolor=black,
			citecolor=red,
			urlcolor=blue,
			pdftitle={Development of energy calibration and data analysis systems for the NUCLEUS experiment.},
			pdfauthor={Giorgio Del Castello},
			pdfkeywords={thesis, sapienza, roma, university,neutrino,cevns,master}
 }

\title{Development of energy calibration and data analysis systems for the NUCLEUS experiment.}
\author{Giorgio Del Castello}
\IDnumber{1755135}
\course[override]{Master's Degree Course in Physics - Curriculum of "Particle and Astroparticle Physics"}
\courseorganizer{Faculty of "Scienze Matematiche Fisiche e Naturali"}
\submitdate{2020/2021}
\copyyear{2021}
\advisor{Prof. Marco Vignati}
\coadvisor{Dr. Claudia Tomei}
\coadvisor{Dr. Nicola Casali}
\authoremail{delcastello.1755135@studenti.uniroma1.it}
\examdate{19 October 2021}
\examiner{Prof.ssa Simonetta Gentile}
\examiner{Prof. Roberto Di Leonardo } \examiner{Prof. Roberto Maoli}  \examiner{Prof. Gianluca Cavoto}  \examiner{Prof. Andrea Crisanti} \examiner{Prof.ssa Maria Chiara Angelini}  \examiner{Prof. Cristiano Palomba} 

\begin{document}

\frontmatter
\maketitle

\begin{abstract}
\lettrine[lines=2, findent=3pt, nindent=0pt]{C}{}\textbf{oherent elastic neutrino-nucleus scattering} (CE$\nu$NS) opens new approaches for the search of new physics beyond the Standard Model at low energies. The NUCLEUS experiment, situated at the Chooz nuclear power plant, aims to use the intense antineutrino flux produced from the reactor cores (with maximum neutrino energy of the MeV order) to perform high precision measurements of the CE$\nu$NS cross-section via gram-scale ultra-low threshold cryogenic detectors. For the first phase of NUCLEUS a 10~g detector using both CaWO$_4$ and Al$_2$O$_3$ as target material will be deployed. An up-scaling to a 1~kg detector is planned for the second phase of the experiment in order to perform a cross-section measurement with an error of the percent order.\\

A common problem with ultra-low threshold cryogenic detectors is the calibration, since most radioactive sources tend to saturate the dynamical range of the sensors. For this reason, in this dissertation, a photon-statistic based optical calibration setup has been developed and tested using kinetic inductance detectors being researched by the BULLKID R$\&$D project. Both the electronic setup for the optical calibration and the relative control software have been developed to automatically perform a full calibration of the detector with minimal user intervention and to be integrated in a wider data acquisition system.\\

Currently the NUCLEUS experiment does not have an official data analysis software. For this reason, during the development of this work, the DIANA analysis framework, developed for the CUORE experiment at LNGS, has been severely upgraded and made compatible with the NUCLEUS data in order to be presented as the new official analysis framework. Besides the upgrade, a new analysis protocol, to be performed with DIANA, has been developed using the data from the NUCLEUS prototype runs. This new protocol aims to be fully automatic and model independent in order to deal with a wide variety of data. The new analysis procedure proved to be extremely accurate at reproducing the results obtained via a model dependent study performed with an almost event by event inspection that was carried out in the early stages of NUCLEUS. Particular focus was given to the development of an event reconstruction procedure in order to accurately measure the energy of events that produce signals with amplitudes outside the linear response region of the cryogenic detector. This procedure is not only necessary to expand the dynamical range of the used sensors but is also mandatory when calibrating the detector with radioactive sources such as $^{55}$Fe which produce events that completely saturate the cryogenic detection system.\\

Finally, the last activity done for the NUCLEUS experiment was to develop a python based toolkit for the study of the discovery potential of the experiment. This toolkit, referred to as PyCEnNS, has to work in parallel with an already existing software (FORECAST) in order to perform cross-checks on the obtained results throughout their development by implementing different calculation algorithms. PyCEnNS was originally developed in the early stages of NUCLEUS by J. Rothe and has been intensely upgraded during the work presented in this thesis in order to reach an agreement below 0.005\% with the FORECAST results. Moreover, the upgrade undergone by PyCEnNS was also aimed at making the code fast and not demanding on the computational resources while handling a large amount of data, which is one of the main weak spots of FORECAST. The resulting performance increase was made possible by making smart use of the already optimized and available python packages along with some basic machine learning techniques regarding data handling.\\

This dissertation serves as a manual for the whole NUCLEUS collaboration when performing data analysis with the newly upgraded software, or during the integration of the optical calibration electronics in the experimental setup or when using the new available toolkit of the sensitivity studies. The promising results in all the three listed departments are shown and justify investing in manpower to further perfect all the newly developed and upgraded tools for the NUCLEUS experiment, which will become essential with the upcoming first phase. Moreover, being the experimental signature of CE$\nu$NS extremely similar to the one expected from WIMP interaction, several of the developed procedures and tools can be used also for future dark matter experiments.

\end{abstract}
\setcounter{secnumdepth}{3}
\tableofcontents

\mainmatter
\chapter{Introduction}
\label{chap:1}
\lettrine[lines=2, findent=3pt, nindent=0pt]{I}{}n 1973 Daniel Z. Freedman theorized the existence of a new interaction process withing the Standard Model regarding the interaction of neutrinos with the surrounding nuclei. This new process, known as \textbf{coherent elastic neutrino-nucleus scattering} or CE$\nu$NS for short, is characterized by a high interaction probability due to the high cross-section (at least two orders of magnitude higher than other neutrino-matter interactions) and by an extremely low detectable signal due to the nuclear recoil. Moreover, since the allowed neutrino energies are below the few tens of MeV, CE$\nu$NS  makes the measurement of extremely low energy neutrinos feasible for the first time opening new doors for the discovery of new physics.\\

Since the signal that arises from CE$\nu$NS is extremely faint and not easily detectable, the neutrino-nucleus coherent elastic scattering was observed for the first time only in 2017 from the COHERENT collaboration at the Oak Ridge National Laboratory using the smallest working neutrino detector so far ($\sim$14~kg of scintillator crystal). Following the COHERENT experimental results, more CE$\nu$NS experiments have been planned for the next years, in particular this dissertation will follow the development of the NUCLEUS experiment. NUCLEUS aims to perform a high precision measurement of the CE$\nu$NS cross-section installing an array of cryogenic detectors at the Chooz nuclear power plant. As many physics experiments, NUCLEUS is divided in phases, in the first one a 10~g cryogenic detector will be deployed and a measurement of CE$\nu$NS dominated by statistical error will be performed with a predicted $\sim10\%$ error. In the second phase a 1~kg array of cryogenic detectors will be used in order to be dominated only by systematical uncertainty  reaching an expected $\sim1\%$ error. In both phases the NUCLEUS detector will be extremely small when compared to other neutrino experiments which allows for easy installation in small environments.\\

This dissertation is essentially divided in two parts, in the first two chapters a general description of the coherent elastic neutrino nucleus scattering will be given followed by an overview of the main features of the NUCLEUS experiment. The last three chapters instead describe in detail the activities performed in the framework of the NUCLEUS experiment during the development of this thesis project. In the \ref{chap:4}$^\text{th}$ chapter an optical calibration system based on photon-statistics for an array of cryogenic detectors will be described along with the development of both the electronics setup and the piloting software. In this chapter the first characterization of the first prototype of the \textit{mirror wafer}, a light splitting device used in the optical calibration setup, is presented as well.\\

In chapter \ref{chap:5} a new protocol for the analysis of the NUCLEUS data is described. This chapter, apart from presenting an automatic and model independent analysis of the data, also serves to show the potentials of the DIANA analysis framework and propose it as the official analysis software. Moreover, in this chapter some of the new upgrades made for DIANA during the development of the new analysis protocol will also be presented. \\

In chapter \ref{chap:6} the development of a new toolkit for the sensitivity studies of the NUCLEUS experiment will be presented along with a brief description on the theory behind the sensitivity calculations. The NUCLEUS collaboration posses two different toolkits: PyCEnNS is a python based toolkit whose development and testing is the main focus of the dissertation, the second toolkit is referred to as FORECAST and is a C++/ROOT based software. The main reason why two different toolkits have been developed is to use different algorithms to compute the same results in order to perform cross-checks along the way. \\

The aim of this dissertation is not only to present the current status of the NUCLEUS experiment and to describe the development of new tools for the collaboration, but it is also meant to be used as a manual from the entirety of the NUCLEUS collaboration, in particular when performing the optical calibration of the detector or the data analysis of the events acquired with the upcoming first phase of the experiment.

\chapter{Coherent Elastic Neutrino Nucleus Interaction}
\label{chap:2} 
\begin{figure}[H]
    \centering
    \includegraphics[scale=0.4]{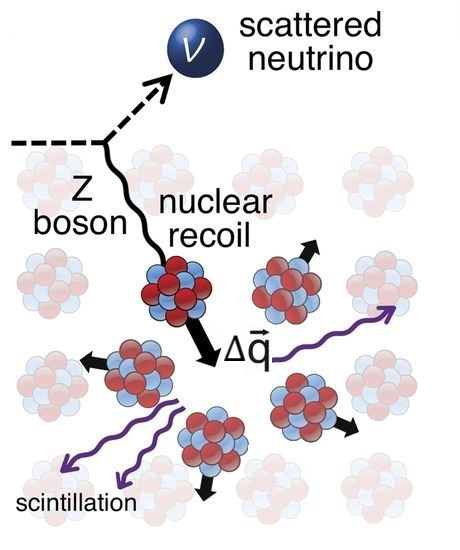}
    \caption{Artistic illustration of the coherent neutrino-nucleus scattering \cite{coherent}.}
    \label{fig:cevns}
\end{figure}
\lettrine[lines=2, findent=3pt, nindent=0pt]{I}{}n 1930 Pauli first proposed the existence of an invisible light neutral fermion, later known as neutrino. Since then it has been clear to all the physics community that this new particle would be quite difficult to directly observe experimentally. This detection difficulty is due to the fact that the neutrino interaction cross-section with matter is mediated by the weak interaction and is thus extremely low, making the interaction probability practically zero. This is why, for example, in most high energy physics experiments the neutrino component of the interactions is mostly measured indirectly using the missing transverse energy, an important result reached with this method is the discovery of the charged weak boson at LEP \cite{w+-}.\\

After more than 20 years from Pauli's proposal, direct experimental evidence of the existence of neutrinos was presented from Clyde Cowan and Frederick Reines. Their experiment detected nuclear reactor antineutrinos using a water tank filled with diluted cadmium cloride to measure the inverse beta decay (IBD) induced from antineutrino-proton interaction. The use of nuclear reactors for neutrino experiments has become common practice throughout the years, this is due to the fact that the typical neutrino cross-section is of order $10^{-42}\text{cm}^2$. This very low cross-section can be partially compensated by using voluminous and massive detectors giving the neutrino a number of targets of the order of Avogadro's number ($10^{23}$). The rest of the cross-section value can be partially compensated using reactor neutrinos since in a reactor core $10^{20}$ electron-antineutrinos are produced for every kilogram of uranium, giving approximately 1 event per day.\\

For most years the main ways of detecting neutrinos consisted in exploiting IBD and neutrino-electron scattering due to their fairly easily distinguishable and measurable event signature. As mentioned previously, Daniel Z. Freedman published an article characterizing a new neutrino-matter interaction predicted by the standard model, the \textbf{coherent elastic neutrino-nucleus scattering} (CE$\nu$NS) (see Figure \ref{fig:cevns}). This new interaction has the peculiarity of having the biggest neutrino-matter cross-section (see \ref{sec:theo}) making it the most common occurring neutrino-matter reaction. The high cross-section is due to the fact that being the scattering coherent the nucleus is seen as one and the contributions of the various nucleons sum up coherently. Moreover, since CE$\nu$NS is an elastic scattering, i.e. no new particles are produced, it does not have a lower energy threshold, unlike the inverse beta decay interaction where the minimum neutrino energy is $1.806$~MeV, and the only limit is given by the sensibility of the detector. This last characteristic is especially interesting since it allows to study extremely low energy neutrinos that have not been studied before.\\

\begin{figure}[H]
    \centering
    \includegraphics[scale=0.3]{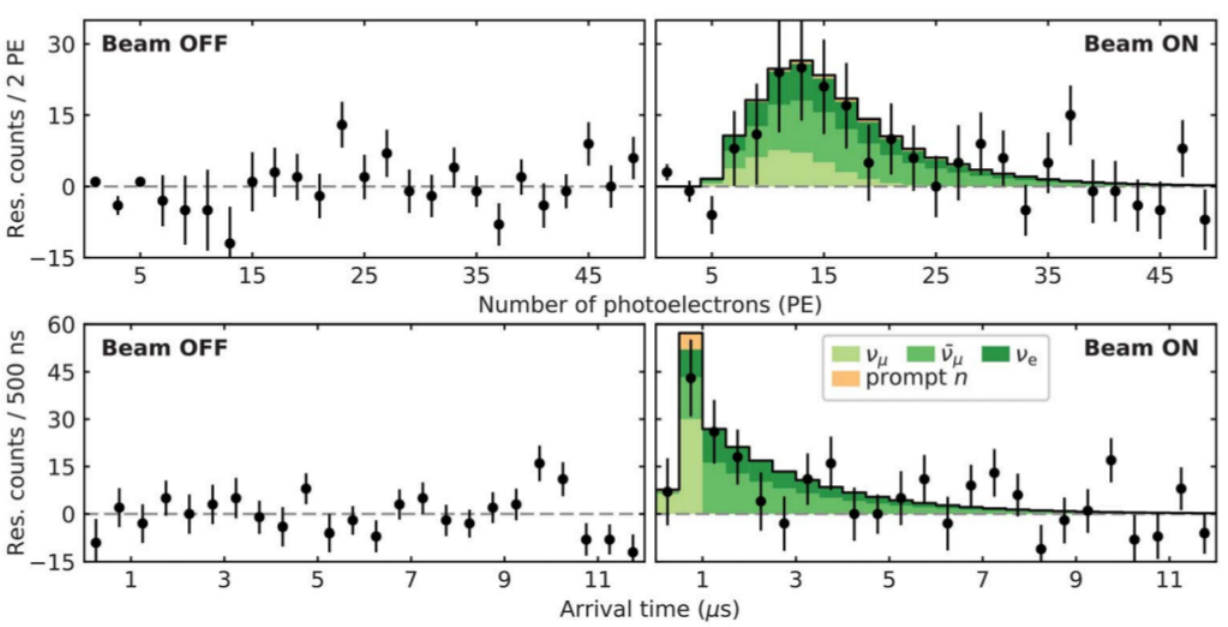}
    \caption{Results on CE$\nu$NS measurements from the COHERENT collaboration. Both top and bottom panels show the residual differences between the pretrigger samples and the data taken after the collision of the proton beam with the mercury target. In panel A their energy distribution is presented (in number of detected photoelectrons) and in panel B their time distribution is shown. As it can be seen, there is very good agreement between the experimental data (data points) and the predictions (green histograms). Figure from \cite{coherent}.}
    \label{fig:coherent}
\end{figure}

Despite having a higher interaction probability, CE$\nu$NS is experimentally challenging to measure. This is due to the fact that the recoiling nucleus, which gives rise to the detectable signal, has, after the scattering, kinetic energies under $\mathcal{O}$(keV) (see section \ref{sec:pheno})  and thus extremely sensitive detectors (at least at $\mathcal{O}$(eV)) are mandatory. Due to this experimental challenge more than 40 years have elapsed between the first theoretical formulation to the first experimental evidence in 2017 from the COHERENT collaboration \cite{coherent}. The experiment conducted by COHERENT at the Oak Ridge National Laboratory used neutrinos produced with a proton beam impinging on a mercury target. The detector used was a $14.57~kg$ sodium-doped CsI crystal paired with a photomultiplier and shielded with various layers of lead, high density polyethylene and water. The collaboration observed an excess of signal whenever the proton bunches hit the mercury target and produced neutrino. The observed excess is completely compatible (under 1$\sigma$) with the Standard Model predictions of CE$\nu$NS (see Figure \ref{fig:coherent}). The COHERENT collaboration not only managed to measure the neutrino-nucleus scattering for the first time but also build the smallest working neutrino detector. For more details on the COHERENT experiment refer to \cite{mia_tesi}.	

\section{Phenomenological Considerations}\label{sec:pheno}
\begin{figure}[H]
    \centering
    \includegraphics[scale=0.45]{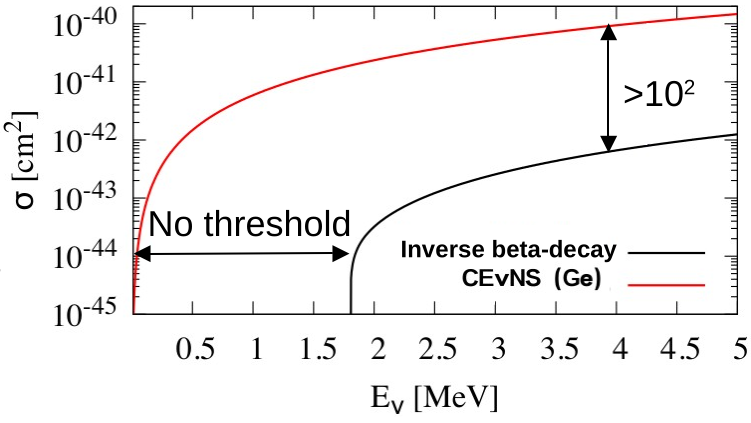}
    \caption{Comparison of the inverse beta decay cross-section with the neutrino-nucleus coherent scattering on a germanium target \cite{vignati}.}
    \label{fig:IBD}
\end{figure}

As mentioned above, the neutrino-nucleus coherent elastic scattering is the neutrino-matter interaction that dominates for neutrinos with energies up to few MeV. Another leading neutrino-matter phenomenon is the inverse beta decay which presents an energy threshold and a cross-section that is more than two orders of magnitude less than the CE$\nu$NS one as it is show in Figure \ref{fig:IBD}. A similar process to the neutrino-nucleus scattering is the neutrino-electron scattering, both interactions do not present an energy threshold but since their cross-sections scale with the mass of the target the neutrino-electron one is more than 2000 times smaller than the neutrino nucleus one.

\subsubsection{Cross-section discussion}\label{sec:xsec}
In order to understand the principles behind the CE$\nu$NS interaction and detection a brief phenomenological discussion of the theoretical results is presented in this section. All the results and quantities presented here should only be used to understand the basics of this interaction and not be used in accurate calculations since they present a great deal of approximations and are not precise enough for the stage in which CE$\nu$NS experiments currently are.

\paragraph{Coherency Condition: } The first important discussion that has to be made in order to understand the interaction's energy scale is regarding the coherency condition. When a composite object, like a nucleus, is considered as one of the interacting parties one must usually take into account its internal structure and how much the projectile (in this case the neutrino) penetrates inside the nucleus and experiences its various constituents. In order to do so the nuclear form factor is usually introduced, it consists of the Fourier transform of the charge distribution inside the nucleus (in this case the weak charge must be used) and is usually normalized to 1. A common approximation of the nuclear form factor for low momentum transfer interactions is:

$$F^2(q^2) = e^{\frac{1}{3}(qR_N)^2}$$

where $q$ is the module of the transferred momentum and $R_N$ is the nuclear radius. In order for the scattering to be coherent the neutrino must not be able to probe the internal structure of the nucleus giving $F^2(q^2)\approx 1$. This condition translates to:
$q\cdot R_N \ll 1$

\begin{figure}[H]
    \centering
    \includegraphics[scale=0.4]{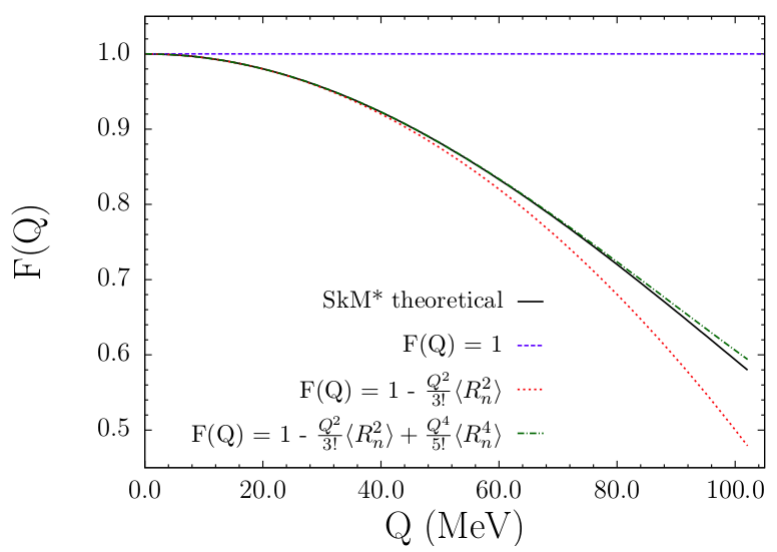}
    \caption{Scaling of the nuclear form factor with the transferred momentum, different approximations are presented (Figure from \cite{form_factor}).}
    \label{fig:form_factor}
\end{figure}

The scattering kinematics gives that the condition that ties together the transferred momentum and the neutrino energy in the laboratory frame is:
\begin{equation}\label{eq:coherency_cond}
\Delta q^2 = 2E_\nu^2(1-\cos(\theta)) \qquad\rightarrow\qquad\Delta q_{max} = 2E_\nu
\end{equation}
this two conditions can be combined in order to have a maximum neutrino energy:
\begin{equation}
E_\nu = \frac{\Delta q_{max}}{2}<\frac{1}{2R_N} \approx 17~MeV (U) \leftrightarrow 60 MeV(He)
\end{equation}
where the numerical estimate was made using a nuclear diameter of $3.351$~fm for helium and $11.71$~fm for uranium. The choice of this two nuclei was made in order to give a complete energy span throughout the elements of the periodic table. The interacting neutrinos thus must have energies at most of a few tens of MeVs in order to respect the coherency condition. Higher energy neutrinos can still undergo nuclear scattering but the coherency condition is lost making the cross-section rapidly decreasing (see Figure \ref{fig:form_factor}). 

\paragraph{Cross-section properties: } The differential cross-section for the neutrino nucleus coherent scattering can be written as:
\begin{equation}\label{eq:lame_xsec}
\frac{d\sigma(\nu N\rightarrow\nu N)}{dT} = \frac{G_F^2}{4\pi} M_N Q_W^2\left(1-\frac{M_N T}{2E\nu^2}\right)F^2(q^2)
\end{equation}
where the usual Fermi constant dependence ($G_F^2$) of weak cross-sections and the anticipated increase with nuclear mass $M_N$ are present. The $Q_W$ term indicates the nuclear weak charge and can be written as:
$$Q_W = N-Z(1-\sin\theta_W^2)$$
considering the low energy value of the Weinberg angle $\theta_W$ ($\sin^2\theta_W \approx 0.24$) the weak charge can be approximated as $Q_W\approx N\approx \frac{A}{2}\propto M_N$, where $A$ is the atomic number of the target and thus scales with the mass of the recoiling nucleus. The $F^2(q^2)$ term is the mentioned nuclear form factor and can be approximated to 1 if the coherency condition is verified. From the coherency condition the maximum recoil energy is estimated to be:
$$T_{max} = \frac{q^2_{max}}{2M_N}=\frac{2 E_\nu^2}{M_N} \approx \mathcal{O}(keV)$$
thus integrating over the nuclear recoil energy one has that the cross section is:
\begin{equation}
\sigma(\nu N\rightarrow\nu N) \approx  \frac{G_F^2}{4\pi}E^2_\nu Q^2_W \approx\frac{G_F^2}{4\pi}E^2_\nu \frac{A^2}{4}\propto M_N^2
\end{equation}
From this last equation it can be seen that the CE$\nu$NS approximately scales with the square of the mass of the target nuclei making the use of high atomic number materials preferable in order to enhance the cross-section. On the other hand, when choosing the target material for a CE$\nu$NS experiment one must consider that increasing the nuclear mass reduces the maximum recoil energy $\left(T_{max}\propto M_N^{-1}\right)$ making materials like silicon, germanium, sapphire and calcium tungstate the go-to choices.

\section{Theoretical Treatment of CE$\nu$NS}\label{sec:theo}
It is useful for the understanding of CE$\nu$NS to see how the interaction is predicted from the Standard Model (SM). Coherent neutrino nucleus elastic scattering is a neutral current electroweak interaction, the Standard Model neutral current (NC) is:

\begin{equation}
J^\mu_{NC}=2\sum_f g_L^f \overline{f_L}\gamma^\mu f_L + g_R^f \overline{f_R}\gamma^\mu f_R
\end{equation}
where $f_{L}$ and $f_R$ are respectively the left and right handed components of the Standard Model elementary fermion $f$ and $g_{L/R}^f$ are the SM quantum numbers of fermions within the $SU(2)_L\times U(1)_Y$ symmetry group. By expliciting the projectors the neutral current can be written as:

\begin{equation}
J^\mu_{NC}=2\sum_f g_L^f \overline{f}\left[\frac{1+\gamma^5}{2}\right]\gamma^\mu\left[\frac{1-\gamma^5}{2}\right] f + g_R^f \overline{f}\left[\frac{1-\gamma^5}{2}\right]\gamma^\mu\left[\frac{1+\gamma^5}{2}\right] f 
\end{equation}
using the commutation properties of the $\gamma$-matrices and the idempotence property of projectors:
$$J^\mu_{NC}=\sum_f g_L^f \overline{f}\gamma^\mu\left[1-\gamma^5\right] f + g_R^f \overline{f}\gamma^\mu\left[1+\gamma^5\right] f=$$
$$=\sum_f \overline{f}\gamma^\mu\left[ (g_L^f + g_R^f) - \gamma^5 (g_L^f - g_R^f)\right] f $$
by defining $g_V^f := g_L^f + g_R^f$ and $g_A^f := g_L^f - g_R^f$ the vector and axial components of the current can be separated into:
\begin{equation}
J^\mu_{NC} =\sum_f \overline{f}\gamma^\mu (g_V^f -g_A^f \gamma^5) f
\end{equation}
Considering that the transferred momentum and the neutrino masses are by far negligible with respect to the Z-boson mass the interaction lagrangian can be written as:
\begin{equation}
\mathcal{L_{NC}} =\frac{G_F}{\sqrt{2}}J_{NC}^\mu J_{NC~\mu}
\end{equation}
which gives rise to the matrix element:
\begin{equation}
i\mathcal{M}^{ss'}(\nu+N\rightarrow\nu+N) = -i\sqrt{2} G_F \bra{N(k_2)}J^\mu_{NC}\ket{N(p_2)}\bra{\nu(k_1)}J_\mu^{NC}\ket{\nu(p_1)}
\end{equation}
where $N,\nu$ are respectively the recoiling nucleus and the neutrino, $s,s'$ are the helicities of the incoming and outgoing neutrinos, the subscripts 1 and 2 indicate the initial and final state, and $p,k$ are the four momentum of the neutrino and the nucleus respectively. The most complex of these terms is the nuclear one which involves the quark content. Considering only the valence quark contents of the nucleus the scalar product can be rewritten as:
$$
\bra{N}J^\mu_{NC}\ket{N}=g_L^u\bra{N}\overline{u_L}\gamma^\mu u_L\ket{N}+g_R^u\bra{N}\overline{u_R}\gamma^\mu u_R\ket{N}+
$$
\begin{equation}
+g_L^d\bra{N}\overline{d_L}\gamma^\mu d_L\ket{N}+g_R^d\bra{N}\overline{d_R}\gamma^\mu d_R\ket{N}
\end{equation}

a large mass number nucleus contains many $u$ and $d$ quarks, one can expect that if the nucleus has total spin equal to 0 statistically it will not violate parity. Considering this, one can write the following relations for respectively the up and down quark content:
$$\bra{N}\overline{u_L}\gamma^\mu u_L\ket{N} = \bra{N}\overline{u_R}\gamma^\mu u_R\ket{N}\qquad \bra{N}\overline{d_L}\gamma^\mu d_L\ket{N} = \bra{N}\overline{d_R}\gamma^\mu d_R\ket{N}$$
moreover since the number of $u$ and $d$ quarks in a nucleus (approximating them with free particles) are respectively $2p+n$ and $ 2n+p$ one can write the following relation:
\begin{equation}
\frac{\bra{N}\overline{u}\gamma^\mu u\ket{N} }{\bra{N}\overline{d}\gamma^\mu d\ket{N}}\propto\frac{2p+n}{2n+p}
\end{equation}
where $p,n$ are respectively the number of protons and neutrons in a nucleus. This ratio holds for free quarks but, since strong interaction cannot distinguish between $u$ and $d$ quarks, it can be assumed that the ratio holds up for the bound quarks in the nucleus as well. So considering that the two scalar products are proportional to one another one can write them (using the Wigner-Eckart theorem):

$$\bra{N}\overline{u}\gamma^\mu u\ket{N} =(2p+n)f^\mu \qquad \bra{N}\overline{d}\gamma^\mu d\ket{N}=(2n+p)f^\mu$$
where $f^\mu$ is a general 4-vector whose components can be determined from the electromagnetic properties of the nucleus and can be written as :
$$f^\mu = (p_2+k_2)^\mu F(q^2)$$
where $F(q^2)$ is the form factor of the nucleus. Thus :

$$\bra{N}J^\mu_{NC}\ket{N}=F(q^2)(p_2+k_2)^\mu \left[(2p+n)g_V^u+(2n+p)g_V^d\right]=$$
$$=F(q^2)(p_2+k_2)^\mu \left[p(2g_V^u+g_V^d)+n(g_V^u+2g_V^d)\right]$$
by defining $g_V^p := (2g_V^u+g_V^d)$ and $g_V^n = (g_V^u+2g_V^d)$ the nuclear scalar product is:
$$\bra{N}J^\mu_{NC}\ket{N}=F(q^2)(p_2+k_2)^\mu [p\cdot g_V^p + n\cdot g_V^n]$$
and the quantum numbers are:
$$g_V^p = \frac{1}{2}-2\sin^2\theta_W \qquad g_V^n = -\frac{1}{2}$$
Reassembling every part in the matrix element:
\begin{equation}
i\mathcal{M}^{ss'}=-i\frac{\sqrt{2}}{2}G_FQ_WF(q^2)(p_2+k_2)^\mu g_L^\nu\overline{\nu}^s(p_1)\gamma^\mu(1-\gamma^5)\nu^{s'}(k_1)
\end{equation}
where $Q_W=\left(-2[p\cdot~g_V^p+n\cdot~g_V^n]=n-(1-4\sin^2\theta_W)p\right)$ is the weak nuclear charge of the nucleus. By averaging over the outgoing neutrino helicities and summing up over the incoming ones the square amplitude of the matrix element is:
$$|i\mathcal{M}|^2 = 32 G_F^2 Q_W^2F^2(q^2)(g_L^\nu)^2M^2 E_\nu^2\left(1-\frac{T}{E_\nu}-\frac{MT}{2E_\nu^2}\right)$$
this matrix element can then be combined with the hypothesis of a non relativistic nucleus to give the differential CE$\nu$NS cross-section:
\begin{equation}
\frac{d\sigma}{dT}=\frac{G_F^2(2g_L^\nu Q_W)^2M}{4\pi}\left(1-\frac{T}{E_\nu}-\frac{MT}{2E_\nu^2}\right)F^2(q^2)
\end{equation}
where $g_L^\nu=\frac{1}{2}$ for a standard model neutrino. This result is extremely similar to the differential cross section presented in equation \ref{eq:lame_xsec}, the main difference is the term proportional to $\frac{T}{E_\nu}$ which is usually omitted due to the fact that it gives a $\sim$0.1\% contribution. \\

This result is valid only for scalar nuclei, if one wants to consider spin $\frac{1}{2}$ nuclei the average value of the neutral current will change to:
\begin{equation}
\bra{N}J^\mu_{NC}\ket{N}=F(q^2) \overline{u}^{r'}(k_2)\gamma^\mu u^{r}(p_2) [p\cdot g_V^p + n\cdot g_V^n]
\end{equation}
where $\overline{u}^{r'}(k_2), u^{r}(p_2)$ denote respectively the final and initial states of the nucleus under the assumption that it behaves as a spin $\frac{1}{2}$ Dirac particle. By performing calculations analogous to the ones described in the case of a spin 0 nucleus one can write the differential cross-section as:
$$\left.\frac{d\sigma}{dT}\right\vert_{spin-1/2} = \frac{G_F^2(2g_L^\nu Q_W)^2M}{4\pi}\left(1-\frac{T}{E_\nu}-\frac{MT}{2E_\nu^2}+\frac{T^2}{2E_\nu^2}\right)F^2(q^2)=$$
\begin{equation}
=\left.\frac{d\sigma}{dT}\right\vert_{spin-0} + \frac{G_F^2(2g_L^\nu Q_W)^2M}{4\pi}\left(\frac{T^2}{2E_\nu^2}\right)F^2(q^2)
\end{equation}
The spin $\frac{1}{2}$ cross-section presents an additional term that is of the order $10^{-6}$ and thus presents an extremely negligible contribution. \\

This discussion on the derivation of the CE$\nu$NS cross-section is presented here in order to highlight all the different approximations and assumptions made through the calculation. Most articles present the CE$\nu$NS cross-section in the form of equation \ref{eq:lame_xsec} without explaining the level of approximation. This approach, while facilitating the phenomenological discussion of CE$\nu$NS, is not extremely helpful when researchers need a more exact result. This paragraph is then presented here to function as a small and simple but detailed guide to the CE$\nu$NS cross-section. The discussion presented here follows the treatment given in appendices A and B of \cite{lindner}, where further details can be found.

\section{Why measure CE$\nu$NS?}
There are several scientific reasons that justify a CE$\nu$NS experiment. First of all CE$\nu$NS is a Standard Model interaction predicted in the '70s and observed only in 2017 and still lacks a high precision measurement in order to compare it with the expected cross-section behaviour. CE$\nu$NS gives access to unprecedented low energy neutrinos and thus by performing a precise observation of such particles deviations from the Standard 
Model predictions might arise, possibly leading to new low energy experiments to explore such non-standard interactions, like the neutrino magnetic moment and the existence of sterile neutrinos.\\

Apart from non-standard interactions, since the CE$\nu$NS cross-section is at least two orders of magnitude higher than the rest of the neutrino-matter interactions, it allows to miniaturize the neutrino detector technology. In fact, experiments like Borexino, SNO, IceCube, Antares all use hundreds of tons of target material to detect neutrinos while the COHERENT experiment, that measured CE$\nu$NS for the first time, already used a detector of only 14~kg and the NUCLEUS experiment, which is the main focus of this dissertation, plans to use a gram-scale detector (see chapter \ref{chap:3}). The detector miniaturization obviously opens new opportunities in the neutrino scientific community but it also gives access to new applications in the military and civil environments allowing for non invasive reactor core monitoring.\\

Moreover, during a supernovae collapse, the main loss of energy is due to neutrinos escaping the collapsing star with energies matching the CE$\nu$NS energy range making coherent neutrino-nucleus elastic scattering based detectors ideal for measuring astrophysical neutrinos generated in these cosmic events.\\

One last reason that motivates the existence a CE$\nu$NS experiment is that neutrino induced nuclear recoils are extremely similar to dark matter nuclear recoils. CE$\nu$NS due to astrophysical and environmental neutrinos will, in fact, prove to be an irreducible background for WIMP detection and an accurate measurement of this neutrino interaction is required in order to discriminate an excess of events due to dark matter.\\

As it can be seen there are a great deal of reason that justify a precise CE$\nu$NS experiment in the near future. In particular, in this dissertation the NUCLEUS experiment will be described and several new protocols, from data analysis to detector calibration will be proposed.

\chapter{The NUCLEUS Experiment}
\label{chap:3}

\lettrine[lines=2, findent=3pt, nindent=0pt]{T}{}he experimental goal of the NUCLEUS project is to provide a precise measurement of the cross-section of the coherent elastic neutrino-nucleus scattering (CE$\nu$NS). In order to do so, an extremely sensitive cryogenic detector will be deployed at the nuclear power plant in Chooz (France) to exploit the high $\overline{\nu}_e$ flux from the reactor cores. The NUCLEUS experiment will proceed in two different phases, in the first phase, starting in 2022, a 10~g cryogenic detector will be deployed in order to measure the cross-section with a 10\% uncertainty. The purpose of this phase, referred to as NUCLEUS-10g due to the detector mass, is to give some preliminary results on the value of the cross-section and most importantly to test all the different aspects of the experiment in order to prepare for the next phase. The second and last phase will exploit most of the technologies and methodologies tested in NUCLEUS-10g in order to perform a precise measurement of the CE$\nu$NS cross-section with an uncertainty at the 1\% level. The planned detector for this last phase is similar to the one of NUCLEUS-10g but is scaled up to 1~kg of absorber material giving the name of NUCLEUS-1kg to this second phase. \\

In this chapter the main features of NUCLEUS-10g will be presented. The first part of the chapter will briefly describe the experimental site, then a description of the instrumental apparatus will follow along with a short discussion on the main background components. Most of the considerations made here are valid also for the NUCLEUS-1kg phase but the experimental setup is still not well defined therefore a detailed description of this second phase is premature.

\section{Very Near Site - Chooz}
The NUCLEUS detector will be installed in between the two pressurized water reactors (PWRs) of the Chooz-B nuclear power plant which are operated by \textit{Electricité de France}. The two N4-type nuclear reactors are separated by 160~m and their respective cores are located about 7~m above the Chooz ground level. Each reactor runs at a nominal power of 4.25~GW$_{th}$ and is switched off for refuelling approximately one month per year. Since 2008 the Chooz power plant already hosts the Double Chooz experiment for neutrino oscillations, and due to this several infrastructures for scientific purposes, such as offices, meeting and storage rooms, are already present on site and can be exploited by the NUCLEUS collaboration. \\

\begin{figure}[H]
\centering
\includegraphics[scale=0.5]{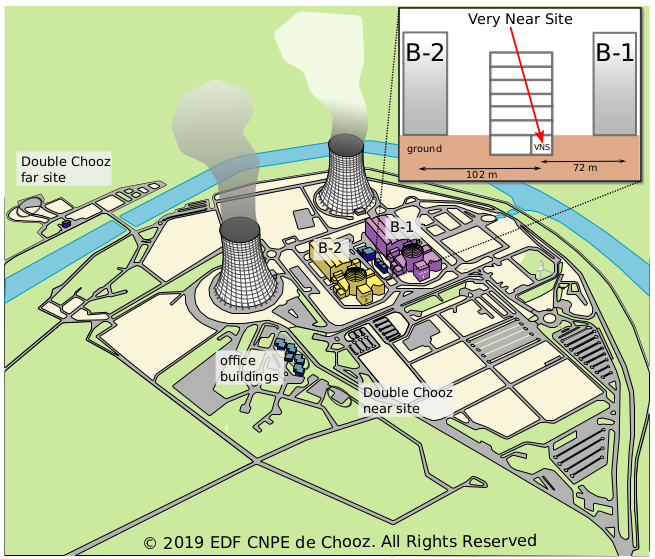}
\caption{Sketch of the Chooz nuclear power plant and of the \textit{Very Near Site} (VNS). The experiment is located at respectively 72~m and 102~m from the reactor cores B1 and B2 \cite{angloher2019exploring}.}
\label{fig:chooz}
\end{figure}

In particular, the deployment site, referred to as \textit{Very Near Site} (VNS), is distant 72~m and 102~m from the two reactor cores respectively (Figure \ref{fig:chooz}). The distance from the reactors cores and the nominal thermal power of operation of the power plant translate to an antineutrino flux of approximately $10^{12}\frac{\overline{\nu}_e}{\text{s cm}^2}$ at the VNS \cite{angloher2019exploring} making all other sources of neutrinos completely negligible. In reactor cores the electron-antineutrinos are produced via the $\beta$-decays of the fission products of $^{235}$U, $^{238}$U, $^{239}$Pu, and $^{241}$Pu with energies below 10~MeV and a mean energy of 1.5~MeV. This energy range is optimal for CE$\nu$NS detection since the neutrinos are expected to be in the fully coherent region for typical nuclear target elements. \\

Logistically speaking, the VNS is a 24~m$^2$ room situated in the basement of a five-story office building at the Chooz power plant. The small size of the room requires careful planning on the volume of the experimental setup, attention must also be payed to the weight of all the instrumentation since it must not exceed the security limit. The VNS is under approximately 3~m.w.e (meters of water equivalent) shielding, thus special care must be taken when addressing the issue of cosmic ray induced background in the detector \cite{angloher2019exploring}.

\section{Detector Overview}
As mentioned in chapter \ref{chap:2}, the nuclear recoil energy produced by neutrinos that undergo coherent elastic scattering is maximum of the order of the tens of keV making CE$\nu$NS detection a challenging task. To observe a wide enough energy range a lower energy threshold of 10~eV is planned for the NUCLEUS-10g phase. In order to detect such faint signals a cryogenic detector is the most reasonable device to be used. In this section the basic working principle of cryogenic devices will be presented along with a general description of the NUCLEUS-10g detector.

\subsection{Cryogenic Detector with Transition-Edge Sensors working principle}
The third law of thermodynamics states:
\begin{displayquote}
\textit{"The entropy of a system approaches a constant value as its temperature approaches absolute zero."}
\end{displayquote}
This principle can be exploited in several ways, one of which is the production of high precision cryogenic detectors. In fact, recalling that a change in entropy is given by:

\begin{equation}
     dS = \dfrac{\delta Q}{T} = C(T)\cdot \dfrac{dT}{T}
\end{equation}

one can immediately notice that, due to the third law of thermodynamics, for temperatures approaching the absolute zero the heat capacity tends to vanish. This behaviour can be easily proven using the Debye model of crystalline solids at low temperatures that estimates a scaling of the heat capacity with the third power of the temperature.  By considering that the energy deposition on a material and a change in temperature are correlated by:
\begin{equation}
    \Delta T = \dfrac{\Delta E}{C}
\end{equation}

if the heat capacity tends to zero, a small energy deposition is then converted into a sizeable change in temperature. This change in temperature is exactly the quantity that needs to be probed and converted into an electric signal in order to build a cryogenic particle detector.

\subsubsection{Transition-Edge Sensors}
Transition-edge sensors, or TES for short, are exactly the device that are up for the task of measuring the change in temperature of a cryogenic bulky object. TESs are composed of a superconducting film that is deposited on the absorber material that acts as the target for the antineutrinos coming form the reactor cores. Once the neutrino undergoes a scattering in the absorber material, the nuclear displacement generates phonons, i.e. a difference in temperature, that are then absorbed by the TES and converted into a difference in the resistance of the sensor. 

\begin{figure}[H]
    \centering
    \includegraphics[scale=0.4]{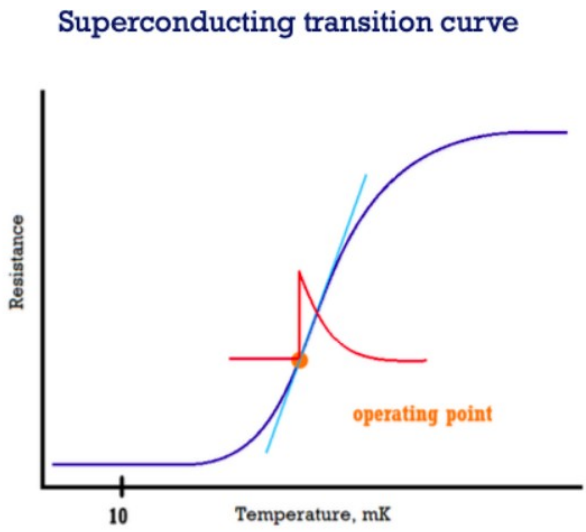}
    \caption{Transition of a superconducting film from nominal to superconducting state. The red point indicated the usual operating point of the TES, the red pulse is an example of a typical nuclear recoil event and in light blue the linear response region of the detector is highlighted. Figure from \cite{bootcamp}.}
    \label{fig:transition_curve}
\end{figure}

 In particular TES owe their name to the way they are operated, in fact when a superconducting film is cooled under the critical temperature it undergoes the superconducting transformation that follows a steep transition curve which brings the resistance of the device to zero (Figure \ref{fig:transition_curve}). TESs are typically operated at the onset of the transition curve, which means that any increase in temperature is translated into a rise of the internal resistance of the sensor allowing it to be used as an extremely sensitive thermometer or phonon detector. \\

\subsubsection{Pulse Generation in TES modules}\label{chap:pulse_shape}
\begin{figure}[H]
    \centering
    \includegraphics[scale=0.45]{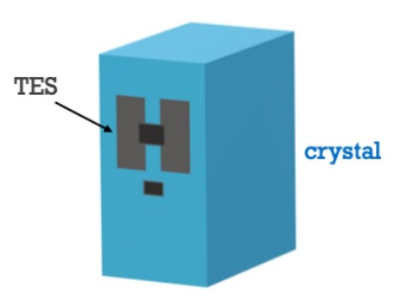}
    \caption{Sketch of the single detector module of NUCLEUS. Figure from \cite{bootcamp}.}
    \label{fig:detector_module}
\end{figure}
As mentioned several times before, the way that a neutrino is detected in the NUCLEUS experiment is via the nuclear scattering that happens in a crystalline cube that is kept at around 10~mK in a cryostat. The nuclear recoil produces phonons that are then absorbed from the TES that is coupled to the absorber crystalline material, the phonon absorption causes a rise in temperature of the transition-edge sensor which is translated into a rise in TES resistance. During the timescale of detection, the nuclear recoil phonons are far more energetic that the thermal phonons from the residual crystal heat, which is why they are referred to as non-thermal phonons. The single detector module consisting of an absorber material cube couple to a TES is shown in Figure \ref{fig:detector_module}.\\

The pulse generated in the TES is caused by a deviation from the equilibrium temperature of the system and the relaxation back to the heat bath temperature. The shape of the recorded event depends on the various heat capacities and thermal couplings, and the temperature rise can be expressed as:

$$\Delta T_e(t)=T_e(t)-T_B$$

where $T_B$ is the heat bath temperature and $T_e(t)$ is the temperature rise due to the nuclear recoil. The temperature rise and decay has the shape of two superimposed pulses with exponential behaviour \cite{Alex_thesis}:

\begin{equation}
\Delta T_e(t) = \Theta(t) \left[A_n\left(e^{-\frac{t}{\tau_{rise}}} - e^{-\frac{t}{\tau_{in}}}\right)+A_t\left(e^{-\frac{t}{\tau_{t}}} - e^{-\frac{t}{\tau_{rise}}}\right)\right]
\end{equation}
the term proportional to $A_n$ is the non-thermal component and is caused by direct absorption of the non-thermal phonons in the thermometer. The term $A_t$, on the other hand, indicates a slower thermal component describing the thermal equalization between crystal and thermometer. The time constants $\tau_{in}$, $\tau_{t}$ and $\tau_{rise}$ are respectively the intrinsic time constant of the thermometer, the thermal relaxation time of the absorber and the time constant of the thermalization of the non-thermal phonons. In particular $\tau_{rise}$ is a function of the thermalization constants of the absorber $\tau_{crystal}$ and the TES film $\tau_{film}$:

$$\frac{1}{\tau_{rise}}=\frac{1}{\tau_{film}}+\frac{1}{\tau_{crystal}}$$
these last two time constants are chosen in order to have the response time dominated by the relaxation time of the absorber material rather than TES module.\\

\begin{figure}[H]
    \centering
    \includegraphics[scale=0.25]{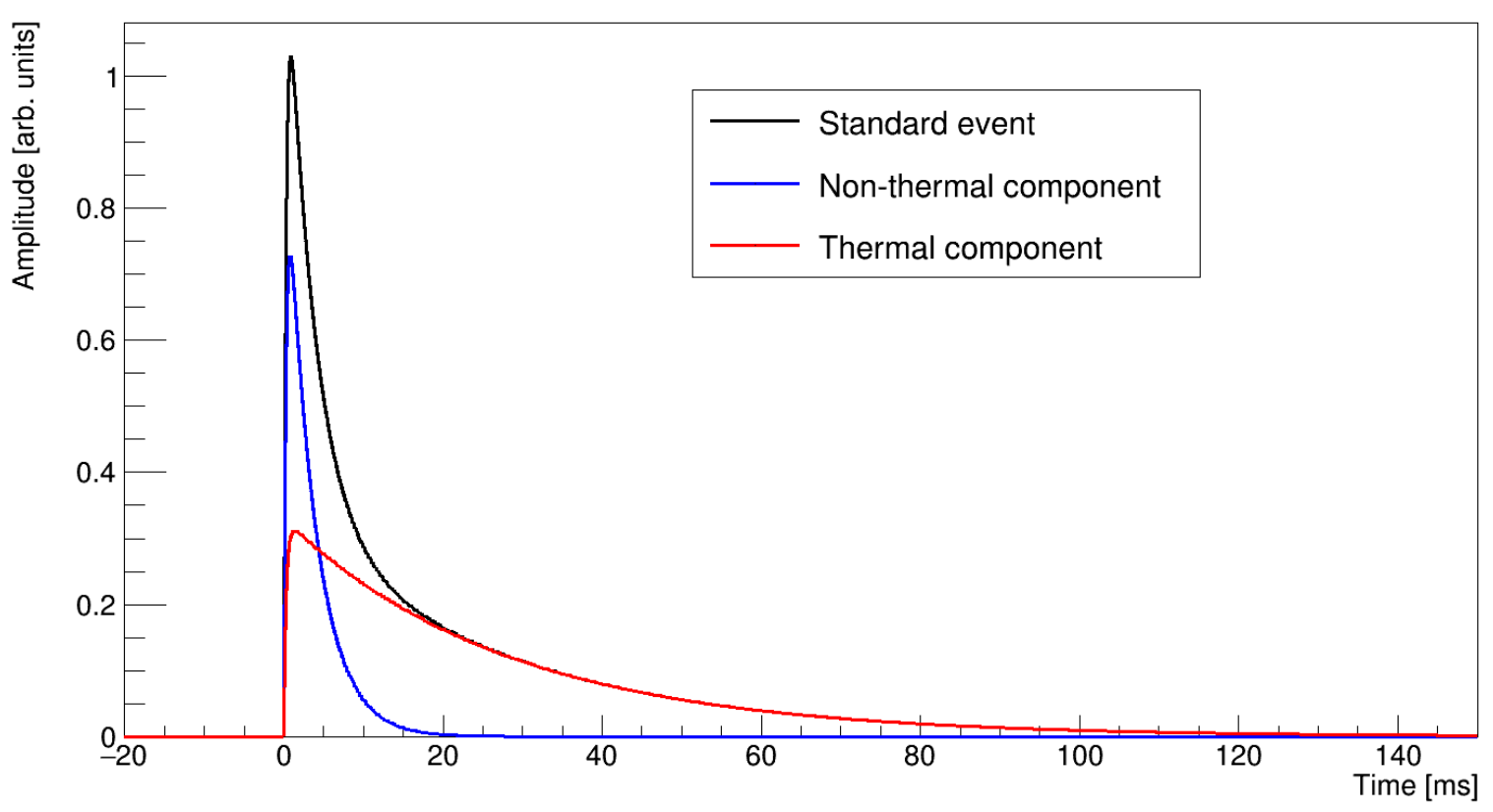}
    \caption{Standard event for a typical NUCLEUS detector module, both the thermal and non-thermal components are presented as well as the total resulting expected signal. Figure from \cite{Alex_thesis}.}
    \label{fig:standard_event}
\end{figure}

The ratio between $\tau_{in}$ and $\tau_{rise}$ controls the operating mode of the detector: if $\tau_{in}\ll\tau_{rise}$ the detector operates in bolometric mode , measuring the non-thermal phonon flux, on the other hand if $\tau_{in}\gg\tau_{rise}$ the detector is in calorimetric mode, where the non-thermal's component amplitude measures the total energy absorbed in the thermometer \cite{Alex_thesis}. The calorimetric mode is the required operation choice for the NUCLEUS experiment. In such operation mode the non-thermal amplitude can be approximated with:
\begin{equation}
A_n\approx -\epsilon\frac{\Delta E}{C_e}
\end{equation}
with $\epsilon$ being the fraction of phonons thermalized in the thermometer, which can be estimated as:

\begin{equation}
\epsilon =\frac{\tau_{crystal}}{\tau_{crystal}+\tau_{film}}
\end{equation}
 and $C_e$ is the heat capacity of the thermometer electron system and scales linearly with temperature following the results of the Sommerfeld electron model in metals. The typical values of the various constants for the NUCLEUS experiment are \cite{Alex_thesis}:
\begin{table}[H]
\centering
\begin{tabular}{cc}
\hline
$\tau_{crystal}$ & 0.34$\sim$ms  \\
$\tau_{film}$    & 2.2$\sim$ms   \\
$\tau_{rise}$    & 0.29$\sim$ms  \\
$\tau_{in}$      & 3.64$\sim$ms  \\
$\tau_{t}$       & 28.17$\sim$ms \\
$C_e$            & 3.9 fJ/K      \\ \hline
\end{tabular}
\end{table}
  which give rise to the so called \textit{standard event} presented in Figure \ref{fig:standard_event}.
  
\subsubsection{Detector Core}

The core of the NUCLEUS detector is composed of a cryogenic array of absorber modules each one coupled with a TES, each module is like the one shown in Figure \ref{fig:detector_module}. The cryogenic setup is hosted inside the commercial LD~400 $^3$He/$^4$He dilution refrigerator produced by Bluefors. The choice of a dry cryostat that uses a pulse tube and gaseous helium in order to reach temperatures of the scale of the tens of mK is due to ease of operation without the requirement of using a complex apparatus needed to handle a liquid refrigerator. A down side of a dry cryostat is the fact that the pulse tube induces vibrations ($\sim 1$~Hz), thus optimal mechanical isolation of the detector core is required.\\

\begin{figure}[H]
\centering
\includegraphics[scale=0.6]{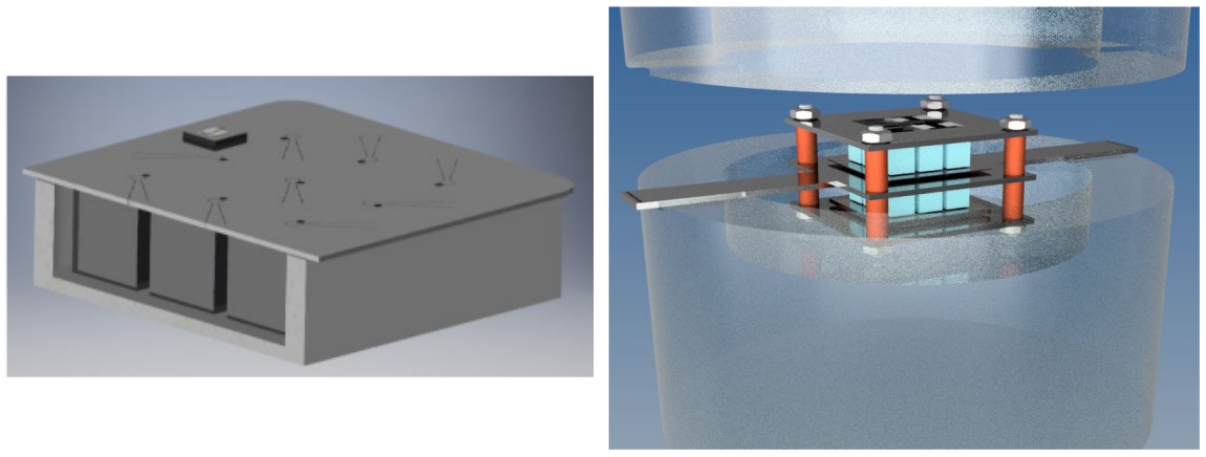}
\caption{Left: sketch of TES array with holder for the NUCLEUS-10g phase (from \cite{mothly_meeting}). Right: Schematic view of the inner setup of the cryodetector for the NUCLEUS-10g phase with 3x3 arrays of CaWO$_4$ and Al$_2$O$_3$ absorbers enclosed in cryogenic veto detectors (from \cite{johannes}).}
\label{fig:TES_ARRAY}
\end{figure}

For the NUCLEUS-10g phase the detector core consists of two different 3x3 arrays of absorber cubes and TESs that are stacked on one another by the means of a supporting structure realized in silicon that also acts as inner veto (Figure \ref{fig:TES_ARRAY}). The mechanical stabilization of each absorber cube is ensured by cutting prongs in the silicon slab. Each prong acts a spring that keeps in place the various detector modules, moreover, in order not to have phonon contamination and dispersion with the mounting structures the prongs' contacts are made to be point-like by carefully etching a small sphere on the prongs' heads.\\

\begin{figure}[H]
\centering
\includegraphics[scale=0.6]{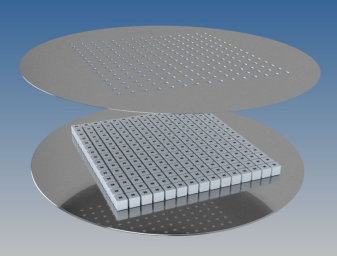}
\caption{Sketch of the planned detector array for the NUCLEUS-1kg phase (from \cite{johannes}).}
\label{fig:nucleus1kg}
\end{figure}

TES modules have resistances of a fraction of an $\Omega$ that need to be measured with a precision under the m$\Omega$ and, due to the heating constraints, the readout currents must be kept at the few tens of $\mu$A. All these constraints make TESs readout quite challenging and feasible only with the use Superconducting Quantum Interference Devices (SQUIDs). A SQUID is essentially an extremely sensitive magnetometer that is magnetically coupled to the TES circuit in order to probe its impedance. Keeping the SQUID in the correct operating conditions is also a difficult but feasible task that must be planned with care. A detailed description of the readout system for NUCLEUS-10g can be found in \cite{johannes}. What this brief discussion serves to highlight is that while using a dedicated SQUID for each absorber module is possible for the NUCLEUS-10g, the up-scaling to the NUCLEUS-1kg array (Figure \ref{fig:nucleus1kg}) is not as smooth as it may seem because TES multiplexing is necessary and further readout techniques need to be developed.\\

One last detail worth mentioning in this paragraph is that the two cryogenic arrays present in the NUCLEUS-10g detector in Figure \ref{fig:TES_ARRAY} are made of different absorber materials, one array will be made of CaWO$_4$ and the other of Al$_2$O$_3$ because of background and CE$\nu$NS decoupling reasons discussed in section\ref{sec:background} and \cite{angloher2019exploring}.

\section{Main Background Sources and Shielding Techniques}\label{sec:background}
NUCLEUS aims to achieve a background count rate of $10^2\frac{\text{counts}}{\text{keV kg d}}$ in the sub-keV detection region. Due to the shallow overburden of the VNS location and the extreme sensibility of the NUCLEUS detector, the background sources must be thoroughly analysed. In this section the main results of the background characterization presented in \cite{angloher2019exploring} are described in order to underline some of the issues that the NUCLEUS experiment has to overcome.\\

A wide variety of radioactive processes can deposit energy in the target detectors and mimic the neutrino signal. Radiation from the reactors, in particular reactor-correlated $\gamma$ and neutron backgrounds, prove to be already negligible thanks to the reactor containment and the tens of meters of soil and concrete that separate the reactor cores from the detector. A second source of background events is due to natural radioactivity, in fact the surrounding environment at the VNS can emit all kinds of radiation, mainly $\gamma$ radiation is the dangerous component since $\alpha$ and $\beta$ particles can be easily shielded. The ambient $\gamma$-rays can be shielded with a high-Z material, such as lead, and the remaining unshielded fraction interacts mostly via multiple Compton scattering inside the detector array generating signals in multiple TES modules and can thus be easily isolated.\\

The $\alpha$ and $\beta$ particles produced by environmental radiation are only a problem when they are generated inside the cryogenic setup. The internal contamination with radioactive isotopes can be controlled using cleanliness procedures in the detector production and assembly. In addition, the inner veto is specifically designed to tag small energy depositions due to decaying nucleides. This particular type of background is mostly due to $^{222}$Rn contamination during the assembly of the inner sections of the detector.\\

\begin{figure}[H]
\centering
\includegraphics[scale=0.5]{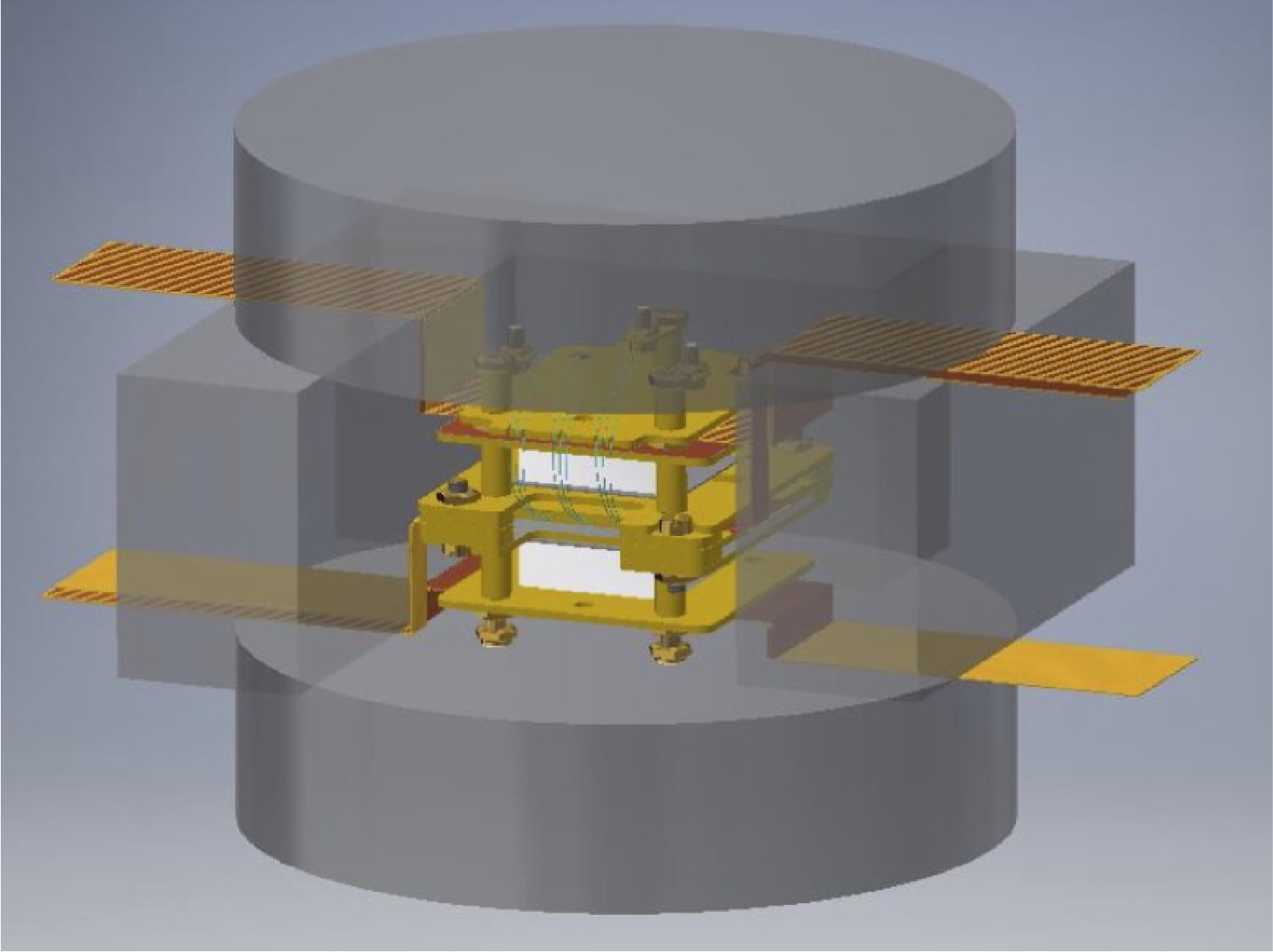}
\caption{Sketch of the cryogenic setup of the NUCLEUS detector. The grey components compose the cryogenic outer veto, while the yellow parts are the cryogenic inner veto. The white spacings between the inner veto planes are the locations in which the TES plus absorber modules will be placed \cite{magnicenns}.  }
\label{fig:vetos}
\end{figure}

In order to further reduce all the ambient radiation, the detector presents both inner and outer cryogenic vetos which consist of properly shaped silicon coupled with TESs (see Figure \ref{fig:vetos}). These vetos will be mostly use to tag background events due to surface contamination. The inner veto will also be used as a support structure for the detector modules. The inner veto, being an active cryogenic veto, can also probe any mechanical stress that the modules undergo that could produce a phonon signal similar to CE$\nu$NS. The designs of such inner and outer veto are still being perfected from the NUCLEUS collaboration.

\begin{figure}[H]
\centering
\includegraphics[scale=0.5]{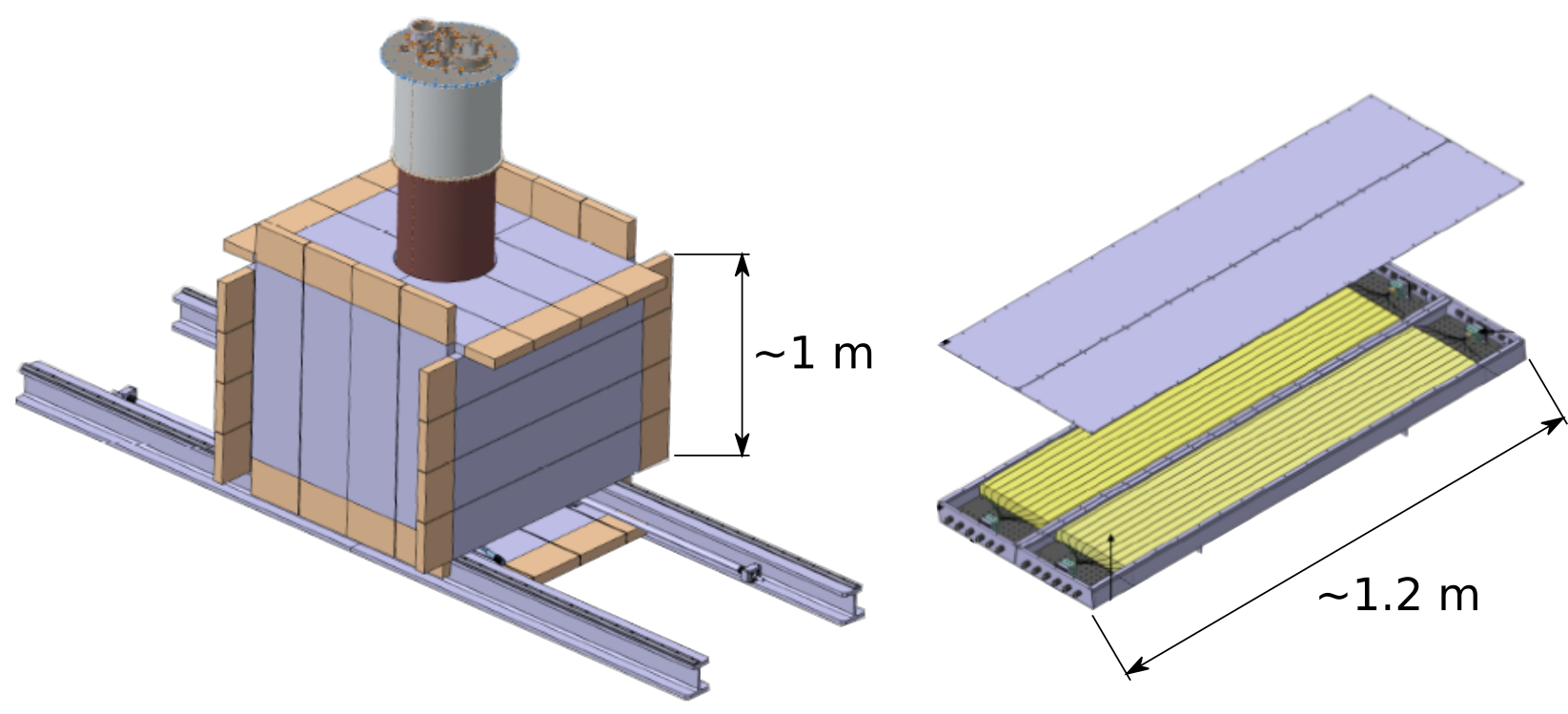}
\caption{Sketch of the external muon veto. Left Panel: Overall sketch of the muon veto (rectangular grey panels) surrounding the cryostat. Right Panel: Detail of the single muon veto panel, in grey the support structure, in yellow the 5 cm plastic scintillator slabs with SiPM and wavelength shifter fiber readout. Sketches from \cite{magnicenns}.  }
\label{fig:muon_veto}
\end{figure}

\begin{figure}[H]
\centering
\includegraphics[scale=0.5]{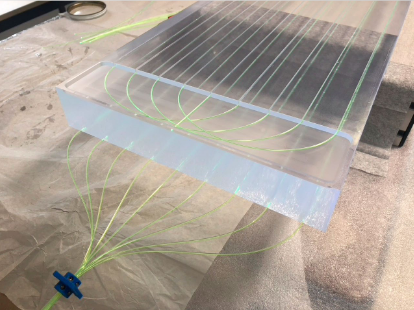}
\caption{Picture of a plastic scintillator slab of the muon veto with the WLS-fiber readout. Picture from \cite{magnicenns}.  }
\label{fig:muon_veto_photo}
\end{figure}

The cosmic-ray induced backgrounds are the most worrisome ones since the VNS presents little overburden ($\mathcal{O}(3~\text{m.w.e.})$). Such shielding, in fact, can efficiently stop protons, pions and electrons produced in extensive air showers (EAS) due to cosmic rays, but proves insufficient when one considers the muon and neutron content of said showers. Muons, in particular, are the most penetrating charged particle species and are going to be dealt using a dedicated anticoincidence veto showed in Figures \ref{fig:muon_veto} and \ref{fig:muon_veto_photo}. Since the muon flux at ground level is approximately 100~$\frac{\text{Hz}}{\text{m}^2}$ particular attention must be paid to the detector response time in order not to have an excessive dead time during the data taking.\\

\begin{figure}[H]
\centering
\includegraphics[scale=0.5]{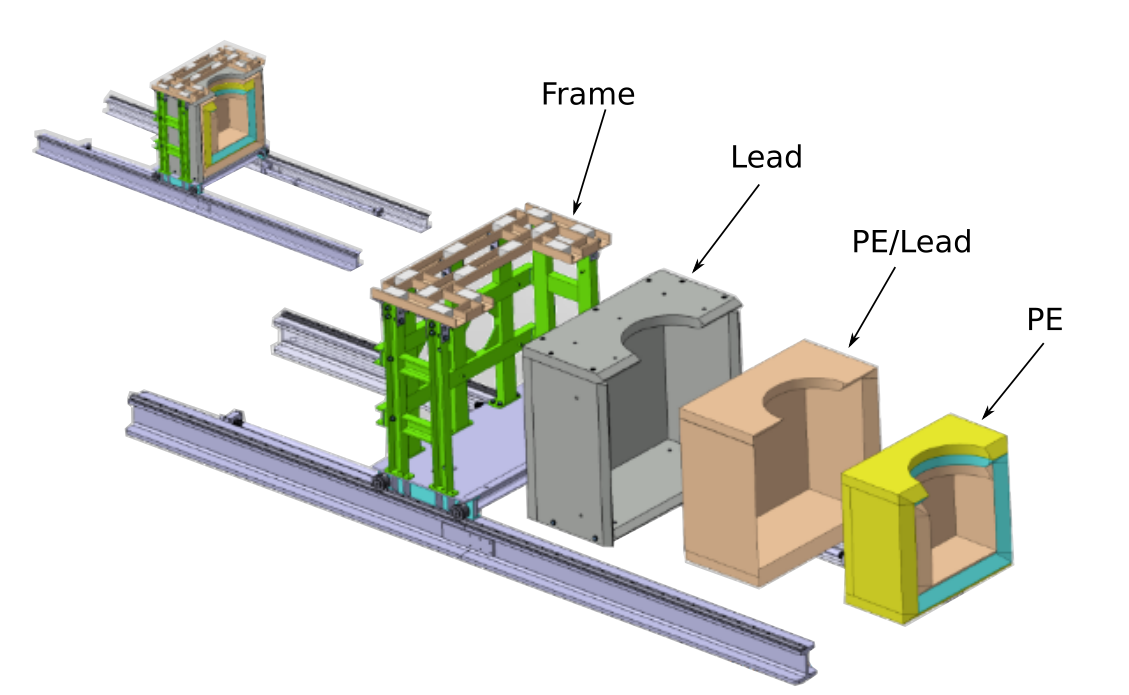}
\caption{Sketch of the various passive shield layers. Picture from \cite{magnicenns}.  }
\label{fig:passive_shield}
\end{figure}

Neutrons, both of atmospheric origin and muon induced, prove to be a particularly dangerous background for CE$\nu$NS detection since the experimental signature of a low energy interacting neutron is a nuclear recoil. In order to lower this background component two different strategies are used, the first one consists of using two different absorber material CaWO$_4$ and Al$_2$O$_3$.  In fact, in CaWO$_4$ CE$\nu$NS is strongly enhanced by the presence of a higher number of neutrons, on the other hand, since the background due to neutron scattering happens mostly on the oxygen nuclei the two materials are expected to have a similar level background counts, thus allowing for a disentanglement between these two components. The second strategy used is specific for lowering the background due to neutrons induced by muons. The neutron production mostly happens in high-Z materials, i.e. the lead shielding, for this reason the most internal layer in the passive shielding will be made of polyethylene (PE) which can efficiently stop the neutrons produced in the lead. The design of the passive shield is presented in Figure \ref{fig:passive_shield}.\\

\section{Overall View}
Summarizing, the NUCLEUS experiment will be operated at the \textit{Very Near Site} (VNS) at the Chooz power plant in France. The detector, distant several tens of meters from the reactor cores, is made from a cryogenic array of absorber crystals that act as target for the antineutrinos emitted during the nuclear reactor processes. The nuclear recoil in the target material is then read through the use of transition-edge sensors.\\

\begin{figure}[H]
\centering
\includegraphics[scale=0.5]{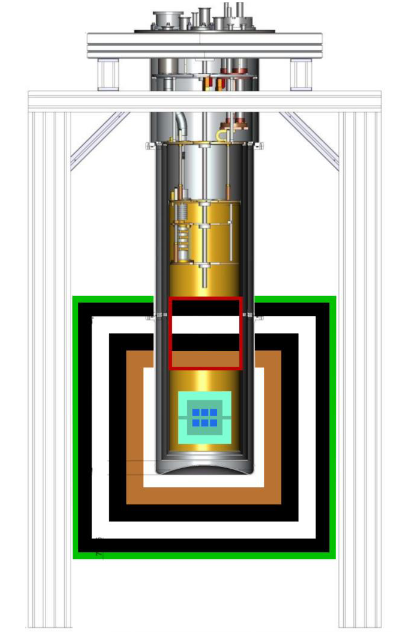}
\caption{Sketch of cryogenic setup for the NUCLEUS experiment from \cite{johannes}. In the central part there is a $29$~cm diameter experimental space for hosting the TES arrays of both the NUCLEUS-10g and the NUCLEUS-1kg phases. Surrounding the setup there is an active muon veto (green). An additional passive shielding and muon veto (red) is present between the mixing chamber of the cryostat and the detector. }
\label{fig:nucleus_detector}
\end{figure}

\begin{figure}[H]
\centering
\includegraphics[scale=0.38]{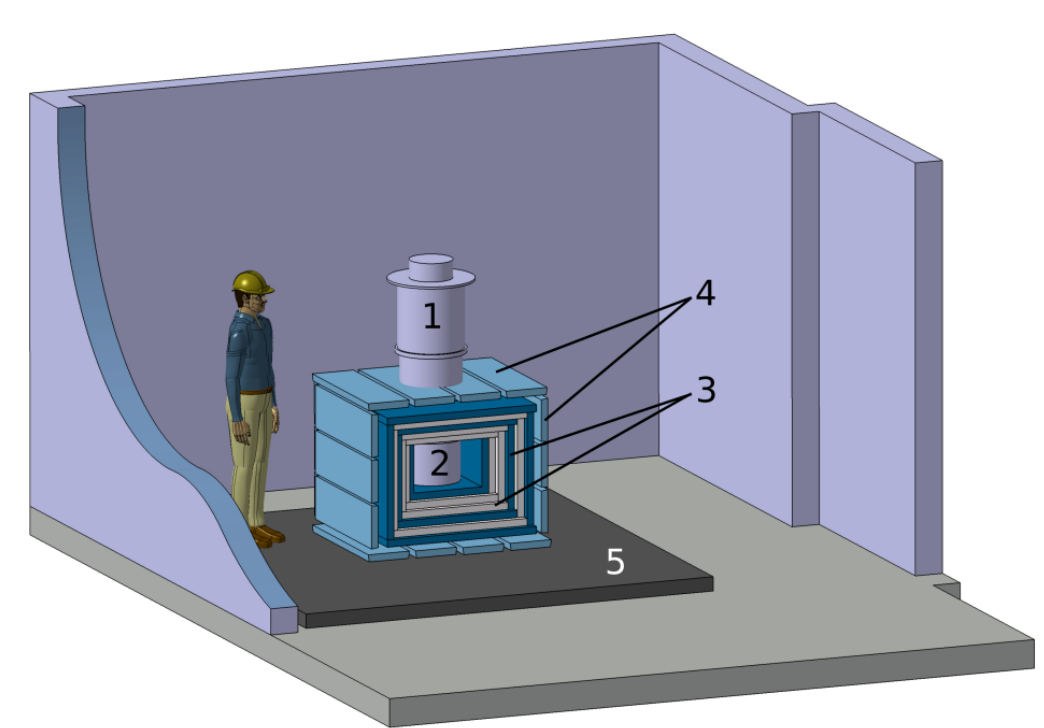}
\caption{Sketch of the VNS room at Chooz with the NUCLEUS detector, the various parts of the picture are correctly scaled. The installation consists of the cryostat (1), the experimental volume (2), the passive shielding (3), the active muon-veto (4) and a weight support platform (5). Picture taken from \cite{angloher2019exploring}.}
\label{fig:vns}
\end{figure}

Apart from the target material, the NUCLEUS detector presents several layers of active and passive shielding in order to lower the background counts coming from both ambient and cosmic ray induced radiation. In particular, some of the active vetos used are cryogenic and also serve as support structure for the absorber modules. A sketch of the various components of the detector is presented in Figure \ref{fig:nucleus_detector} and a correctly scaled sketch of the VNS with the detector is presented in Figure \ref{fig:vns}.

\begin{figure}[H]
\centering
\includegraphics[scale=0.5]{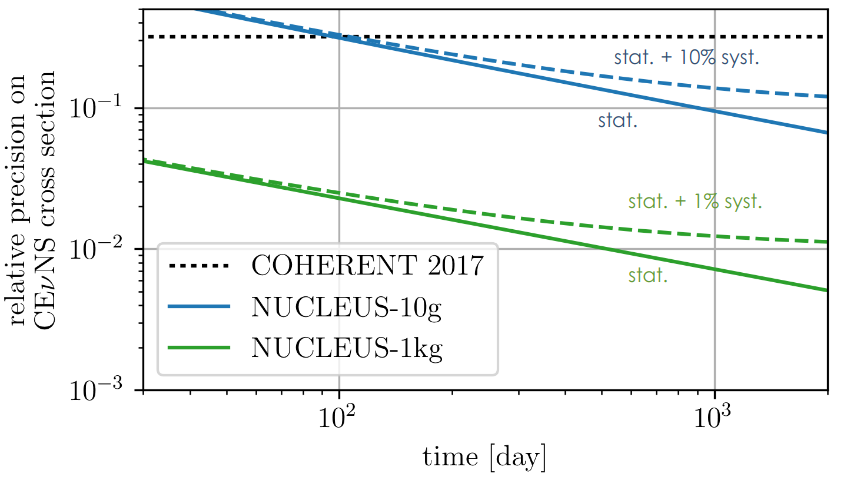}
\caption{Worst case scenario for CE$\nu$NS detection in both NUCLEUS phases \cite{johannes}. }
\label{fig:phases}
\end{figure}

With this setup the NUCLEUS collaboration expects to measure the CE$\nu$NS cross-section with a 10\% uncertainty in less than a year of data taking, depending on the background present at the VNS, using a 10~g neutrino detector (the smallest neutrino detector so far). In the next phase, NUCLEUS-1kg, the collaboration wants to upscale the detector to 1~kg of material maintaining more or less the same technology used for the first phase. In this way, with a data acquisition run lasting more or less like the one of the first phase, the CE$\nu$NS cross-section will be measured with a 1\% error. In Figure \ref{fig:phases} a preliminary study on the expected error on the cross-section is presented for both NUCLEUS phases, this study corresponds to the worst case scenario of an extremely high uncertainty of the background model which reflects in a longer period of acquisition time ($\mathcal{O}$(1~year)) in order to reach the desired uncertainties.

\chapter{NUCLEUS detector optical calibration}
\label{chap:4}
\lettrine[lines=2, findent=3pt, nindent=0pt]{A}{} central problem in low energy experiments such as NUCLEUS is the detector calibration, using standard calibration sources like $^{55}$Fe proves difficult and prone to mistakes or excessive approximations, as described in chapter \ref{chap:5}. In order to overcome these issues and guarantee maximum energy resolution a newly developed calibration procedure custom made for NUCLEUS is described in this chapter. This new setup allows to perform direct characterization of the detector performances down to extremely low energies, giving a more precise description of the cryodetector response.

\section{Absolute Calibration}\label{sec:absolute_cal}

The \textbf{absolute calibration} is a type of detector calibration that exploits the Poisson statistics of photon counting. It consists of shining with nearly monochromatic electromagnetic radiation the absorber material of the cryodetector depositing a variable number of photons, such quantity follows Poisson statistics. The power of the absolute calibration consists in being able of characterizing the detector response without requiring the knowledge of the exact number of photons delivered, which makes it experimentally simple to perform.\\

When a certain number of monochromatic photons $N_{ph}$ is delivered on a detector in a period that lasts less than the reaction time of the device, the various energy depositions are integrated and a single signal that scales with the number of delivered photons is produced. Since the measurement of the amplitude of the produced signal is basically a counting experiment, it will follow poissonian statistics. Assuming a high number of delivered photons, the distribution of the energy deposition can be approximated with a gaussian of mean $\mu \propto N_{ph}$ and standard deviation $\sigma\propto\sqrt{N_{ph}}$. If one defines the responsivity of the detector as the measured signal for each deposited photon:

\begin{equation}
r=\dfrac{\mu}{N_{ph}} \qquad \left[\frac{\text{mV}}{\text{photon}}\right]
\end{equation}

the standard deviation of the signal amplitude distribution can be written as:
\begin{equation}\label{eq:sigma}
\sigma = \sqrt{N_{ph}}\cdot r = \sqrt{\dfrac{\mu}{r}}\cdot r = \sqrt{\mu \cdot r}  \qquad [\text{mV}]
\end{equation}

\begin{figure}[H]
\centering
\includegraphics[scale=0.43]{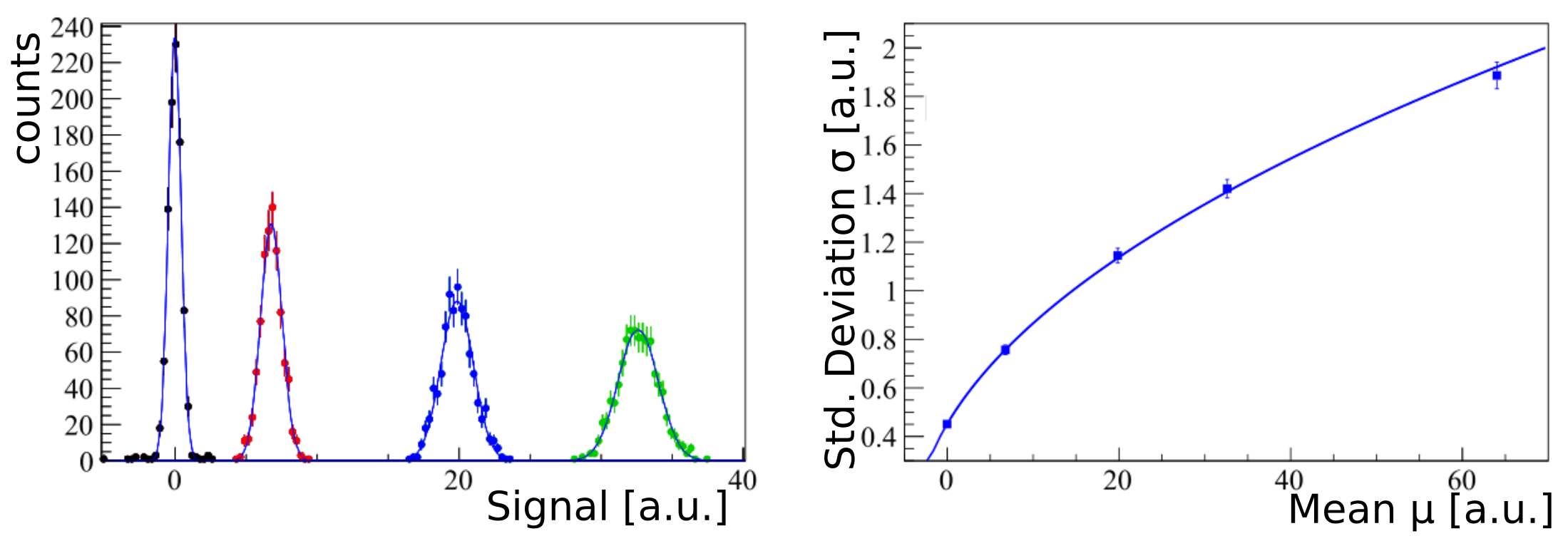}
\caption{Right: Sequence of signal amplitude distributions for different energy depositions (re-adapted from \cite{absolute_calibration}). Left: Absolute calibration experimental distribution in the $\sigma-\mu$ plane (re-adapted from \cite{absolute_calibration}).}
\label{fig:absolute_cal_gaussians}
\end{figure}

By changing the energy deposited on the detector, thus measuring multiple distributions (Figure \ref{fig:absolute_cal_gaussians}), one can plot the standard deviations and the averages of the various deposition on the $\sigma-\mu$ plane as shown in Figure \ref{fig:absolute_cal_gaussians}. 

Considering equation \ref{eq:sigma}, these points can then be fitted with the function:
\begin{equation}
\sigma = \sqrt{\sigma_0^2 + r\cdot \mu}
\end{equation}
where $\sigma_0$, the baseline resolution, and $r$, the responsivity mentioned above, are the fitting parameters. This formula assumes that the poissonian term from the photon counting and the baseline resolution are independent terms. With this fit it is then possible to extract the responsivity of the detector and its baseline resolution thus characterizing the whole response of the bolometer.\\

From the definition of responsivity one has that $r$ is in $\frac{mV}{\text{photon}}$, these units are not particularly experimentally friendly and require the knowledge of the amount of photons delivered on the detector and their wavelength in order to quantify the measured energy. By dividing $r$ with the energy of the single photon used for the calibration this setback is fixed and one can measure the energy responsivity of the detector in \textit{signal amplitude} over \textit{energy deposited} (in the NUCLEUS case $\frac{mV}{eV}$).\\

In \cite{absolute_calibration} the whole process of the absolute calibration is described and used for the characterization of Kinetic Inductance Detectors (KIDs) in the framework of the CALDER collaboration \cite{calder}. In the article, the absolute calibration is compared with a more classical calibration made using the x-rays from $^{55}$Fe decay and the results appear to be in good compatibility with one another thus proving the potential of this new type of calibration.\\

The advantage of using the absolute calibration is the fact that one can actually probe the response of the detector for low energy events without having to infer it from more high energy calibration sources as $^{55}$Fe which can saturate the cryodetector dynamical range (see section \ref{sec:truncated_fit} for an example). Besides saturating the detector's response or being in energy regions outside the desired one, using radioactive sources negates all those advantages that come with using an electronics based calibration system, such as real time monitoring of the detector's status and being able to label the calibration acquisitions with an appropriate trigger type.

\section{Calibration Electronics}\label{sec:cal_electr}
In order to perform the absolute calibration on every TES plus absorber module of the NUCLEUS detector, additional new electronics was acquired. The main components are listed and described below.\\

\paragraph{Light Source :} 

The most important element of the calibration setup is the light source. This element must have some simple but mandatory features in order to be usable for the absolute calibration efficiently. The first and most important one is that it must have a way of choosing the amount of light to be shed in a period of time shorter than the reaction time of the detector. Another important property is that the light must be fairly monochromatic in order to have a precise energy calibration. One last property is that it must be triggerable with electronics in order to be integrated in a wider and automatized calibration setup.\\
\begin{figure}[H]
\centering
\includegraphics[scale=0.5]{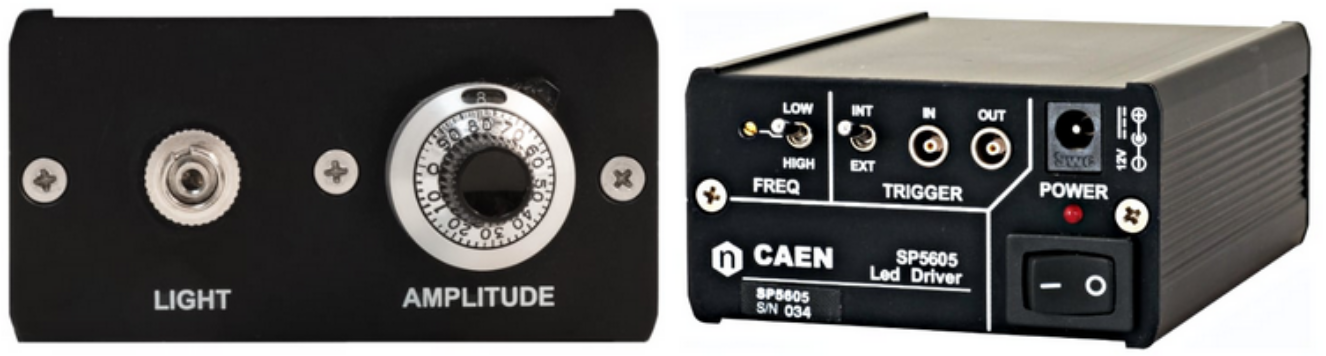}
\caption{\textbf{UV LED Driver SP5605, Caen} - light source for the absolute calibration of the NUCLEUS detector.}
\label{fig:led_driver}
\end{figure}
All these features point to the use of the \textit{UV LED Driver SP5605} from \textit{Caen} \cite{led_driver} show in Figure \ref{fig:led_driver}. The LED driver has an output of $(248\pm 8)~nm$ ($(5.00\pm 0.08)$~eV) photons and a minimum time interval between consecutive light emissions of approximately $3.3\mu s$ ($\approx 1/300~kHz^{-1}$) which is several orders of magnitude quicker than the rise time of cryogenic bolometers which is of the order of the microsecond.\\

As it can seen from Figure \ref{fig:led_driver}, the \textit{LED driver} also presents a gear with which the intensity of light output can be regulated manually. This gear is meant to be regulated once during the detector commissioning (for this dissertation it was set to the maximum value). Later the output intensity is regulated using a trigger signal sent to the back panel of the device (see \ref{sec:trigger}).

\begin{figure}[H]
\centering
\includegraphics[scale=0.3]{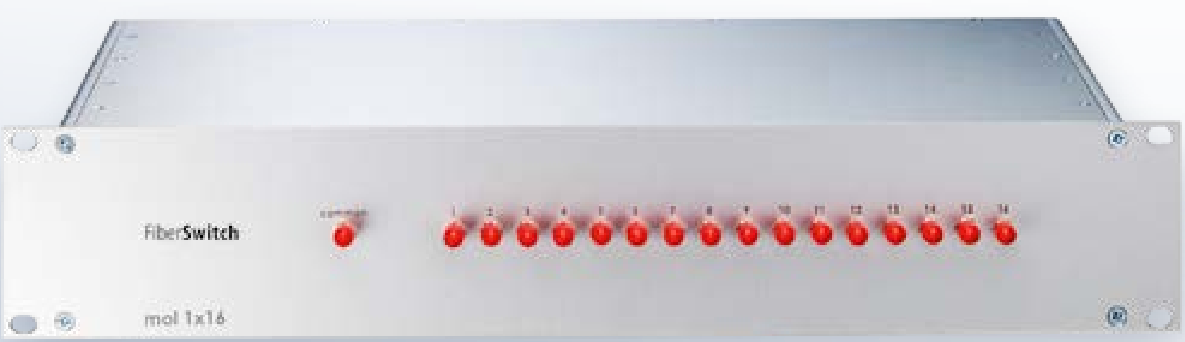}
\caption{\textbf{Fiber Optical Switch mol1x16, Leoni}.}
\label{fig:fiber_switch}
\end{figure}

\paragraph{Fiber Optical Switch :} As mentioned above the core of the NUCLEUS detector is made of multiple modules and each one of them must be separately calibrated. In order to have the light from the \textit{LED driver} reach all the parts of the detector multiple light guides must enter the cryostat and be pointed at the different absorber modules. Using multiple optical fibers with the \textit{LED driver} described above is not possible unless an additional device is used. Such device is the \textit{Fiber Optical Switch mol~1x16} from \textit{Leoni} \cite{fiber_switch} and is shown in Figure \ref{fig:fiber_switch}.

This device is used to connect one input optical fiber to another optical channel in output that can be chosen by the user via serial communication. The \textit{fiber switch} thus allows to connect the optical fiber coming from the \textit{LED driver} to one of the multiple output fibers (up to 16 different light guides) that are pointed to the various sectors of the cryodetector.\\

This device is fairly simple to operate but some care must be taken on the timing of the output channel change. In fact, a typical delay time for the a channel change is about $6$~ms and the switching status must be maintained for at least $40~\mu s$ to avoid problems. Thus the \textit{fiber switch} is not suitable for extreme high speed usage but it is suitable for the calibration of cryodetectors that have reaction times of $\mathcal{O}(1~ms-10~ms)$.

\begin{figure}[H]
\centering
\includegraphics[scale=0.5]{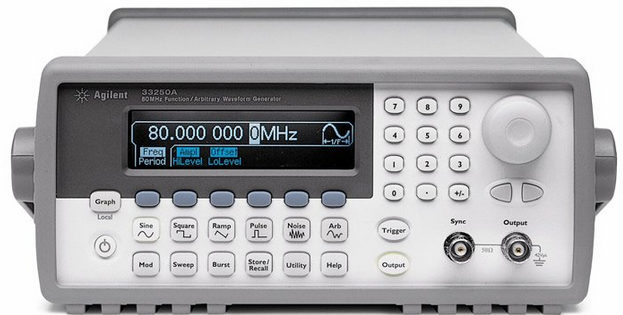}
\caption{\textbf{Agilent Technologies 33250A} signal generator.}
\label{fig:33250A}
\end{figure}

\begin{figure}[H]
\centering
\includegraphics[scale=0.25]{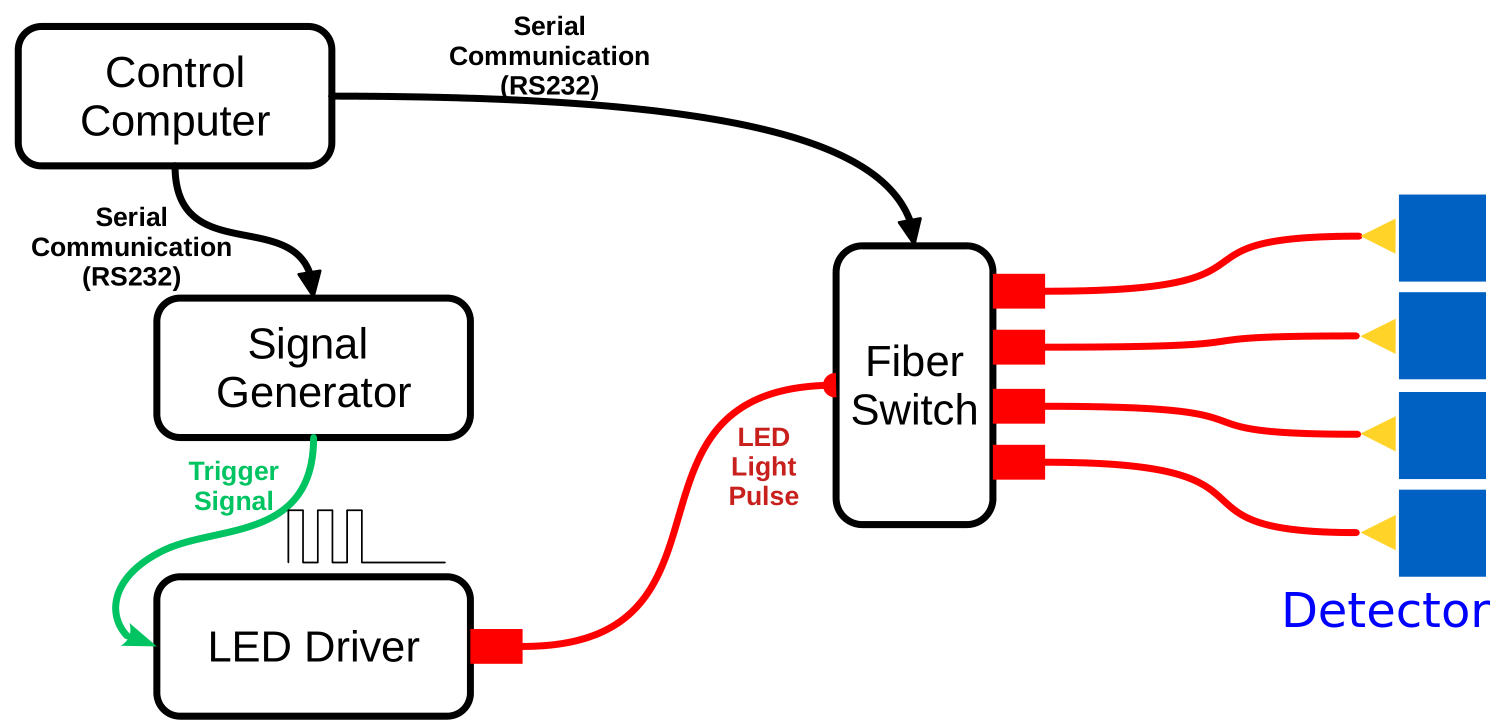}
\caption{Sketch of the laboratory setup of the calibration electronics.}
\label{fig:cal_setup}
\end{figure}

\paragraph{Signal Generator :} The \textit{LED driver} described above needs a TTL trigger signal to be activated. This signal can be generated using an appropriate signal generator like \textit{33250A} from \textit{Agilent Technologies} \cite{33250A}. This signal generator, shown in Figure \ref{fig:33250A}, has several remote connections that can be used to pilot it via computer, in particular the serial connection will be extensively used in the following sections of this dissertation.
\paragraph{Calibration Electronics Setup}
As mentioned above, both the \textit{signal generator} and the \textit{fiber switch} can be remotely piloted via serial communication. This feature was then exploited in order to build the setup shown in Figure \ref{fig:cal_setup}, where a dedicated computer pilots the two devices simultaneously.

This setup was designed to perform various types of studies on the cryodetector and in particular it allows to perform the absolute calibration fully automatically. 
\subsubsection{Triggering Signal of LED Driver}\label{sec:trigger}
Part of the study for the calibration electronics setup was devolved to the identification of a trigger pulse for the \textit{LED driver} suitable with both the desired timing and energy deposition requirements.\\

In order to have the energy regulation needed for the absolute calibration the approach of \textbf{pulse-width modulation} (\textbf{PWM}) was followed. In fact, if one shines the cryodetector with a certain amount of light burst in a period of time lasting much less than the reaction time of the detector the contributions of the single light bursts gets integrated and thus appears (to the detector) like a single pulse with higher energy. This approach then allows to vary the energy deposited on the cryodetector by simply varying the amount of short light pulses. This type of energy regulation then requires to have a very fast commutation time for the production of the light bursts in order to have the highest possible flexibility when choosing the amount of deposited energy.\\

The triggering signal for the \textit{LED driver}, in order to produce a variable energy deposition, must be composed of a train of short-lived square waves (because the \textit{LED driver} follows the transistor-transistor logic) that produce several brief light bursts that are integrated by the detector. The single short-lived square wave must have a period equal or greater than $3.3~\mu s$ since this is the minimum dead time of \textit{LED driver} between one light pulse and the next. The last requirement is that between one train of triggers and the next, so between one integrated energy deposition and the next, enough time must elapse for the signal to fully develop and decay in the detector. This dead time should also be long enough to perform changes on the settings of the electronics, like changing the output channel of the fiber switch and the number of single light bursts.

\begin{figure}[H]
\centering
\includegraphics[scale=0.33]{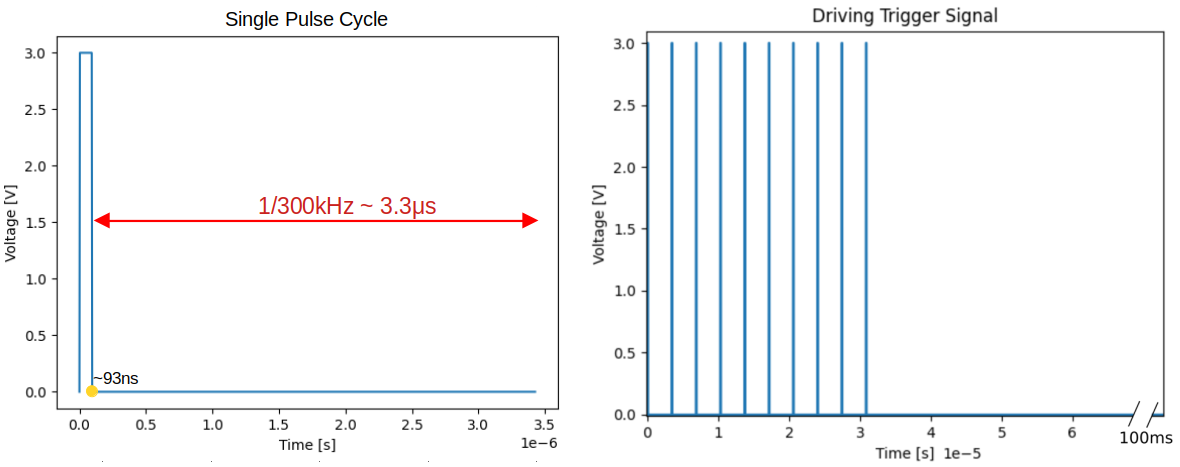}
\caption{Sketch of the trigger signal for the \textit{LED driver}. On the left a single square wave period for the triggering of a single light burst is shown. On the right one period of the total triggering signal is shown, it is comprised of a train of square waves like the one presented in the left panel and a long dead time.}
\label{fig:trigger_signal}
\end{figure}

All these requirements point to the triggering signal showed in Figure \ref{fig:trigger_signal}. In the right panel of the figure the train of square waves is shown, the overall signal has a period of $100$~ms which is more than enough time for the signal to develop in the detector and to make any changes in the electronics' settings. In the left panel the trigger of the single light burst is presented, it consists of a square wave of frequency $\sim 290~kHz$ (because of the \textit{LED driver} dead time requirement) and duty cycle of $\sim 2.3\%$. The value of the duty cycle was chosen for the trigger signal to be compatible also with the $400~nm$ \textit{SP5601 LED Driver Caen} \cite{SP5601} that was also tested and is used for detector characterization in the framework of the BULLKID \cite{bullkid} experiment (a very similar electronics setup is used in this experiment as well).\\

The $\sim 2.3\%$ duty cycle is not a requirement of the NUCLEUS experiment since the \textit{UV LED driver} is \textit{up edge triggered} and will be relaxed in the final experimental setup to an up-time of $\mathcal{O}(1~\mu s)$. This, in fact, allows the replacement of the \textit{signal generator} with a NUCLEUS custom-made data acquisition (DAQ) board that can thus flag all the events produced from the \textit{LED driver} and can perform the calibration periodically throughout the experimental runs.

\section{Calibration Software}
One additional part of the calibration setup is the software for piloting the different devices remotely. This software has been newly developed and is currently being used by the BULLKID R$\&$D experiment \cite{bullkid} for detector characterization.
\\

The developed software is written in Python 3 (v 3.6.12 and above) and makes use of standard Python libraries such as \textit{Numpy}\cite{numpy}, \textit{PySerial}\cite{pyserial}, \textit{Pandas}\cite{pandas}, \textit{TQDM} \cite{tqdm}. One additional package used is \textit{Python-usbtmc} \cite{usbtmc}, that is not as standard as the previous ones but is used in order to make the code compatible with the new \textbf{USBTMC} serial communication protocol \cite{usbtmc_protocol} that is employed by newer models of \textit{Agilent} signal generators, replacing the more classic \textit{RS232} port.\\

\begin{figure}[H]
\centering
\includegraphics[scale=0.33]{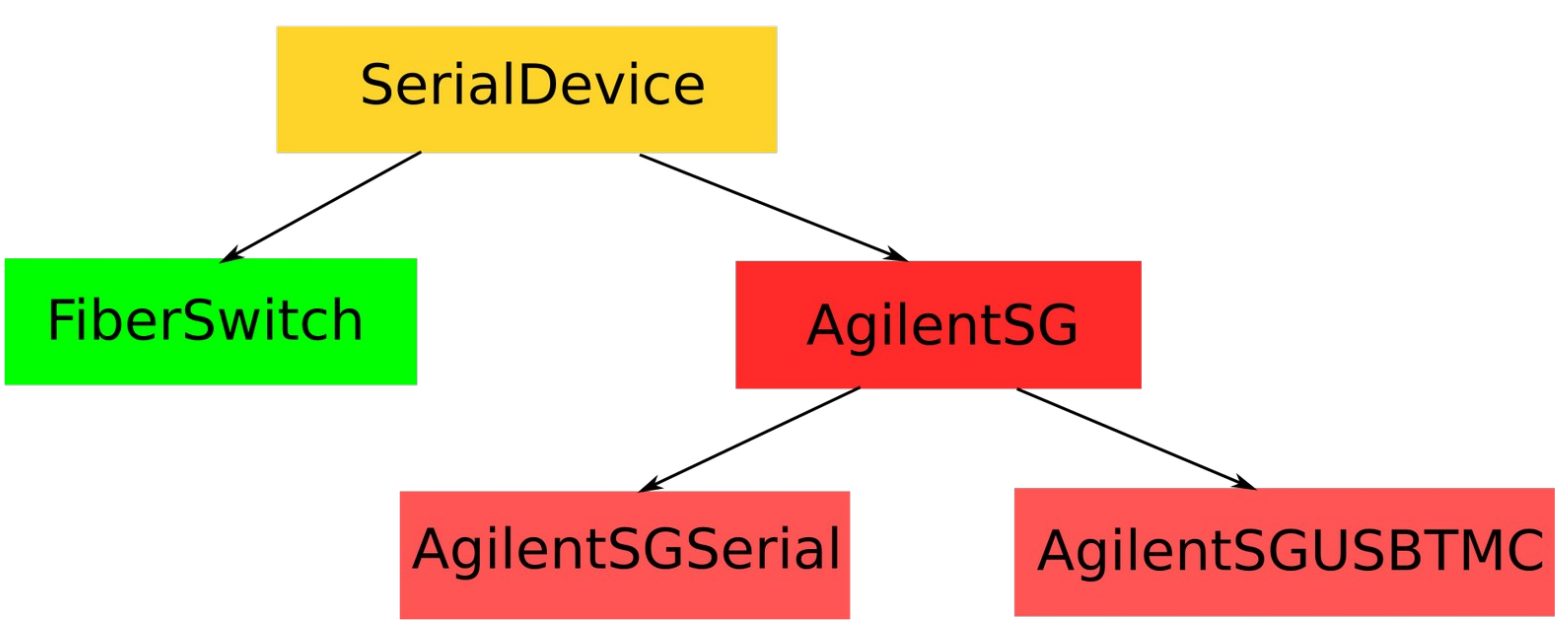}
\caption{Schematic view of the various classes in the Python software toolkit developed for the calibration electronics.}
\label{fig:software_scheme}
\end{figure}

This software is a toolkit comprised of several python classes that take care of the communication with the different devices used for the absolute calibration, namely the \textit{fiber switch} and the \textit{signal generator}. The central class around which the code revolves is \textit{SerialDevice}, this class contains the generic methods for serial input/output and status control of the device. The rest of the toolkit is divided into two main branches, one dedicated to \textit{fiber optical switches} and the other to \textit{Agilent signal generators}. As it can be seen from Figure \ref{fig:software_scheme}, the code is built in a modular fashion and more devices can be easily added in the future possibly making it useful for several experiments and research collaborations. Moreover the choice of building Python classes and not directly a custom made program performing the absolute calibration is due to the fact that these classes can be later integrated into a wider data acquisition software making, for example, the detector calibration completely automatized. \\

Further discussion of the inner workings of the classes is out of the scope of this dissertation but everything is explained in detail in the documentation that is distributed along with the code.

\subsubsection{Graphic User Interface}\label{sec:GUI}

\begin{figure}[H]
\centering
\hspace*{-0cm}
\includegraphics[scale=0.3]{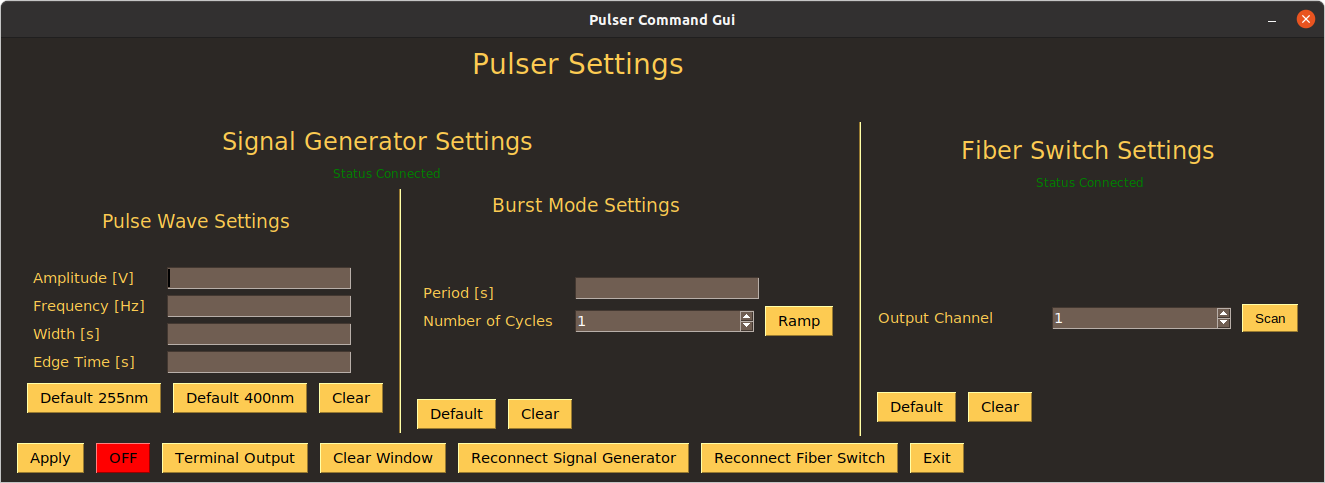}
\caption{Graphic User Interface for piloting the calibration electronics.}
\label{fig:GUI}
\end{figure}

Besides the different classes for single device control, a graphic user interface (GUI) was also developed and has been tested by the BULLKID experiment \cite{bullkid}. The main idea behind this GUI, shown in Figure \ref{fig:GUI}, is to provide a user friendly experience for the experimenter while making smart use of the properties of the classes and devices presented above to fully exploit their potential.\\

\begin{figure}[H]
\centering
\includegraphics[scale=0.33]{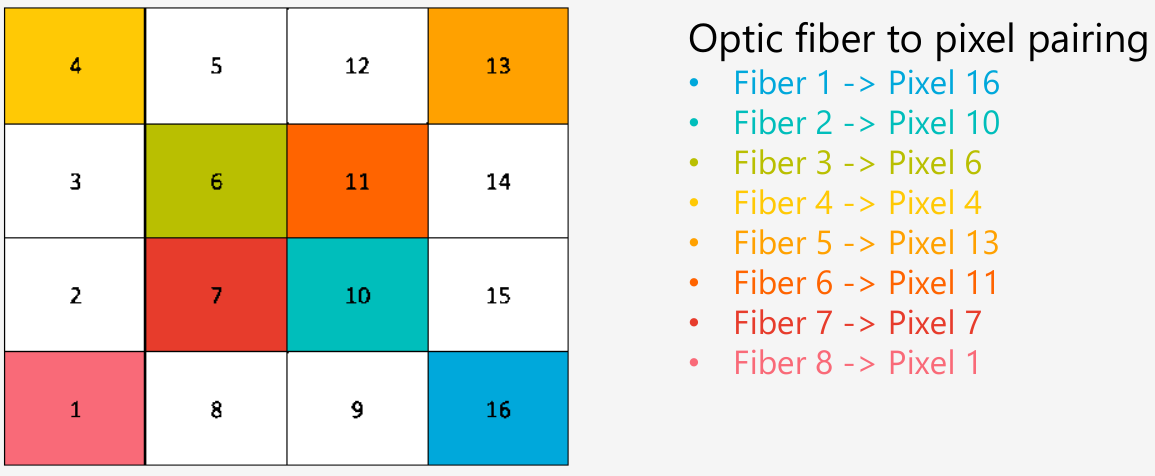}
\caption{Setup of the optical fibers with respect to the pixels of the KID array. Left: kid array with the pixels numbered from 1 to 16. Right: List of the optical fibers and their position on the KID array. In both panels each color is specific for a different optical fiber(Figure from \cite{daniele}).}
\label{fig:kid_setup}
\end{figure}

The data presented in this section was taken with an array of kinetic inductance detectors (KIDs) deposited on a silicon wafer used as energy absorber, during an acquisition run for the detector characterization of the BULLKID experiment \cite{bullkid}. Since KIDs have a very delicate calibration procedure and discussion of such procedure is outside the scope of this dissertation, only the raw data from the data acquisition will be used for the following plots. The cryodetector array, as shown in Figure \ref{fig:kid_setup}, is composed of 16 pixels and 8 different optical fibers coming from the \textit{fiber switch} were used. The optical fibers are displayed in a cross configuration over the array in order to reach every sector of the detector, the correspondence of the fiber id number with the pixel index is presented in the panel next to the figure.

\begin{figure}[H]
\centering
\includegraphics[scale=0.6]{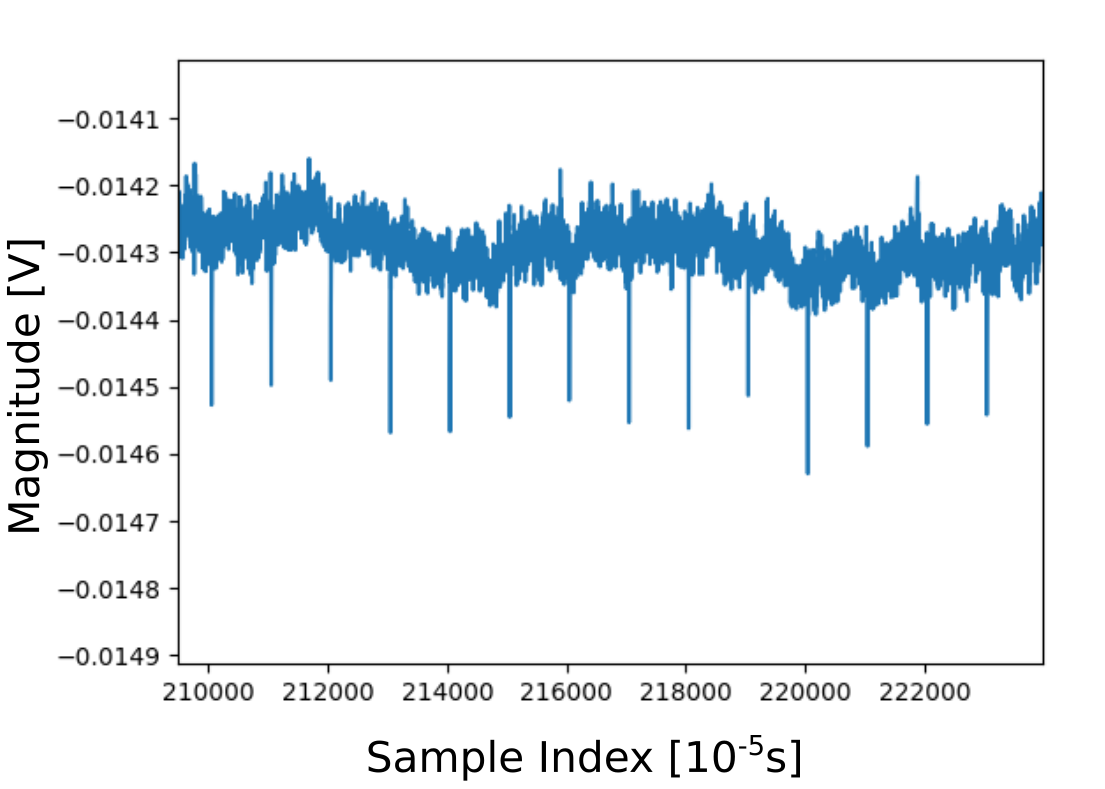}
\caption{KID response to the basic usage of the GUI. The settings used are the one discussed in section \ref{sec:trigger} with 50 light pulses per energy deposition.}
\label{fig:basic}
\end{figure}

\paragraph{Basic Usage:} The most simple way to use the GUI and the Python toolkit is to deliver always the same amount of energy on the same KID pixel. An example of this type of operation is shown in Figure \ref{fig:basic} where the response of KID is shown. It is clear that, besides a slight modulation in the baseline the different spikes in the detected energy are more or less of the same amplitude (keep in mind that this is the raw data from the detector and in order to be usable in a quantitative way it needs to undergo a series transformations typical of KID analysis).\\

This mode obviously does not make use of any of the properties of the \textit{fiber switch} nor the ability of the signal generator to quickly change its settings. In this way the absolute calibration can be performed with a much simpler electronics, only \textit{LED driver} and \textit{signal generator} are required, but with the necessity of constant user intervention.

\begin{figure}[H]
\centering
\includegraphics[scale=0.6]{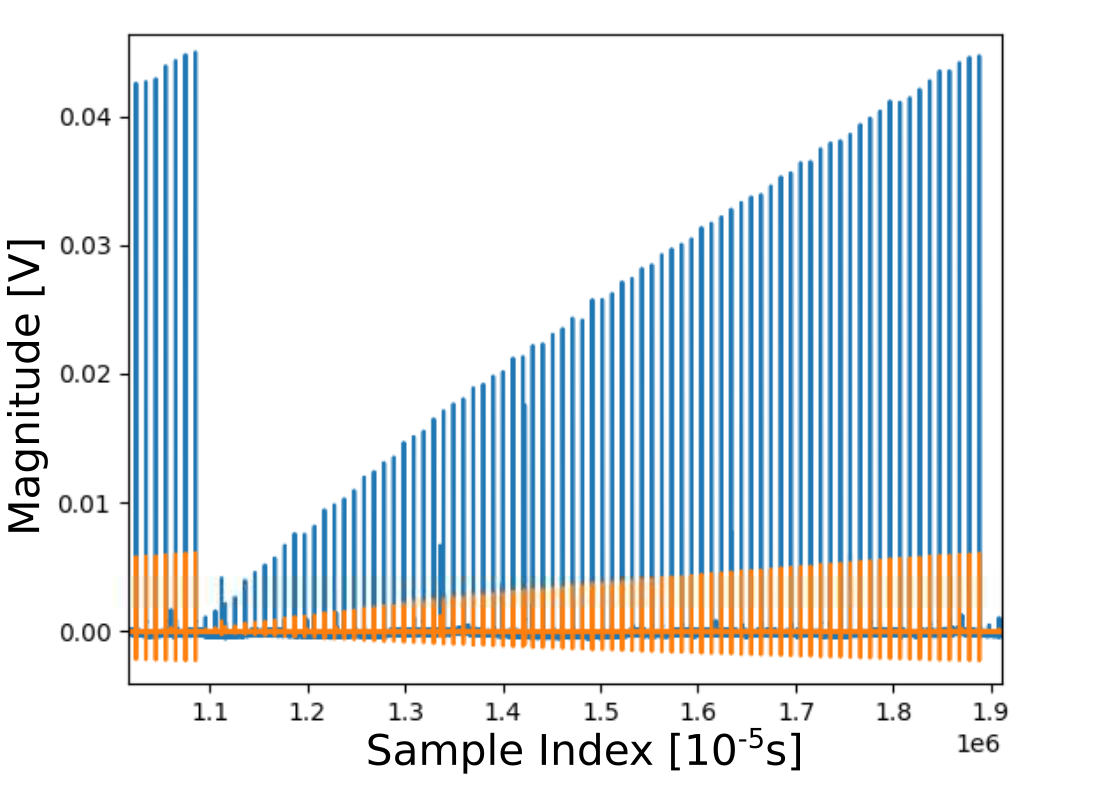}
\caption{KID response to the energy ramp usage mode of the GUI. The settings used are the one discussed in section \ref{sec:trigger}. In blue the raw data from the KID pixel is presented while in yellow the trigger of the data acquisition (DAQ) is shown (the triggering algorithm is similar to the one presented in \ref{sec:peak_identifier}).}
\label{fig:energy_ramp}
\end{figure}

\paragraph{Energy Ramp:}  A more advanced feature that can be used to automatically perform energy scans on a single optical channel. As mentioned before this can be done exploiting the long dead time between one energy deposition and the next to increase the number of light pulses by changing the number of trigger pulses in the signal generator's settings. Increasing the number of trigger pulses in this way produces an increasing energy deposition, referred to as energy ramp. Once the maximum number of light pulses, which is set by the user, is reach the energy ramp starts again from the beginning. By letting the loop go over several ramps one can quickly acquire enough statistics in order to perform the absolute calibration of the selected pixel with minimal user intervention. An example of the resulting signal magnitude increase with this operation mode is presented in Figure \ref{fig:energy_ramp} where the linear increase of the deposited energy can be easily seen, on the left and right edges of the plot the final and initial depositions of the neighboring energy ramps are noticeable as well.

\paragraph{Fiber Loop:} An additional advance feature implemented in the GUI is the \textit{fiber loop}. This feature allows to have a user defined number of energy depositions on multiple optical channels (selected by the user) in a sequential manner. This function exploits the features of the \textit{fiber switch} to probe multiple sections of the detector one after the other. As for the \textbf{energy ramp}, the change of the optical channel is done during the dead time between one energy deposition and the next.\\

\begin{figure}[H]
\centering
\includegraphics[scale=0.45]{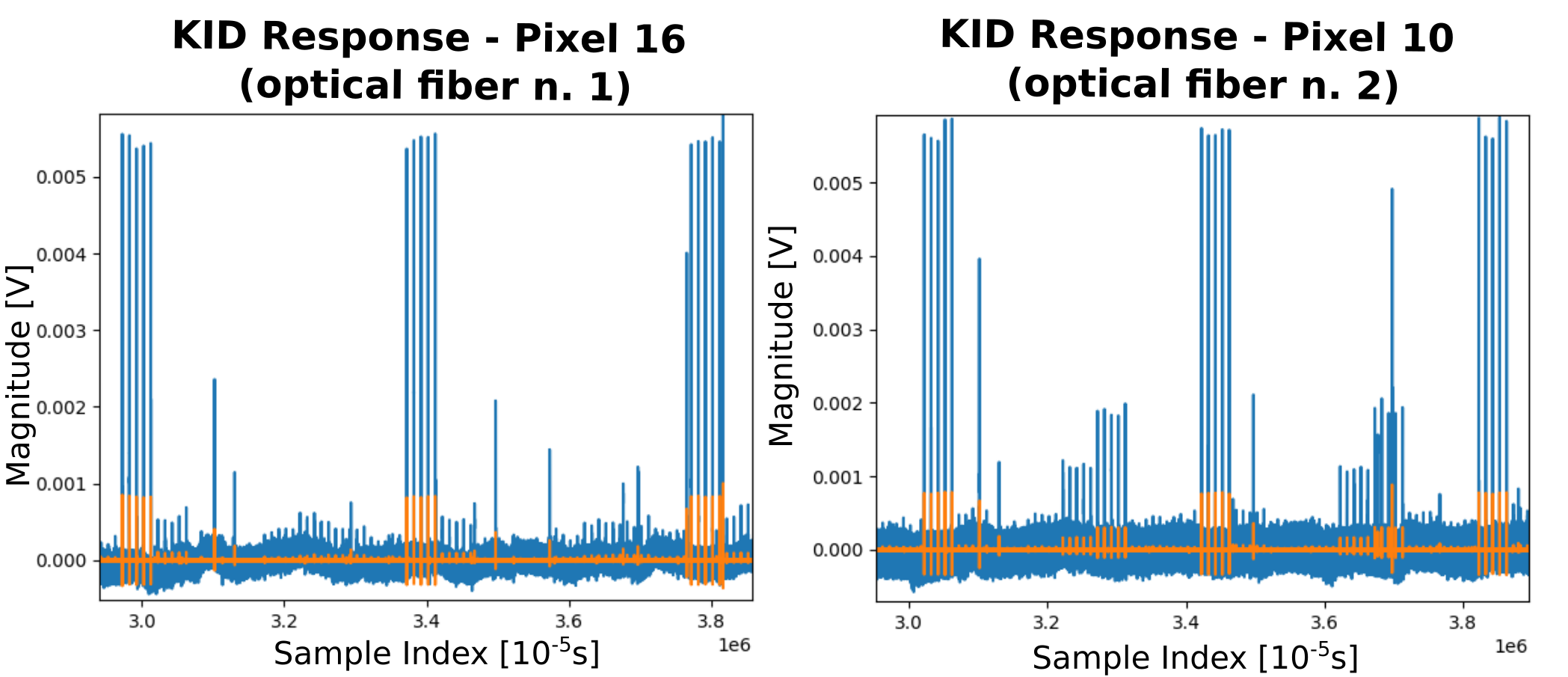}
\caption{KID response to the fiber loop usage mode of the GUI. The settings used are the one discussed in section \ref{sec:trigger}. Left: response recorder from pixel 16 corresponding to optical fiber n. 1. Right: data acquired from pixel 10 which is placed right under optical fiber n. 2. In both plots the color blue is used for the raw data from the KID pixel while in yellow the trigger of the DAQ is shown (the triggering algorithm is similar to the one presented in \ref{sec:peak_identifier}).}
\label{fig:fiber_loop}
\end{figure}

The results produced by this way of operating the electronics are shown in Figure \ref{fig:fiber_loop} where the responses of two neighbouring KID pixels (pixel 10, optical fiber n. 2, and pixel 16, optical fiber n. 1) are presented. From the two plots it can be seen how the first train of energy depositions in fiber 1 is immediately followed by the depositions in fiber 2, this can be both noticed by looking at the sample indexes or by looking at the plot corresponding to the first optical fiber and noticing that after the first 5 energy depositions another 5 impulses with lower magnitude follow. This lower amplitude signals are due to the cross-talk between the two KID pixels.

This function thus allows to probe sequentially different pixels of an array of cryodetectors, and can be thus used for verifying the correct status of the different sections or to perform cross-talk studies.
\paragraph{Fiber Loop Energy Ramp:}One last feature that was later added to the GUI is the \textbf{fiber loop energy ramp}. As the name suggests it is a combination of the two previous operation modes and allows to perform energy scans of the various optical channels automatically. This, in particular, is the feature that allows to perform automatically all the data taking to later perform the absolute calibration on the whole cryodetector array. The use of this operation mode will then allow for minimal user intervention and time consumption when performing the absolute calibration of the NUCLEUS detector, which is an important feature considering the difficult access to the NUCLEUS experimental site (VNS, see chapter \ref{chap:3}).\\

This mode was added after the end of the acquisition run of the previously presented data and thus no plot of the response of the KID array is available. The proper functioning of such mode was verified by inspecting the status of the electronics throughout its usage. Proper testing of this function and further testing of the rest of the toolkit will be performed in the BULLKID data acquisition runs in the end of 2021 and beginning of 2022.

\section{Germanium Mirror Wafer : first prototype characterization}
As previously shown in Figure \ref{fig:TES_ARRAY}, the NUCLEUS detector is composed of multiple modules made of a TES deposited on a cube of absorber material. Each of these modules must be thoroughly calibrated using the procedure described in section \ref{sec:absolute_cal} which requires each absorber cube to be reached by the light emitted from the \textit{LED driver} (see section \ref{sec:cal_electr}). There are several possible setups to perform the calibration, the first that comes to mind is having a dedicated optical channel from the \textit{fiber optical  switch} for each module but this solution has two major disadvantages, the first one is due to the limited output channels of the \textit{fiber switch}, in fact, the detector designed for the NUCLEUS-10g phase would already need more channels than the ones available and this is even more true for the NUCLEUS-1kg phase. A second important disadvantage is the thermal isolation of the various stages of the cryostat, in fact, any connection between the various temperature stages of a cryostat worsens the thermal isolation and makes the cooling and the operation of TES arrays more difficult. \\

\begin{figure}[h]
\centering
\includegraphics[scale=0.3]{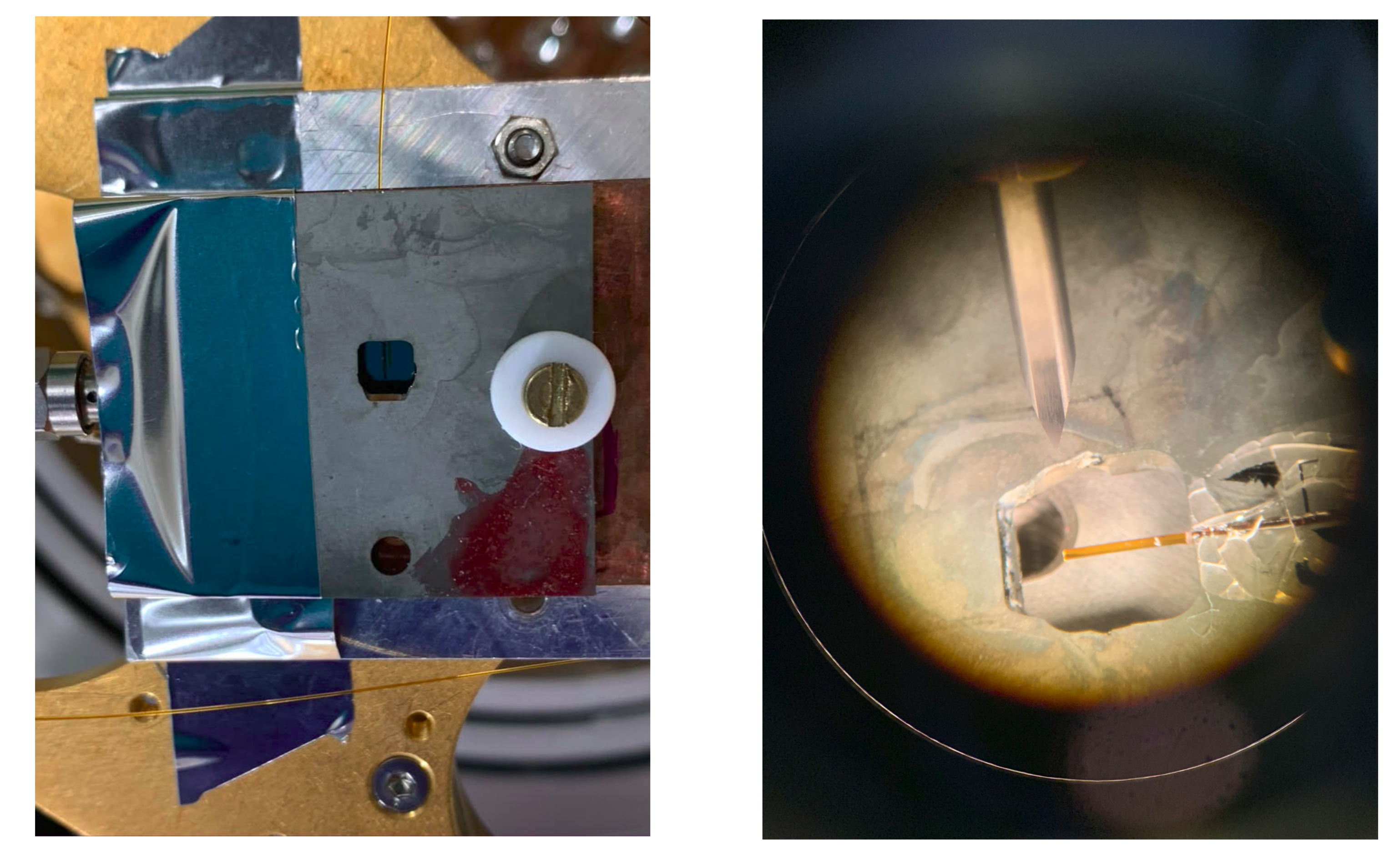}
\caption{Photos of the mirror wafer prototype. Left: mirror wafer mounted on top of the KID. Right: mirror wafer's light splitting section (picture taken with the help of a microscope). }
\label{fig:mirror_wafer}
\end{figure}

To overcome the issues that using a large number of optical fibers causes a new device was designed for the NUCLEUS experiment, the \textit{mirror wafer}. This device, shown in Figure \ref{fig:mirror_wafer}, serves as a light splitter allowing to simultaneously shine light on two different detector modules with one single optical fiber. The prototype shown and characterized in this section is made out of a germanium wafer and has an incoming optical fiber aligned with a prism-shaped cut that splits the light from the \textit{LED driver} in two different rays.

\begin{figure}[h]
\centering
\includegraphics[scale=0.43]{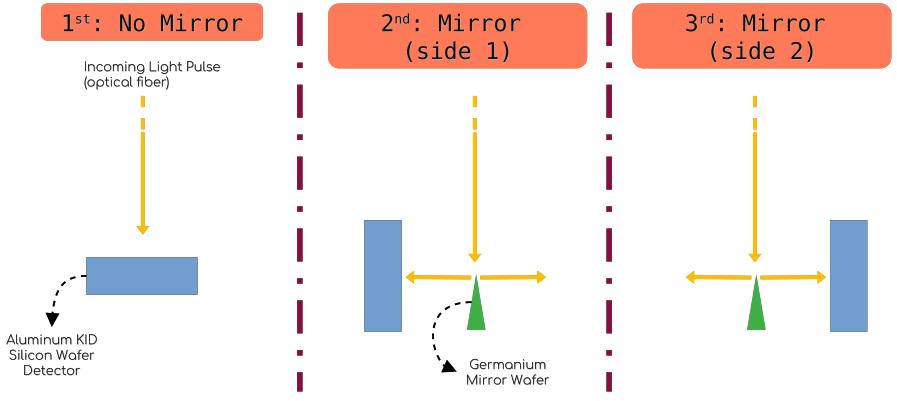}
\caption{Schematic description of the three steps performed to characterize the mirror wafer. }
\label{fig:mirror_wafer_steps}
\end{figure}

The characterization of the mirror wafer prototype was performed using the absolute calibration with the aid of an aluminum KID deposited on top of a silicon wafer used as light absorber. The setup for the calibration is similar to the one described in section \ref{sec:cal_electr} with the only variation being that two different \textit{LED drivers} where used, the UV one \cite{led_driver} and a $400~nm$ one \cite{SP5601}. Moreover since only one optical fiber was used the \textit{fiber switch} was not needed for the following measurements. The quantification of the light splitting abilities of the mirror wafer was performed in three different steps summarised in Figure \ref{fig:mirror_wafer_steps}, and consist of:
\begin{itemize}
\item \textbf{Step 1:} Characterization of the KID used without the mirror wafer.
\item \textbf{Step 2:} Probing the characteristics of one side of the mirror wafer.
\item \textbf{Step 3:} Probing the other side of the mirror wafer to see the light splitting abilities of the device.
\end{itemize} 

The idea of this analysis is to perform the calibration after each step and compare the energy depositions on the detector with respect to the number of light pulses (see section \ref{sec:trigger}) used to build up the signal in the cryodetector. In order to do so the calibration electronics was used following the \textit{basic mode} described in section \ref{sec:GUI} because the number of light pulses has to be known for each energy deposition. Moreover, as mentioned, two different wavelengths were used ($\sim 255~nm$ and $\sim 400~nm$) to probe different energy ranges and to verify the behaviour of the light splitter at different light frequencies.\\

\begin{figure}[H]
\hspace*{-2.5cm}
\includegraphics[scale=0.6]{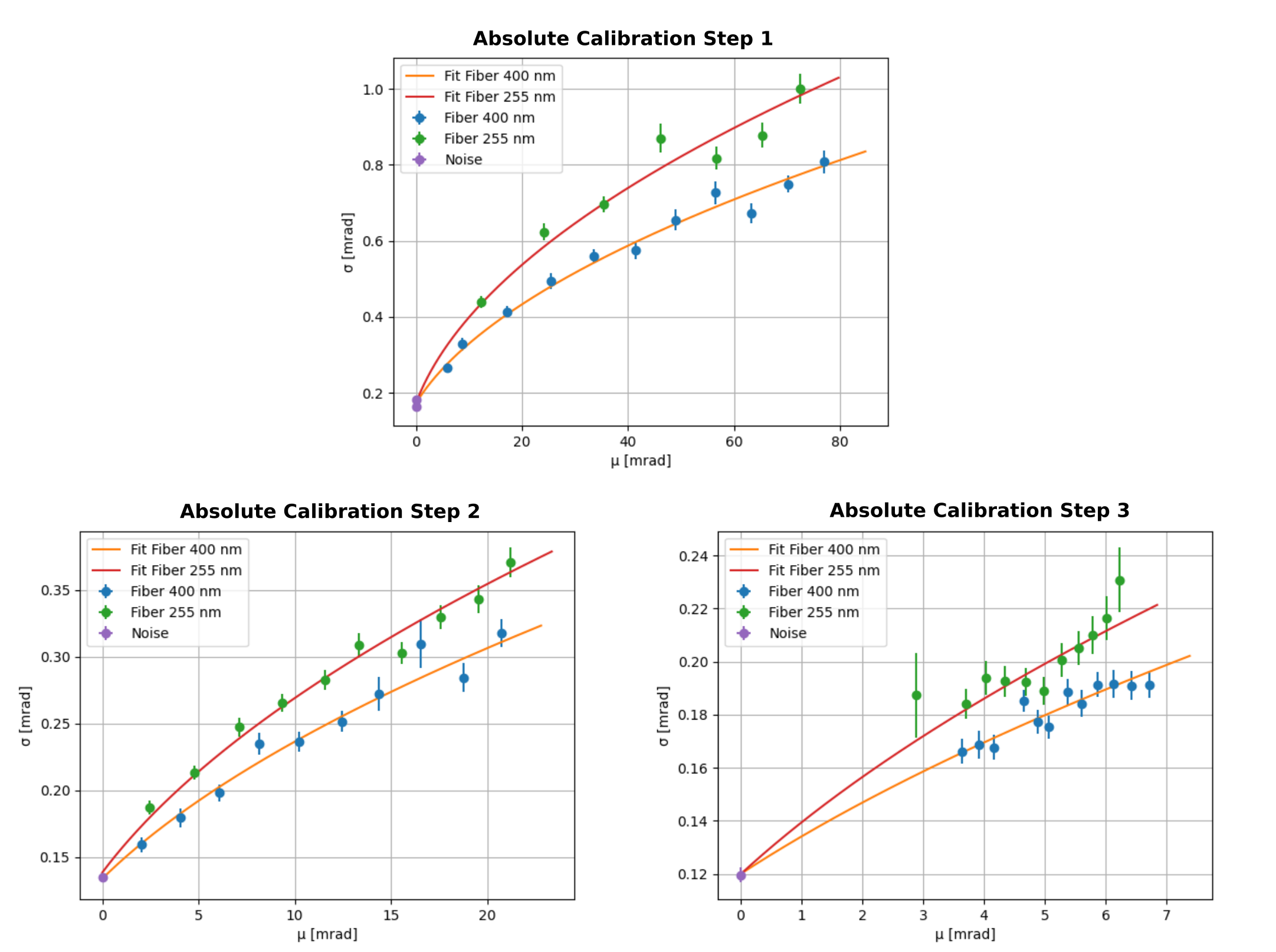}
\caption{Plots of the absolute calibration distributions in the $\sigma - \mu$ plane for each step and the respective fit (see text for more details). }
\label{fig:mirror_wafer_cal}
\end{figure}

The energy responsivity of the detector evaluated by means of the absolute calibration are presented in Table \ref{tab:mirror_wafer_cal} and the characteristics of each energy distribution are plotted in the $\sigma-\mu$ plane in Figure \ref{fig:mirror_wafer_cal}. As mentioned in section \ref{sec:absolute_cal} one important step to evaluate the various energy distributions is to correctly measure the amplitude of each acquired signal, this is done using the \textit{optimum filter} procedure (a similar procedure using the optimum filter and same data analysis framework is described in section \ref{sec:optimum_filter}). Moreover in Figure \ref{fig:mirror_wafer_cal} the absolute calibration fit for the two different wavelengths are also present in the plots. 

\begin{table}[H]
\centering
\caption{Energy responsivity extrapolated with the absolute calibration performed after each step of the characterization procedure.}
\begin{tabular}{lccc}
\hline
                            & Step  1        & Step 2         & Step 3         \\
\hline
\hline

\makecell{Energy Responsivity\\($400$~nm) $\left[\frac{\text{mrad}}{\text{keV}}\right]$} & $2.54\pm 0.08$ & $1.23\pm 0.05$ & $1.16\pm 0.05$ \\
& & & \\
\makecell{Energy Responsivity\\($255$~nm) $\left[\frac{\text{mrad}}{\text{keV}}\right]$} & $2.65\pm 0.13$ & $1.10\pm 0.04$ & $1.04\pm 0.04$\\
\hline
\end{tabular}
\label{tab:mirror_wafer_cal}
\end{table}

As it can be seen from the values in Table \ref{tab:mirror_wafer_cal} the energy responsivities of the detector between the two different wavelengths appear to be compatible which is a good indicator of the success of the absolute calibration procedure, since this quantity should be, in first approximation, independent of the energy of the single photons. Moreover, it is clear that between step 2 and step 3 the detector maintained the same characteristics (which is not always true since they strongly depend on the quality of the superconducting transition that can be easily affected by magnetic fields) thus allowing a correct comparison of the results between the two stages. On the other hand, it is also clear that between step 1 and step 2 there was a change in operating conditions of the KID resulting in approximately a factor 2 loss in energy responsivity. This can be due to several reasons but it is most likely due to the insertion of the mirror wafer in the apparatus since the same loss is present between step 1 and step 3 as well. \\

Once the energy responsivities have been measured, the energy deposition per trigger pulse must be calculated from the data in order to highlight the splitting abilities of the mirror wafer. This was done by extracting the point in the $\sigma -\mu$ plane that presented the least deviation from the absolute calibration fit. The average signal response $\mu$ corresponding to this point was then divided by the energy responsivity to convert it in an energy deposition. This energy deposition was then divided by the number of light pulses used to create it in order to extract the amount of energy deposited in each light pulse:
\begin{equation}
\Delta E_d = \frac{\mu}{n_p\cdot k}\qquad\left[\frac{keV}{\text{pulse}}\right]
\end{equation}
where $\Delta E_d$ is the energy deposited in each light pulse, $k$ is the energy responsivity and $n_p$ is the number of light pulses used. The results of this estimations are presented in Table \ref{tab:energy_burst}.

\begin{table}[H]
\centering
\caption{Energy deposited in each light pulse for both the $255~nm$ and the $400~nm$ \textit{LED drivers} in each step of the characterization procedure.}
\begin{tabular}{lccc}
\hline
                            & Step  1        & Step 2         & Step 3         \\
\hline
\hline
\makecell{$\Delta E_d(400~nm) \left[\frac{keV}{\text{pulse}}\right]$} & $3.42\pm 0.11$ & $1.65\pm 0.07$ & $(13.09\pm 0.54)\cdot 10^{-2}$ \\
& & & \\
\makecell{$\Delta E_d(255~nm) \left[\frac{keV}{\text{pulse}}\right]$}& $18.68\pm 0.91$ & $8.93\pm 0.29$ & $(30.8\pm 1.3)\cdot 10^{-2}$\\
\hline
\end{tabular}
\label{tab:energy_burst}
\end{table}

From the values in Table \ref{tab:energy_burst} it is clear that in \textbf{step 3} there was a steep drop in energy deposition for each light pulse that can only be attributed to the presence of the germanium mirror wafer. In fact, the energy deposition per pulse in \textbf{step 2} corresponds to about 48\% the one observed in \textbf{step 1} for both wavelengths, while in \textbf{step 3} the relative energy deposition per light pulse corresponds to $\sim 4\%$ for the $400$~nm photons and $\sim 2\%$ for the $255$~nm photons. The values for \textbf{step 3} also suggest a wavelength dependence that requires further investigation.\\

Besides the underwhelming numbers just presented, the characterization and usage of this device proves successful and indicates that it is worthy of further perfecting. In fact, the difference in the energy deposited by the two sides is most likely due to a misalignment of the optical fiber with respect to the vertex of the germanium prism, this mismatch is probably caused by the glue used to fix the optical fiber in place which cracked when cooled at sub-kelvin temperatures. Moreover, this was the first test performed on the first prototype of such a device and it served as proof of concept of the usage of a mirror wafer to double the amount of cryodetector modules that can be shined with optical fibers. Further testing of the next generation of mirror wafer prototypes will follow in the end of 2021 and beginning of 2022.

\chapter{Data Analysis}
\label{chap:5}
\lettrine[lines=2, findent=3pt, nindent=0pt]{I}{}n this chapter a new analysis protocol for treating the NUCLEUS prototype data will be described. The aim of this analysis is to build an algorithmic procedure for event analysis and detector calibration specific for the NUCLEUS experiment. Moreover, this chapter is meant to serve as a data analysis guideline in sight of the upcoming detector commissioning. Article \cite{RUN1} presents a first analysis of the data taken with a prototype pixel of the designed TES array for NUCLEUS showing a lower energy threshold of $\sim 20$~eV without the use of any shielding and in a surface laboratory. This article will be considered for comparison for results obtained in this section since it is based on the same data considered in the following analysis. 

\section{Importing Data in the analysis framework}
\subsection{DIANA: A low energy analysis framework}
In this section the main software used for data analysis is the \textbf{DIANA} analysis framework. DIANA was initially developed for the CUORE experiment \cite{cuore} from Marco Pallavicini and Marco Vignati, and is completely based on ROOT \cite{ROOT} classes. It also comes equipped with a graphic user interface that allows for interactive data analysis without the requirement of any coding knowledge. The framework is build in modules and new elements have been integrated for the sake of compatibility with the NUCLEUS data analysis procedure. \\

The basic elements of DIANA are called \textit{modules} and they perform basic or specific calculations to be used throughout the analysis. The different \textit{modules} can then be combined in one or more \textit{sequences} in order to build a complete analysis protocol. A peculiarity of DIANA is that the \textit{modules} are the only part that requires coding knowledge, while the \textit{sequences} are just text files that contain the \textit{modules'} input parameters and interconnections.
\subsection{Data Format and Importing}
The format of NUCLEUS prototype data is modelled after the CRESST experiment \cite{cresst}. Each acquisition is composed of two different files, a text file with the \texttt{.par} extension that contains all the global characteristics of the acquisition such as sampling frequency, pretrigger duration, number of events, etc... A second binary file, with \texttt{.rdt} extension, contains all the ADC values of the events from the data acquisition plus some additional quantities specific for each event, like the time stamp or the acquisition channel.\\

In order to use DIANA for the analysis of the NUCLEUS prototype data a \textit{module} for conversion and upload of the data had to be written. This new \textit{module}, built specifically for the NUCLEUS experiment, is divided in two different main sections following the structure of the input data files. The first section takes care of reading the \texttt{.par} file in order to save the global data of the acquisition and save the parameters describing the structure of the \texttt{.rdt} file. The second part uses the information read from the \texttt{.par} file to correctly interpret the bits read from the \texttt{.rdt} binary file.\\

Once all the events and the global data are imported in DIANA, a different \textit{module}, independent of the initial data format, takes care of saving the data in a \texttt{.root} file. From this point on all the rest of the \textit{sequences} and \textit{modules} in DIANA will work with the newly generated \texttt{.root} file adding trees and leaves for every new \textit{sequence} that is run on the data. This is one of the main strengths of DIANA, the analysis is mostly independent of the data format and already existing and tested \textit{modules} can be used without any compatibility concern.

\subsection{NUCLEUS Prototype Data - RUN 1 description}
The data considered for the following analysis was recorded using a single TES especially designed for the NUCLEUS experiment, coupled to a 125~mm$^3$ Al$_2$O$_3$ cube and the acquisition software is the same one developed for the CRESST experiment \cite{cresst}  (version 9.0). Inside the cryostat a sample of $^{55}$Fe was positioned as a 0.6~Bq source of X-rays used for calibration purposes (see section \ref{sec:energy_cal}). The total data acquisition lasted approximately 12 hours but after about 5 hours the detector left the operating conditions due to technical issues with the cryostat and thus more than half the acquisition duration must be discarded.\\

Each event in RUN 1 is composed of 8192 samples acquired with a 25000~Hz sampling frequency. This sampling frequency is more than enough for the time constants of cryogenic and phononic detectors that have reaction times greater than 0.1~ms. The first 2048 samples of each event are pretrigger samples that are used to establish the baseline level and eventual harmonics present in the background noise.\\

As mentioned above, the waveforms saved in the \texttt{.rdt} files are preceded from several important quantities, like the acquisition channel, the time-stamp of the event and the triggering sample. One extremely important label listed in these pre-waveform quantities is the trigger type. There are 4 main types of triggers in the acquisition software:
\begin{enumerate}
    \item \textbf{Noise trigger}: Noise waveforms are acquired to be later used for noise reduction protocols and the optimum filter algorithm for extracting detailed event characteristics.
    \item \textbf{Signal trigger}: Signal triggers are the actual data that undergoes the analysis and contains particle events. This is the portion of data under test in this chapter.
    \item \textbf{Test pulse trigger}: Signals used for detector calibration (see text).
    \item \textbf{Control pulse trigger}: Energy depositions used to probe the TES operating conditions (see text).
\end{enumerate}
  The last two types of triggers are the most peculiar and are typical to the NUCLEUS experiment and TES operation. Test pulses are heater signals of varying amplitude sent to the TES to characterize the detector response to various energies by sampling the superconducting transition curve. Control pulses operate in a similar way but serve to monitor the stability of the TES, they are high energy heater pulses that completely saturate the TES and thus measure the operating point and available dynamical range of the sensor in the resistance vs temperature plane.\\

The data taken for RUN 1 was the first tryout of a NUCLEUS prototype detector and minor issues were encountered. Despite the issues this data still gave rise to 3 papers on detector performances \cite{Strauss2017GramscaleCC}, new experimental procedures for CE$\nu$NS detection \cite{Strauss2017The} and new limits on sub-GeV dark matter searches \cite{RUN1}.\\

\begin{figure}[H]
\centering
\includegraphics[scale=0.4]{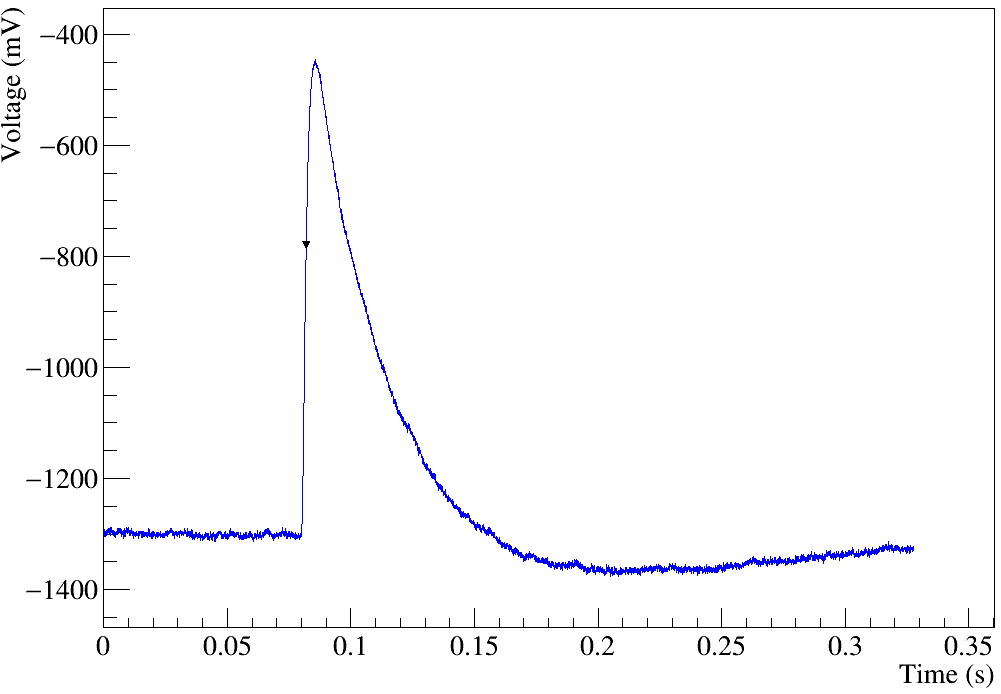}
\caption{Test pulse example showing an undershoot in the descending slope.}
\label{fig:test_pulse}
\end{figure}

One of the issues of this run was the shape of the test and control pulses, these in fact were extremely different compared with particle events, in particular they showed an undershoot in the descending slope, an example can be seen in Figure \ref{fig:test_pulse}. As will be later discussed, an extremely central part of the NUCLEUS analysis protocol is building and usage of an average pulse as an event template, since the test pulses have quite a different waveform they will not be used in the following analysis. Moreover since the analysis framework is still unripe to take advantage of all the features of the NUCLEUS prototype data the control pulses will not be used in the analysis as well but this does not affect majorly the results of the analysis protocol.

\section{Single Event Analysis}
Once the \textit{sequence} for importing the data into DIANA has been run, some basic quantities and characteristics of the pulses are calculated with the "Preprocess" sequence. This sequence performs an event by event analysis and calculates the following quantities:
\begin{itemize}
\item \textbf{Baseline}: mean value of the baseline of the signal recorded from the data acquisition calculated on the pretrigger samples;
\item \textbf{BaselineRMS}: variance of the baseline calculated on the pretrigger values;
\item \textbf{BaseslineSlope}: Average derivative of the pretrigger window;
\item \textbf{MaxBaseline}: Difference between the maximum nearest to the trigger sample and the \textbf{Baseline} value;
\item \textbf{MaxMinInWindow}: Difference between the absolute maximum and minimum in the acquisition window;
\item \textbf{MaxMinInWindowRMS}: same as the \textbf{MaxMinInWindow} variable but expressed in units of \textbf{BaselineRMS};
\item \textbf{Rise Time}: Pulse rise time;
\item \textbf{Decay Time}: Pulse decay time;
\item \textbf{Number of Pulses}: this quantity estimates the number of pulses in the same acquisition window. It is used to prevent events with piled-up pulses from contaminating the analysis.
\item \textbf{RightLeftBaseline}: Absolute value of the difference between \textbf{Baseline} and the average value of the last samples in the acquisition window (samples on the right). The number of samples taken on the right is symmetric to the one taken in the \textbf{Baseline} calculation;

\end{itemize}
the listed quantities are a fundamental basis for the whole analysis protocol and will be used in more advanced sequences for event selection and characterization and are thus presented here as a legend to be consulted while reading the following sections.\\

\subsection{Number of pulses estimation}\label{sec:peak_identifier}
The estimation of the number of pulses in one acquisition window is one of the most complex steps performed in the "Preprocess" sequence and deserves to be analysed in detail, especially because a very similar strategy is employed by the new data acquisition system developed in the framework of the BULLKID R$\&$D experiment \cite{bullkid}. On the other hand the two present substantial differences since the algorithm described here is meant to be used offline and thus does not have extremely stringent requirements on execution time and resource exploitation.\\

The main goal of this filter is to count the number of events present in a single acquisition window and thus the preservation of the pulse characteristics is not a requirement. The main idea is to make the signal-to-noise ratio as good as possible in order to make pulses stand-out from the baseline noise present in every detector. Ideally one would reach the situation in which the pulse presents Dirac's deltas only where the events start and is constant otherwise.\\

\paragraph{Step 1 - High Pass Differentiating Filter:} The first step of this algorithm is to eliminate low frequency components from the signal, like a low frequency modulation of the baseline and the slow decay slope of the pulse. Additionally for ease of the next steps it is desirable to eliminate constant offsets of the recorded waveforms.\\

\begin{figure}[H]
\centering
\hspace*{-1.7cm}
\includegraphics[scale=0.33]{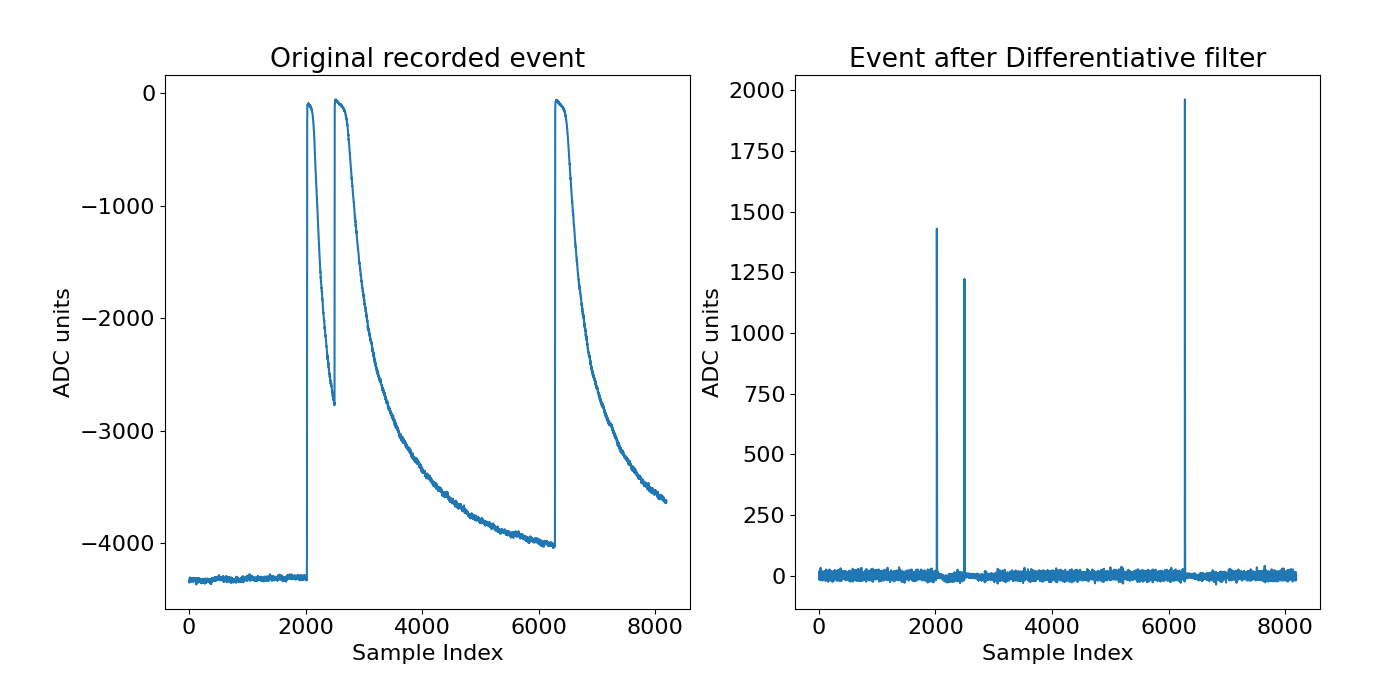}
\caption{Differentiating filter effect on a sample event of the RUN 1 data where 3 piled-up pulses are present. In the left panel the original event is presented while on the right panel the waveform after the employment of a differentiating filter is shown.}
\label{fig:differentiative_filter}
\end{figure}

\begin{figure}[H]
\centering
\hspace*{-1.7cm}
\includegraphics[scale=0.38]{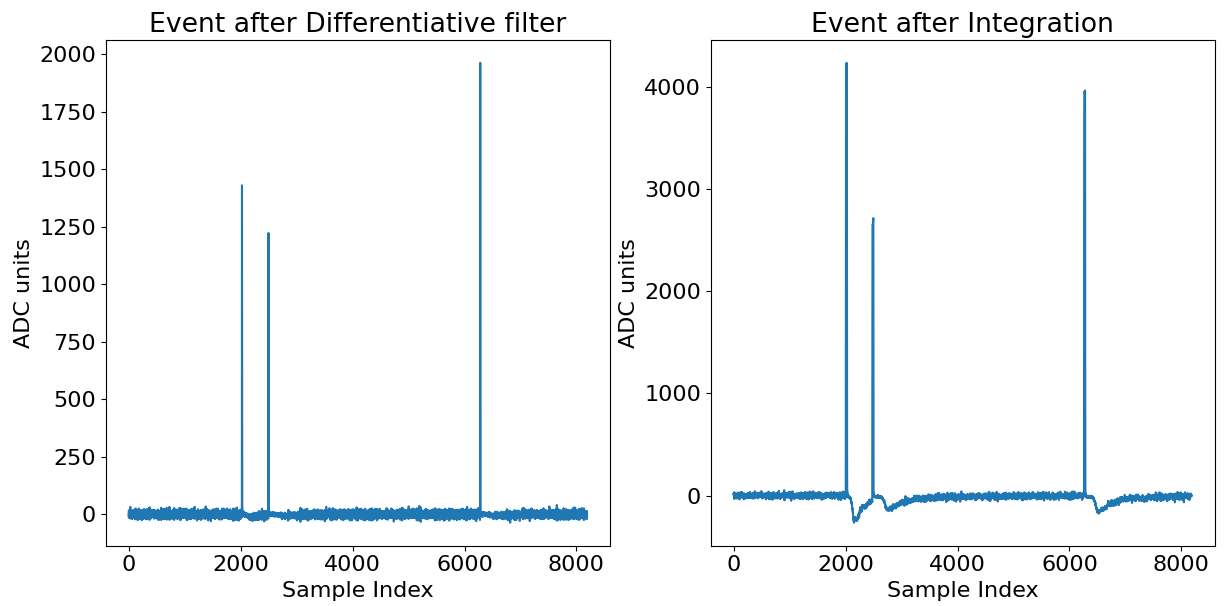}
\caption{Effect of the local integral (right panel) on the differentiated signal (left panel).}
\label{fig:local_integral}
\end{figure}

All of these steps point to the necessity of the usage of a high-pass filter. The easiest and less computationally demanding high-pass filter that can be realized is a differentiating filter, which consists of performing the difference between two neighboring samples. An example of the effect of the differentiating filter on the NUCLEUS prototype data is presented in Figure \ref{fig:differentiative_filter} where a rare event with 3 piled-up pulses is shown.
\paragraph{Step 2 - Event Integration:} The second step is a discrete integration. The effects of the local integration performed on the differentiated event are presented in Figure \ref{fig:local_integral}. This step is meant to highlight some features that were hidden by the differentiating filter without completely changing the obtained result which was already close to the desired situation. It can be also be noticed that the discrete integration steps has another important feature: it also serves as a signal amplification step and enhances the overall signal-to-noise ratio.

\paragraph{Step 3 - Low Pass Filter:} The next step is to pass the integrated pulse through a low pass filter in order to eliminate all the high frequency noise. Moreover, taking away the high frequency components of the event peaks makes them wider and less resolute which is an advantage for the last stage of the peak counting algorithm.\\

The low pass filter used here is a moving average filter. This choice was made for several reasons, the main one is ease of coding into the DIANA analysis framework and the fact that already made and tested subroutines could be used. Moreover, making a sum of values requires much less computational resources, time and coding effort to be performed with respect to more complex low pass filters like the common \textit{Butterworth filter}. Despite not being an online filtering procedure these are all requirements when large amounts of data have to be processed.\\

For better user interface and communication with other modules already present in the DIANA framework, the choice of the number of points to use for the moving average is translated into a cutoff frequency via the formula derived in \cite{moving_average}:
\begin{equation}
n=\text{Floor}\left[\left(\sqrt{0.196202+\left(\frac{f_0}{f_s}\right)^2}\right)\cdot \frac{f_s}{f_0}\right]
\end{equation}

where $f_0$ is the cutoff frequency in input from the user and $f_s$ is the sampling frequency. The "Floor" operation is there to convert the result of the operation into an integer, this could also have been done rounding the number but the "Floor" operation was chosen to keep a conservative approach since a smaller number of points translates to a higher cutoff frequency. This formula is thoroughly analysed in \cite{moving_average2} and it is stated to be asymptotically correct for a large number of samples and has about 2\% error for $n=2$. A more detailed discussion of this formula is beyond the scope of this dissertation since a high accuracy in the filtering is not required.\\

\begin{figure}[H]
\centering
\hspace*{-1.7cm}
\includegraphics[scale=0.32]{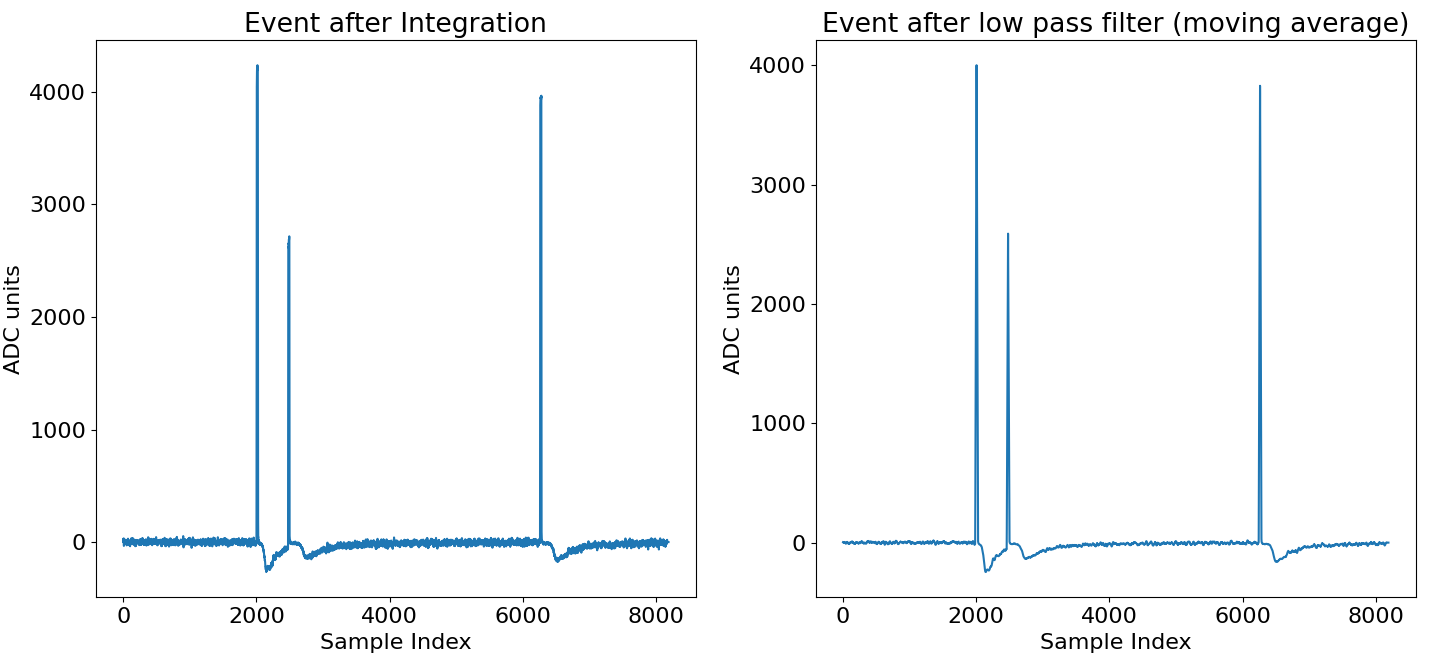}
\caption{Moving average filter effect with a cutoff frequency of $f_0=0.02218\cdot f_s \approx 554.5$~Hz that corresponds to 20 points. In the left panel the integrated pulse before the moving average filter is shown, while on the right the effects of the low-pass filter are presented. }
\label{fig:low_pass}
\end{figure}

The effects of the low pass filter are presented in Figure \ref{fig:low_pass} and it can be clearly seen how the signal-to-noise ratio gets better. To a careful eye it is also visible how the moving average filter spreads a little the pulse peaks without any major loss in amplitude.

\paragraph{Step 4 - Evaluation of minimum rms deviation of the signal and identification of number of peaks:} The last step remaining for this procedure is to establish a methodology with which to count the number of events in a single acquisition window. The underlying idea of this step is to find a threshold above which the filtered section of the recorded waveform is considered an event. In order to correctly identify the threshold without having to make a case by case analysis is to define it in units of the minimum rms deviation of the window.\\

\begin{figure}[H]
\centering
\hspace*{-1cm}
\includegraphics[scale=0.3]{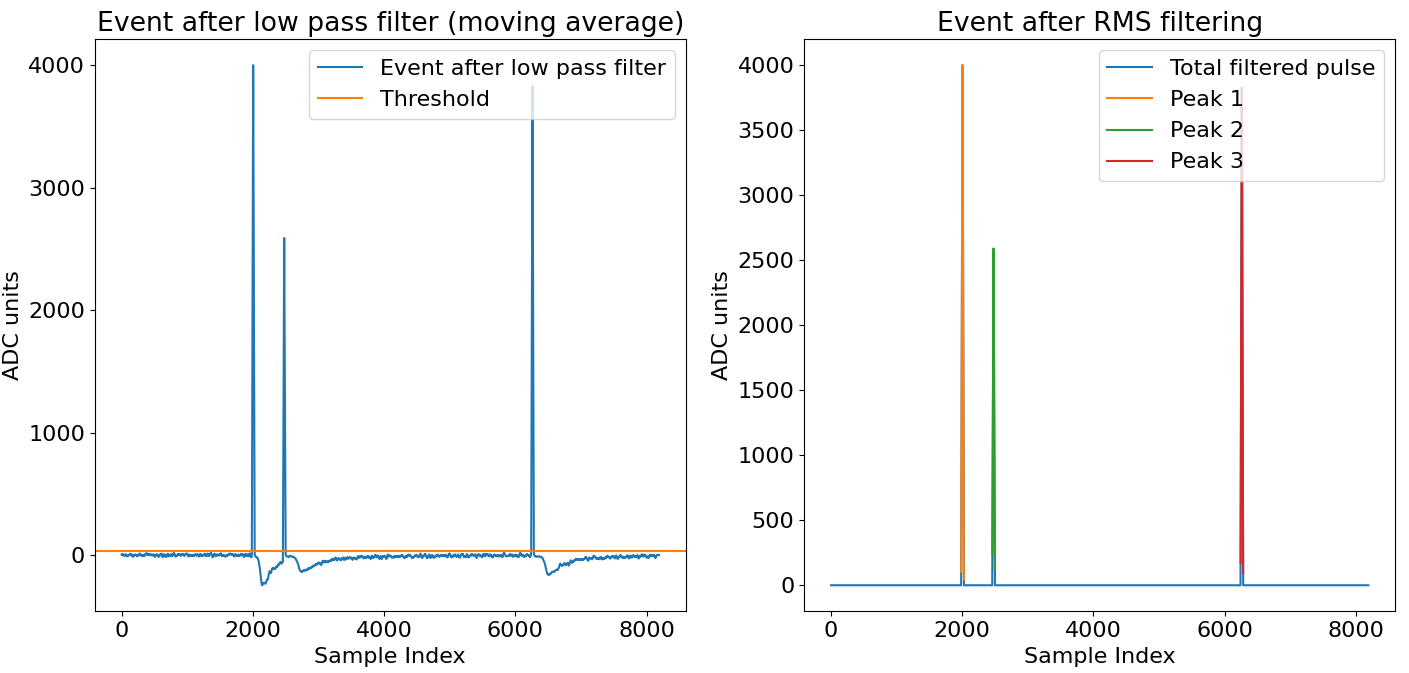}
\caption{Last stage of the algorithm of number of pulses evaluation. On the left the signal after the low-pass filter is shown in blue and in yellow the threshold line is also presented. In the right panel the result of the peak counting is shown, each color represents a different peak individuated from the algorithm.}
\label{fig:rms_eval}
\end{figure}

To define the minimum rms deviation, the filtered waveform is divided in segments of half the length the pretrigger samples and the variance of each segment is evaluated, then the minimum is taken of all the calculated deviation values. The division in segments and taking the minimum rms deviation is necessary not to take event peaks into account, since in most cases at least the pretrigger values should only have background noise of the data acquisition. After the minimum variance is identified, the user specifies, in rms deviation units, the before-mentioned threshold. For ease of computation all the samples of the waveform that are under said threshold are set to zero as shown in the right panel of Figure \ref{fig:rms_eval}. In order to identify something as a valuable peak at least 3 neighboring values must be above the threshold and separated sets of points are considered as different peaks as illustrated by the colors in Figure \ref{fig:rms_eval}. This algorithm performs extraordinarily well especially when compared with a previous existent protocol present in the DIANA framework.\\

From this example all the steps following the differentiation might appear redundant since the signal is already in a state similar to a comb of Dirac's deltas where the events are. On the other hand it must be kept in mind here that the showed event is a high energy event and the pulse counting yields good results quite easily but the NUCLEUS experiment is meant to measure low energy events that could be completely covered by the background and thus the signal-to-noise ratio must be improved as much as possible before attempting the peak counting in order not to miss events.

\paragraph{Formal analysis of the filtering procedure:} The previous description of the filtering procedure was done using an example event but a more filter-like brief description helps to clarify the inner workings of the algorithm. This analysis consist of generating a chirp-like signal that is then passed through steps 1 to 3 of the previous protocol. The chirp-like signal consist of a sinusoidal wave whose frequency is increased during its development in order to scan a continuous range of frequencies with a single signal.\\

\begin{figure}[H]
\centering
\hspace*{-2cm}
\includegraphics[scale=0.3]{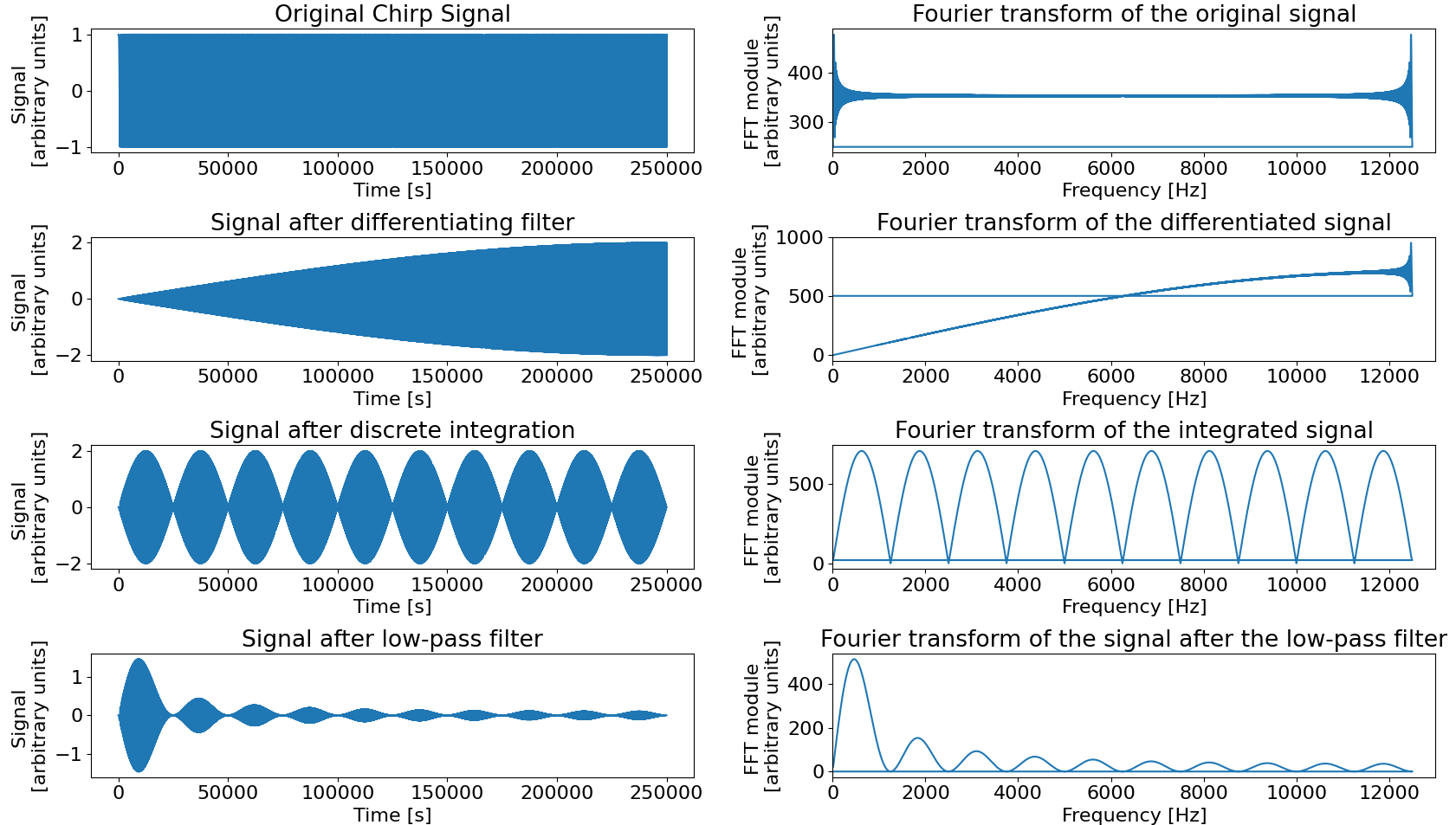}
\caption{Step by step study of the peak finding algorithm. On the left side of the panel all the transformations that the signal undergoes are presented while on the right the Fourier transform of such transformations are presented. From top to bottom the following situations are presented: original starting signal (chirp generated with the \textit{scipy} python package), signal after differentiation (step 1), signal after discrete integration (step 2), signal after low-pass moving average filter (step 3).}
\label{fig:step_by_step_fft}
\end{figure}

In Figure \ref{fig:step_by_step_fft} a step by step presentation of the transformations the signal undergoes is presented, on the left part of the panel the shape the signal is presented at each step while on the right the module of the Fourier transform of the signal is showed. The generated chirp has increasing frequency with time and the effects of the differentiation as a high-pass filter and moving average as a low-pass filter are clearly visible from both the signal shape and the Fourier transform. The behaviour of the integration filter is not as easy to understand but the important features are the fact that it lowers the high frequency components of signal while amplifying the low frequency ones. This works extremely well when coupled with the differentiating filter and the moving average filter that practically eliminates all undesired frequency components. \\

\begin{figure}[H]
\centering
\hspace*{-0cm}
\includegraphics[scale=0.37]{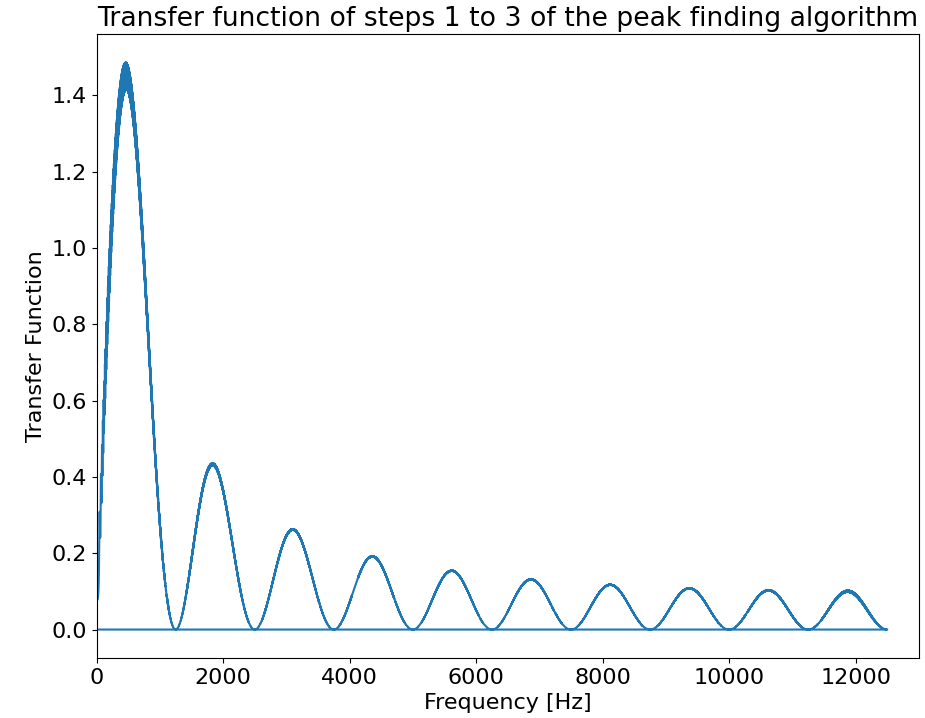}
\caption{Module of the cumulative transfer function of steps 1 to 3 of the peak finding algorithm.}
\label{fig:transfer_function}
\end{figure}

A more comprehensive view of the workings of this protocol is given in Figure \ref{fig:transfer_function} where the transfer function of steps 1 to 3 is presented. From this plot, in fact, it can be seen that the protocol works as a pass-band filter that must be carefully tuned around the typical event rise frequencies of the data. Moreover it can be seen that due to the integration step it also presents an amplification factor in the focused frequency region.

\subsection{Decay Time - Preliminary Analysis}\label{sec:decayTime}
A preliminary test that was performed in order to study the correct behaviour of the importing and preprocess \textit{sequences} and \textit{modules} is the estimation of the average decay time of the pulses by simply imposing cuts on the data using only the graphic user interface of DIANA. The cuts required for this study will also be relevant later for a more complex analysis.

\begin{figure}[h]
\centering
\includegraphics[scale=0.7]{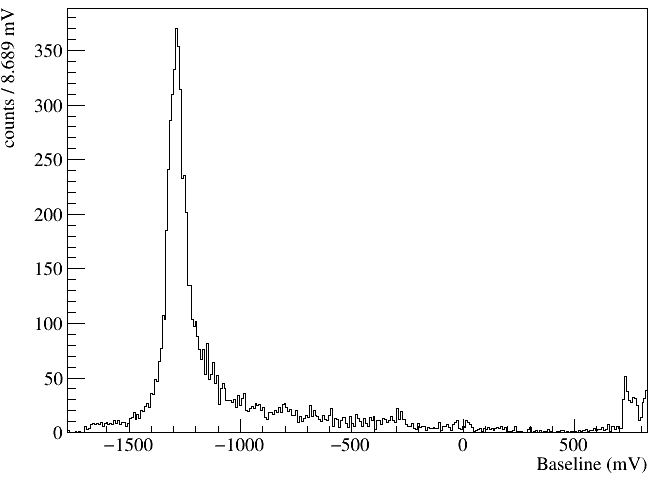}
\caption{Baseline distribution for the signal trigger type of RUN 1.}
\label{fig:baseline}
\end{figure}

\paragraph{Step 1 - Baseline and Time cuts:}
After the "Preprocess" \textit{sequence} was run on the data several distributions of different characteristics can be plotted, the first taken into account in this study is the \textbf{Baseline} distribution shown in Figure \ref{fig:baseline}.

From Figure \ref{fig:baseline} it is noticeable that several signal events have a baseline above $0$~mV, these events correspond to the period in which the detector lost the working conditions. Since the majority of the events are distributed below $-500$~mV a conservative cut of $Baseline < 0$~mV was imposed to take out most of these events. 
\\
Since not all of the events corresponding to the loss of ideal working conditions of the detector are eliminated from the analysis using simply a baseline upper limit an additional cut on the timestamp of the events was also necessary (time elapsed from the beginning of the acquisition must be less than $15000$~s).

\paragraph{Step 2 - MaxBaseline Spectrum Study:}
After the \textbf{Baseline} and timestamp cuts, the \textbf{MaxBaseline} spectrum for the RUN 1 data was plotted. This spectrum is particularly important because the \textbf{MaxBaseline} variable can be seen as a proxy for the event energy before more advanced algorithms are used.\\
\begin{figure}[H]
\centering
\includegraphics[scale=0.65]{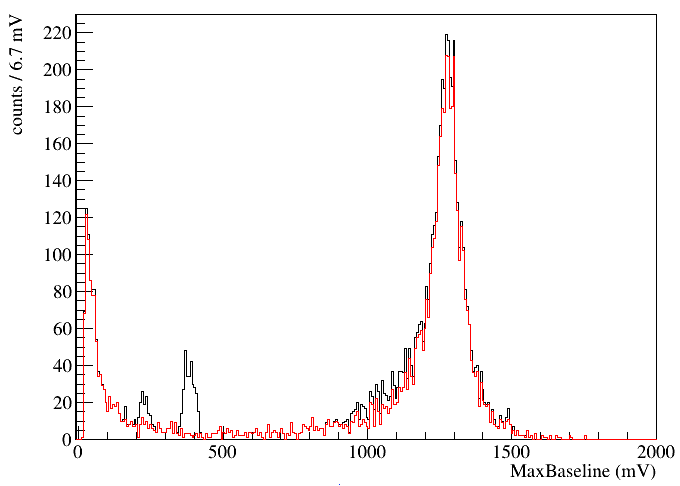}
\caption{\textbf{MaxBaseline} distribution for the signal trigger type of Run 1. In black the full spectrum without cuts is presented, in red the spectrum with the time and baseline cuts is shown (time elapsed form the beginning of the acquisition less than $15000$~s and \textbf{Baseline} lower than $0$~mV.}
\label{fig:maxbaseline}
\end{figure}
As shown in Figure \ref{fig:maxbaseline}, the spectrum can be divided in multiple regions, and each region corresponds to events with different amplitudes. In particular, it can be seen how the previous cuts on \textit{Baseline} and \textit{time stamp} eliminate 2 distribution peaks from this spectrum. The two remaining peaks are between $0$~mV and $200$~mV, and $1000$~mV to $1500$~mV; the first one is comprised of mostly unsaturated events and noise mistriggers, while the latter contains all the pulses that saturate the dynamical range of the TES and thus appear to have the same amplitude at this stage of the analysis. In order to correctly evaluate the decay time of the pulses only non-saturated events must be selected, unfortunately the \textbf{MaxBaseline} spectrum does not present a clear separation between the two regions and a clean cut was impossible to make. As mentioned, there are a certain amount of noise events that were mislabel as signal events so a minimum \textbf{MaxBaseline} threshold set at $25$~mV allowed to correctly eliminate these mistriggers.

\paragraph{Step 3 - Decay Time Spectrum Study:}

\begin{figure}[H]
\centering
\includegraphics[scale=0.65]{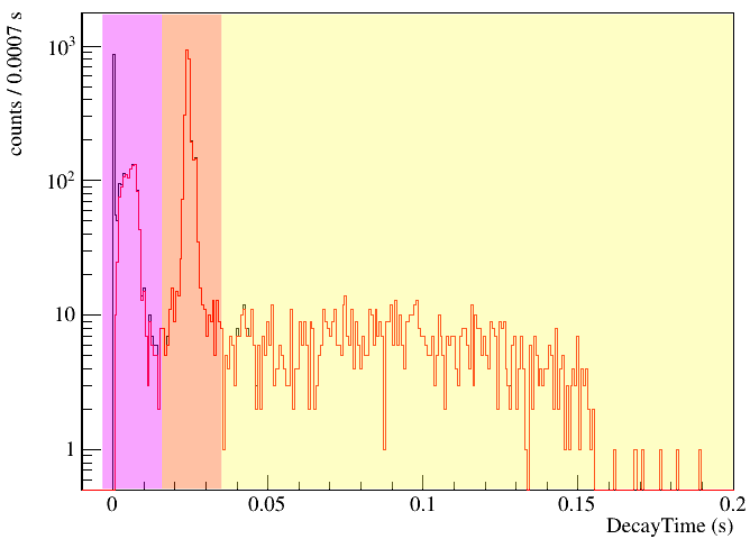}
\caption{Decay Time distribution for the signal trigger type of RUN 1. In black the spectrum corresponding to the timestamp (elapsed time $<15000$~s) and \textbf{Baseline} ($<0$~mV) cuts is shown, while in red the spectrum with the additional \textbf{MaxBaseline} ($>25$~mV) threshold is presented. The colored areas are: purple, non-saturated pulses; orange, mildly saturated events; yellow, extremely saturated events.}
\label{fig:decaytime}
\end{figure}

Finally, an analysis of the decay time spectrum was used to conclude this preliminary study. In Figure \ref{fig:decaytime} the decay time spectrum is shown, in particular in black the spectrum with the timestamp and baseline cuts is presented while in red the one with the additional \textbf{MaxBaseline} threshold is shown. In this plot it is clear that the minimum \textbf{MaxBaseline} threshold eliminates all those events that have a decay time that is extremely close to $0$~s and can thus be interpreted as fluctuations of the noise that were mistriggered as signal events.\\

The overall distribution can be divided in three different regions: the yellow one from $33$~ms to $200$~ms is comprised of very saturated pulses that must be necessarily discarded from the analysis. The region from $17$~ms to $33$~ms is somewhat of a transition region from saturated events to non-saturated events. In this region the pulses present a less prominent saturation and are mostly due to the calibration source, but due to the saturation they must be discarded since they do not give a correct evaluation of the decay time. Finally the region between $0$~ms and $17$~ms consists of mostly non-saturated pulses and can thus be used to estimate the average decay time.\\

\begin{figure}[H]
\centering
\includegraphics[scale=0.55]{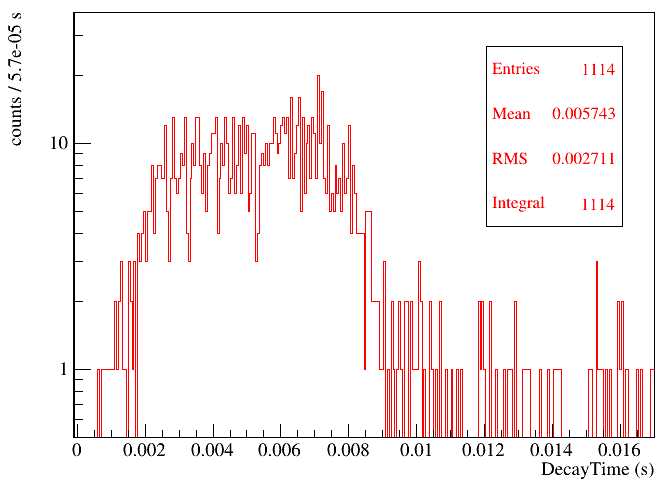}
\caption{Decay Time distribution for the signal trigger type of Run 1 (only non-saturated pulses).}
\label{fig:decaytime_zoom}
\end{figure}

Zooming in on the non-saturated region it can be seen, from Figure \ref{fig:decaytime_zoom}, that the distribution has a mean value of $(5.7\pm 2.7)$~ms. This value is compatible under one standard deviation with the one estimated in \cite{RUN1} of $(3.64\pm 0.01)$~ms. This result can be considered a success of the preliminary analysis since the desired outcome only differed of approximately a factor 2. The difference between the two values is due to the fact that the presented study was performed only imposing cuts on the results from the "Preprocess" sequence. On the other hand, \cite{RUN1} uses a much more complex analysis such as the optimum filtering and event reconstruction to evaluate this result.

\section{Average Pulse Identification}\label{sec:av_pulse}
After some basic parameters were identified in section \ref{sec:decayTime} in order to isolate the non-saturated events, a more detailed characterization of the detector response must be performed. The first step in such analysis is to determine the linearity response region of the detector with the procedure described in section \ref{sec:lin_reg}. In order to do so, it is mandatory to build a template event that must be later compared to the acquired events. The template event is built by the means of the average pulse procedure which consists of averaging non-saturated events to have an almost noiseless characterization of a typical event. Subsequently said average pulse is fitted to each event in order to reconstruct saturated events via the truncated fit method (see section \ref{sec:truncated_fit}) and the event amplitude distribution is built.\\

\subsection{Event selection for average pulse construction}

The construction of the average pulse is by far the most fragile part of all the analysis. This is due to the fact that if the average pulse's shape is not well representative of the events all the following analysis will fail and it will be close to impossible to perform the detector's calibration with radioactive sources.\\

Since the scope of this section is to build an analysis protocol for the NUCLEUS experiment an algorithm as general as possible that can be applied to runs acquired in different experiment conditions must be identified. In section \ref{sec:decayTime} a cut on the timestamp was used in order to find non-saturated pulses, obviously this is a cut that is extremely dependent on the acquisition and thus a different strategy must be found. On the other hand many of the cuts already discussed will be used to identify the correct pulses that must be included in the average pulse construction.\\

\begin{figure}[H]
\centering
\includegraphics[scale=0.45]{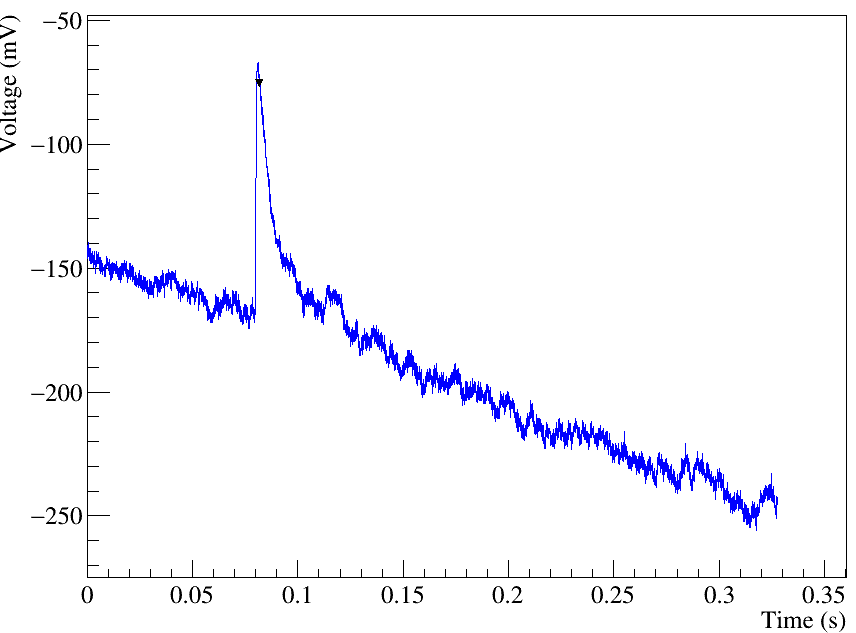}
\caption{Example of pulse that presents a particularly evident slope in the pretrigger samples that would be considered in the average pulse construction without if the \textbf{BaselineSlope} cuts were not present.}
\label{fig:pulso_storto}
\end{figure}

The first step of the event selection is to consider events that present only one peak in the acquisition window using the protocol discussed in section \ref{sec:peak_identifier}. The optimal configuration of the input parameters of the algorithm were a 554.5~Hz (20 points) cutoff frequency for the low-pass filter and a threshold corresponding to 5 times the minimum rms deviation. The next considered cut is on the average derivative or slope of the pretrigger window that must be in the interval spacing from $-0.01\frac{\text{mV}}{\text{s}}$ to $0.01\frac{\text{mV}}{\text{s}}$. These values themselves were identified via a trial-and-error procedure but the necessity of the cut was evident because including pulses like the one shown in Figure \ref{fig:pulso_storto} when building the average of pulses would produce a resulting waveform which is incompatible with more straight pulses that form the majority of the population.

The following step in the pulse selection is to use the cuts identified in section \ref{sec:decayTime}. The negative baseline requirement remains unaltered while the \textbf{MaxBaseline} cut is slightly modified. Since a good signal-to-noise ratio in the average pulse is reached with just a few tens of pulses, more stringent requirements than the ones used in the decay time study can be used in order to achieve the optimal shape. This is the reason why the \textbf{MaxBaseline} minimum threshold was raised from $25$~mV to $60$~mV in order to eliminate completely noise mistriggers. A higher limit of $1000$~mV is also used for the \textbf{MaxBaseline} variable to eliminate in bulk the majority of the saturated pulses (see Figure \ref{fig:maxbaseline}). A more refine filtering of the saturated pulses was done by requiring the decay time to be in the interval from $0$~s to $0.015$~s as it can be understood from the results in Figure \ref{fig:decaytime_zoom}.\\

Furthermore a $60$~mV to $1000$~mV limit was also imposed on the \textbf{MaxMinInWindow} variable. This might seem redundant with the \textbf{MaxBaseline} cut but it is actually mandatory because it gets rid of a small number of pulses (less than 10) that have two events in the acquisition window that were misidentified as 1 from the peak finder algorithm. These pulses typically present two peaks: the one nearest to the trigger sample passes the \textbf{MaxBaseline} requirements, but the second is much greater and is outside the chosen cuts.\\

\begin{figure}[H]
\centering
\includegraphics[scale=0.45]{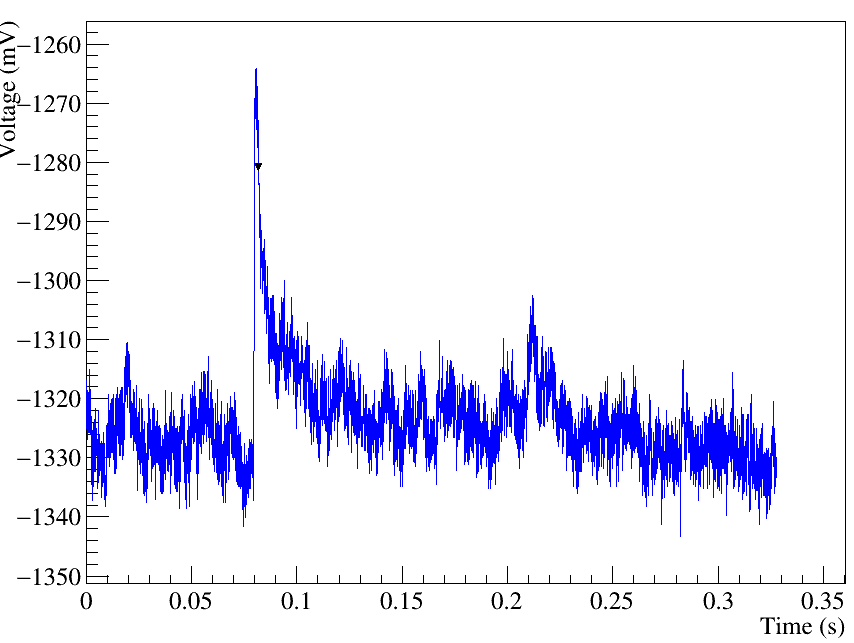}
\caption{Event outside the boundaries set for baseline variance.}
\label{fig:noisy_pulse}
\end{figure}

Two more cuts were used to clean the pulse samples from spurious events that would ruin the average pulse. The first one was a cut on the variance of the baseline (\textbf{BaselineRMS}$<4.2$~mV) in order to select events that do not present an excessive amount of noise like the one presented in Figure \ref{fig:noisy_pulse}. Finally the last cut used was on the absolute value of difference between the baseline calculated on the pretrigger values and the values at the end of the acquisition window (\textbf{RightLeftBaseline}$<100$~mV) where, theoretically, the pulse should have gone back to the baseline level once again. This last cut is necessary to avoid events that do not fully decay in the acquisition window or whose baseline value changes during the event and thus decay to a value different from the pretrigger level. These pulses in fact distort the average pulse making it not a good event template to be used in the next sections of the analysis.\\

\begin{figure}[H]
\centering
\includegraphics[scale=0.5]{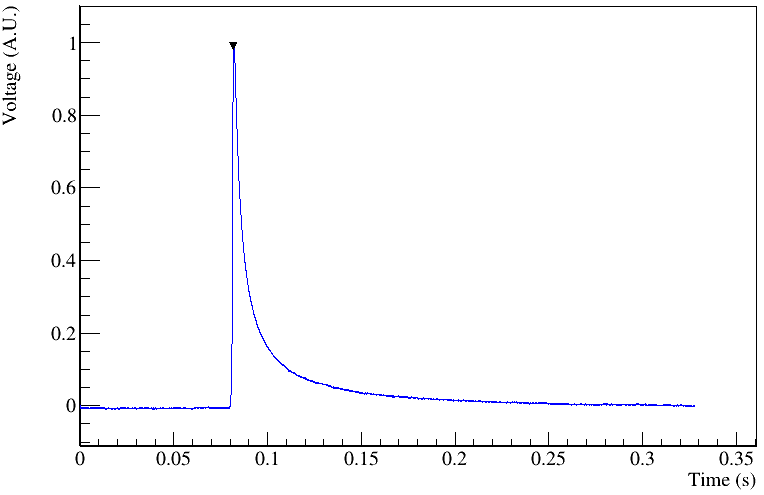}
\caption{Average pulse built from the RUN 1 of the NUCLEUS prototype data.}
\label{fig:averagepulse}
\end{figure}

All the cuts and limits just described select 122 events out of the total 7688 signal events acquired. This event are then normalized and averaged with one another in order to build the average pulse presented in Figure \ref{fig:averagepulse}. As it can be seen, the average pulse presents very little noise, a basically flat baseline and the pulse is well contained in the length of the acquisition window. Moreover, it decays to the average value it had in the pretrigger samples and thus has all the desired characteristics.

\section{TES linearity response region extrapolation}\label{sec:lin_reg}

After the identification of a viable template pulse the next step is the characterization of the linearity response region of the TES under test. The understanding of linearity threshold of the detector is a necessary ingredient for the truncated fit procedure which is central to this study and to the NUCLEUS analysis.\\

The idea behind this section is to perform an event by event scaling of the average pulse in order to have it overlap as much as possible the acquired pulse. Mathematically this translates to adding an appropriate offset to the average pulse so that the baseline of the event and the template coincide, to properly amplify the average pulse in order to match the event's amplitude and to add a time jitter (or shift) in order to have them completely overlap. All of this is done by minimizing a cost function that depends on the three degrees of freedom mentioned before, the function in question is just the square of the difference of the two waveforms:
\begin{equation}
\chi(A,\Delta t, D) = \sum_{i=1}^{N} \left[s_i-p_{i+\Delta t}(A,\Delta t, D)\right]^2
\end{equation}
\begin{equation}
p_{i+\Delta t}(A,\Delta t, D) = A\cdot ap_{i+\Delta t} + D
\end{equation}
 where $s_i$ are the samples of the event under study, $p_i$ is the average pulse ($ap_i$) multiplied by the fitting parameter $A$ (amplitude), time shifted of the fitting parameter $\Delta t$ and vertically shifted of the parameter $D$ (offset). The value $N$ corresponds to the number of points considered in the sum which corresponds to approximately 97.5\% of the points in the whole acquisition window. The missing 2.5\% of the points (200 points) is due to the fact that the very extremes of the acquisition window must not be counted in the sum because they might present a discontinuity (for example \textbf{RightLeftBaseline} might not be exactly zero). The minimizing method used here is the \textit{MIGRAD} algorithm already present in the ROOT toolkit.\\
 
The fitting procedure just described yields good results on non-saturated events but is quite sensitive to the chosen average pulse and the starting values of the fitting parameters. It is thus mandatory, before the fit of each event, to properly initialize the fitting parameters. The newly added DIANA module for this task initializes the values to:

$$ A_0 = \text{MaxBaseline}\qquad \Delta t_0=\text{Index of the maximum value} \qquad D_0=\text{Baseline}$$

\begin{figure}[H]
\centering
\includegraphics[scale=0.55]{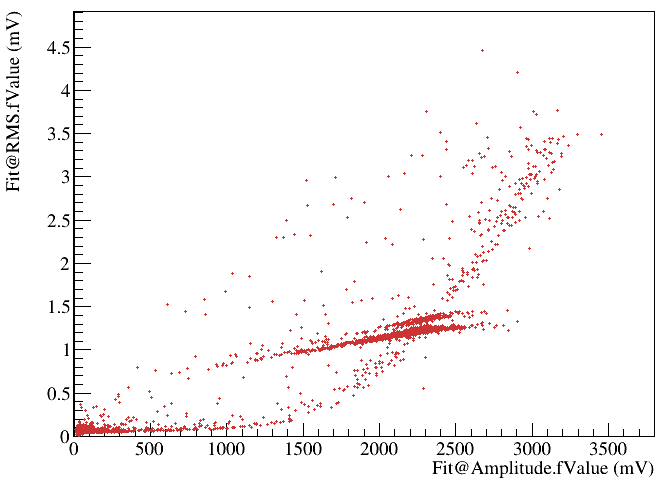}
\caption{Result of the average pulse fit for the identification of the linear region of response of the detector.}
\label{fig:linearityregion}
\end{figure}

\begin{figure}[H]
\centering
\includegraphics[scale=0.6]{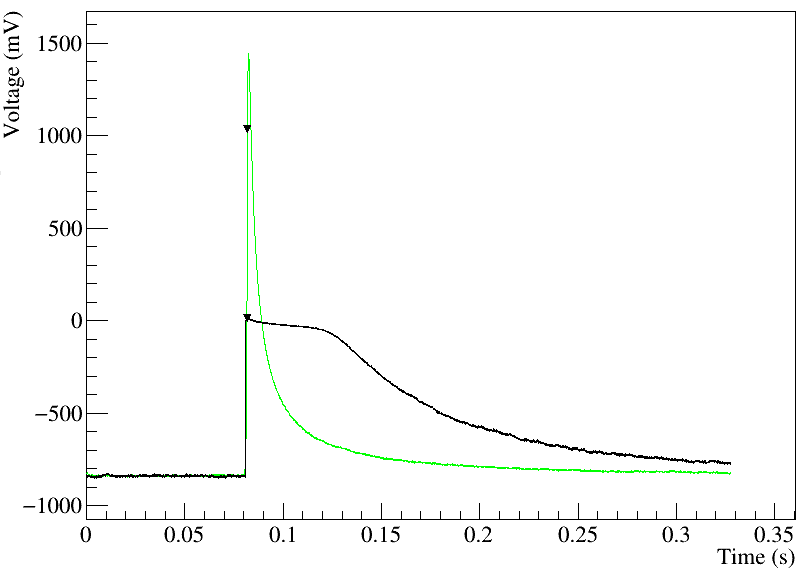}
\caption{Example on the failure of the fitting procedure on extremely saturated pulses. In black the recorded event is shown, in green the best fit of the average pulse with the event is presented.}
\label{fig:fit_fail}
\end{figure}

This strategy of fitting the average pulse to each event allows to estimate the minimum difference of the two in rms with the square-root of the minimum of the function $\chi$ divided by the number of considered points. When an event is saturated the fit will not converge properly since it has a drastically different shape with respect to the average pulse and the value of the rms difference will be high. On the other hand, if the event is not saturated the average pulse and the acquired pulse will overlap giving a low rms difference and an accurate amplitude measurement. This feature of the rms difference is what allows the determination of the linear response region of the TES, in fact, once all the events have been fitted one can plot the rms difference versus the fitted amplitude as shown in Figure \ref{fig:linearityregion}. From the plot it can be seen that for low enough amplitudes the rms difference stays approximately constant at a low value but as soon as the amplitude reaches values above $600$~mV the rms difference starts to grow. At first (between $600$~mV and $1000$~mV) the difference grows slowly, this is due to the fact that events are outside the linearity response region of the detector but are still inside the dynamical range of the TES and thus do not present saturation. On the other hand above of $1000$~mV the rms difference starts to rise much more rapidly and this is due to the fact that the fit fails with saturated pulses and the rms grows with the saturation (and thus the amplitude) since the shape of the events differs drastically from the one of the average pulse as it is show in Figure \ref{fig:fit_fail}.\\

\begin{figure}[H]
\centering
\includegraphics[scale=0.5]{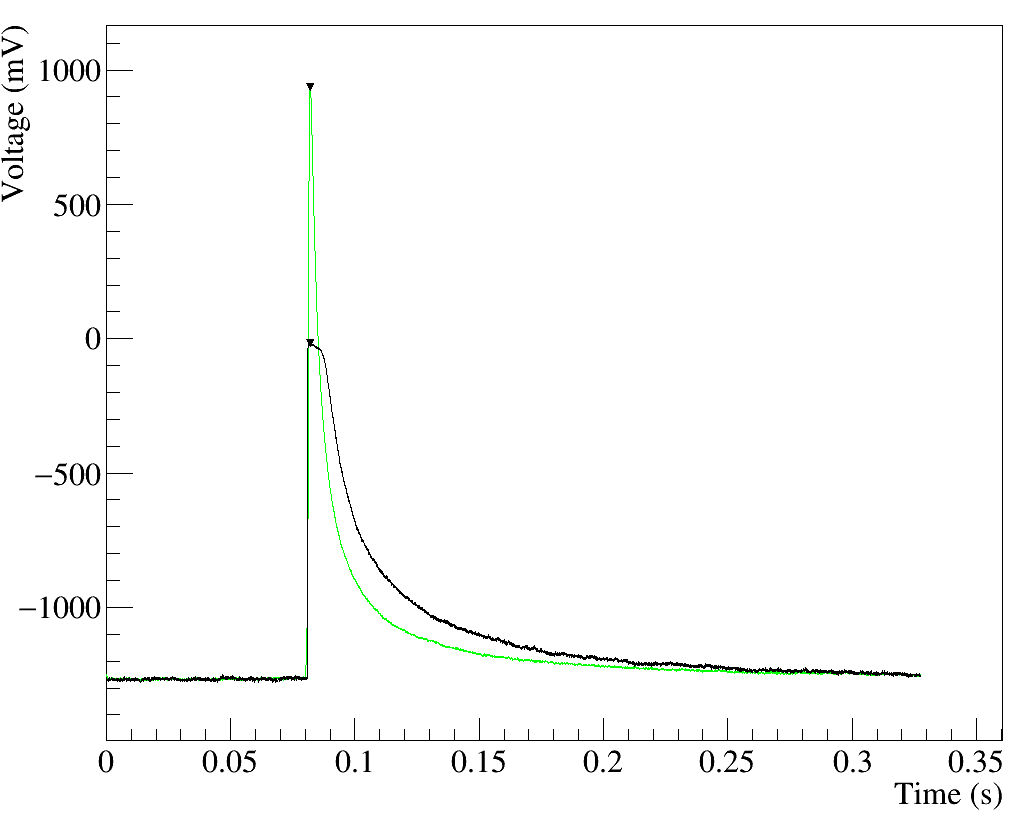}
\caption{Example pulse of the mentioned populations in Figure \ref{fig:linearityregion}. In black the recorded event is shown, in green the best fit of the average pulse with the event is presented. It can be seen that the reconstruction failure is not as bad as the one presented in Figure \ref{fig:fit_fail}.}
\label{fig:population}
\end{figure}

One detail worth mentioning in Figure \ref{fig:linearityregion} is the clusters of points that are not distributed on the ascending curve of the rms difference. These points are composed by events that are somewhat saturated but still retain a lot of the characteristics of the average pulse and thus whilst having a high rms difference the amplitude fit still gives a reasonable (but incorrect) result as shown in Figure \ref{fig:population}. This is the reason why they form a population of events with high rms but whose increase is weaker than the completely saturated pulse distribution.

\section{Truncated Fit Method for Pulse Reconstruction}\label{sec:truncated_fit}

As mentioned several times before, the NUCLEUS prototype data, being recorded with a TES, presents a saturation region that corresponds to the high end of the superconducting transition of the sensor. The next step in the analysis is thus to try and recover the part of the pulse that is lost in the saturation of the TES. This is not just an exercise in style but it is actually a true necessity for the NUCLEUS experiment due to the fact that the detectors can also use as calibration source a sample of $^{55}$Fe ($0.6~Bq$) that emits photons at $5.899$~keV and $6.490$~keV \cite{iron_peaks} that give rise to pulses with energies beyond the TES dynamical range.\\

In order to reconstruct the saturated pulses the procedure referred to as \textbf{truncated or severed fit} is used. The procedure works in a very similar way to the one described in section \ref{sec:lin_reg} and basically consists in fitting the average pulse built before in section  \ref{sec:av_pulse} to each event. A crucial difference between the two methods is that the truncated fit method does not consider all the points of the in the acquisition window but only the ones that differ from the baseline of less than the linearity region threshold of $600$~mV derived in section \ref{sec:lin_reg}. The idea is, in fact, to use only the information derived from the portion of the event in the linear response region of the TES to fit the three parameters of the template pulse (offset, temporal jitter and amplitude) and thus reconstruct the event. This method has the downside that it assumes that the shape of the events produced (if the TES had an infinitely extended dynamical range) is independent of the event energy. The remaining details of the fit such as the minimizing method or the cost function remain unaltered from the case in section \ref{sec:lin_reg} and will not be re-discussed. \\

\subsection{Truncated Fit Results}
\begin{figure}[H]
\centering
\includegraphics[scale=0.35]{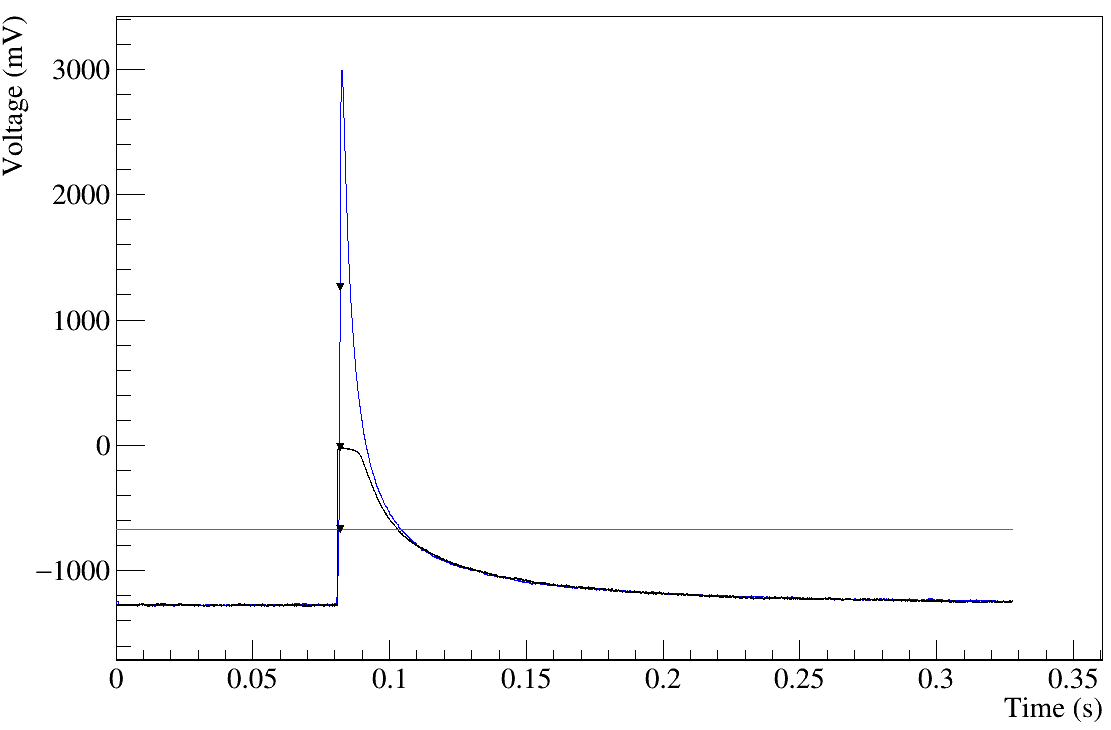}
\caption{Example of pulse reconstruction with the \textbf{truncated fit} method. In black the original event is presented, in blue the fitted template pulse and in magenta the upper limit of the linearity response region of the TES is plotted.}
\label{fig:truncatedfit}
\end{figure}

The truncated fit methodology is able to correctly reconstruct reasonably saturated events like the one shown in Figure \ref{fig:truncatedfit} that most likely is generated from an x-ray emission of the calibration source. This means that the described procedure effectively extends the dynamical range of the detector giving access to otherwise forbidden energy regions. 

\begin{figure}[H]
\centering
\includegraphics[scale=0.28]{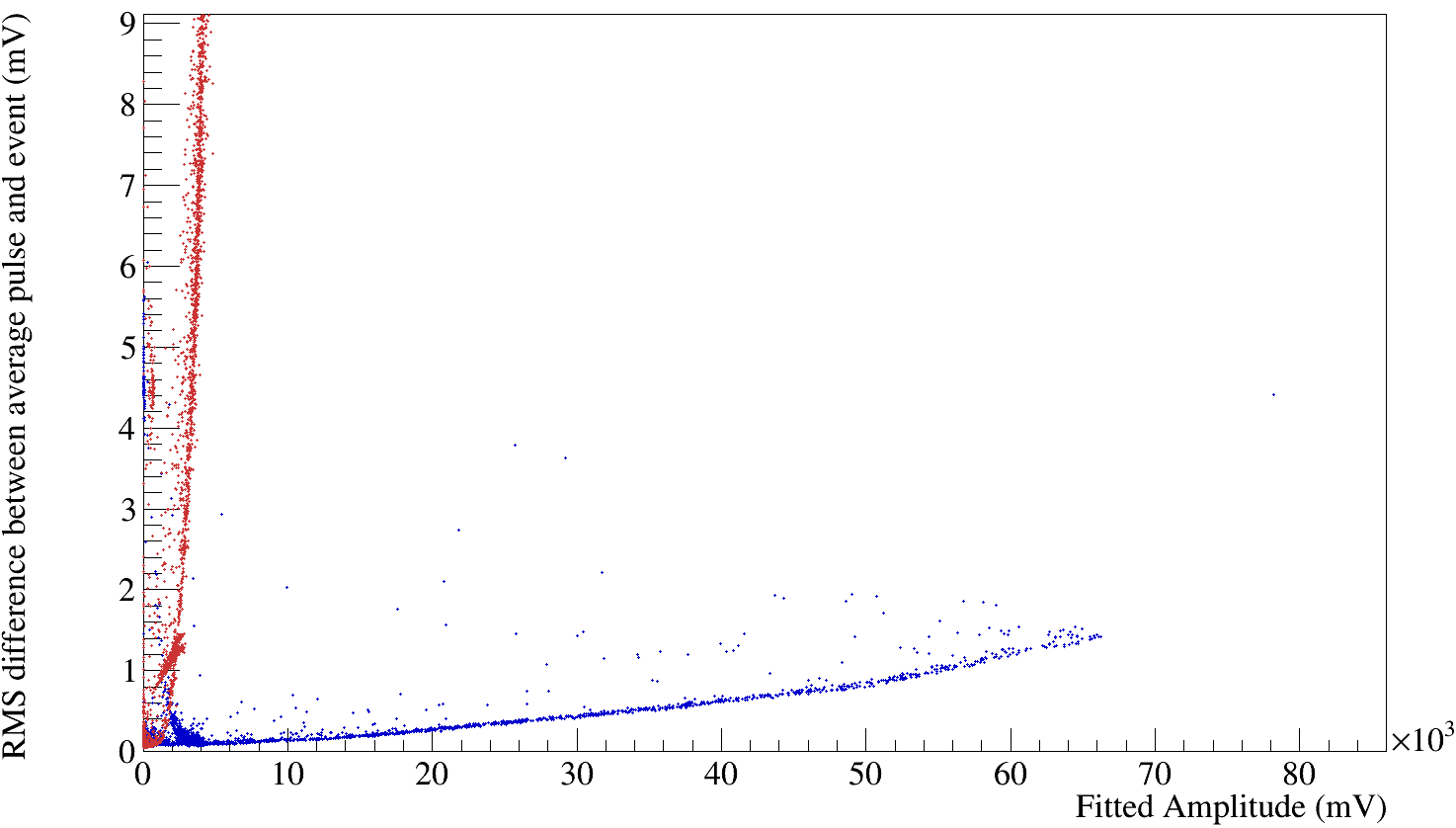}
\caption{Distribution of the rms difference versus the fitted amplitude for the normal fit (red) and the truncated fit (blue).}
\label{fig:rms_vs_amp}
\end{figure}

Besides just inspecting single pulses, a very convincing plot on the capabilities of extending the dynamical response region of the detector is presented in Figure \ref{fig:rms_vs_amp}. In this scatter plot the points in red are the same points of Figure \ref{fig:linearityregion}, which means the ones resulting from the normal fit, while in blue the same kind of distribution but generated from the results of the truncated fit method is shown. It is clear how the rms difference is in general much lower in the truncated fit case and it grows much more slowly that normal fit distribution. Moreover, the amplitude range (and thus the effective dynamical range) is extended from approximately $3000$~mV to almost $70000$~mV with an rms difference under $1.6$~mV throughout the whole range.

\section{Energy Calibration with Truncated fit method and calibration source}\label{sec:energy_cal}
\begin{figure}[H]
\centering
\includegraphics[scale=0.55]{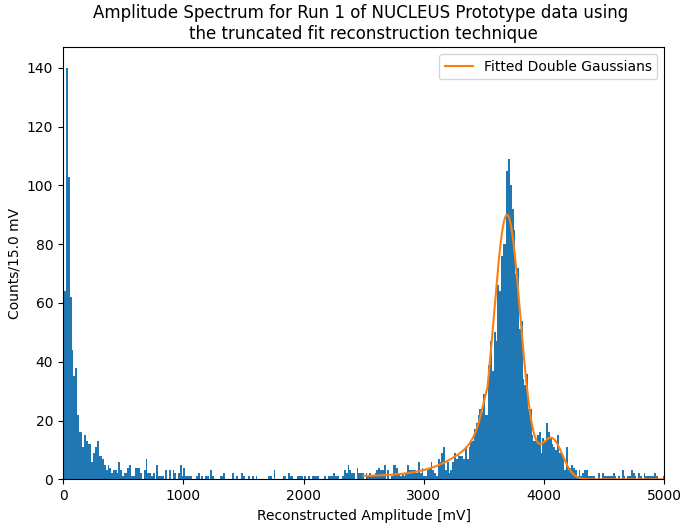}
\caption{Amplitude spectrum for the Run 1 NUCLEUS prototype data calculated with the \textbf{truncated fit} method.}
\label{fig:truncated_spectrum}
\end{figure}

Once the \textbf{truncated fit} is performed and all the pulses are reconstructed it is possible to carry on with the calibration and analysis of the amplitude spectrum of the RUN 1 acquisition. The spectrum is shown in Figure \ref{fig:truncated_spectrum}. Since the aim of these section is to use the events produced by the $^{55}$Fe sample to perform the energy calibration of the detector the amplitude spectrum is only plotted up to $5000$~mV since the histogram peaks by the x-ray source are already clearly visible around $4000$~mV. \\

In order to perform the energy calibration the center of the peaks due to the iron must be extrapolated, this has been done fitting the data between $2500$~mV and $5000$~mV with two uncorrelated gaussians. The results of the fit are presented in Table \ref{tab:fit_results} below. The results for the second gaussian have higher uncertainty with respect to the first due to the less amount of events (as expected) which makes the peak's signal-to-noise ratio of the histogram less favourable. A simple test that can be performed to check the quality of the fit is to check the ratio of the centers of the gaussians, this ratio is in fact independent of the units used and is well known from x-ray spectroscopy studies:

$$\text{Center Ratio} = \frac{4063~mV}{3691~mV} = 1.1005\pm 0.0051$$

the expected value is $1.1002$ and is thus compatible with the measured one. Besides the compatibility in terms of standard deviation the two values differ from each other of about $0.033\%$ and thus the fit can be considered successful.

\begin{table}[H]
\centering
\caption{Results of the amplitude spectrum fit with two independent gaussians.}
\begin{tabular}{ccc}
\hline
Parameter               & 1$^{st}$ Gaussian & 2$^{nd}$ Gaussian \\
\hline
\hline
Center (mV)             & $3691\pm2.8$      & $4063 \pm 17$     \\
Amplitude               & $90.0 \pm 1.9$    & $13.8\pm 2.1$     \\
Standard Deviation (mV) & $112.0 \pm 3.1$   & $88\pm 17$       	\\
\hline
\end{tabular}
\label{tab:fit_results}
\end{table}

A detail worth mentioning are the data cuts applied to produce a clear amplitude spectrum:
\begin{itemize}
    \item \textbf{BaselineSlope} in the range from $-0.005\frac{\text{mV}}{\text{s}}$ to $0.01\frac{\text{mV}}{\text{s}}$: The importance of this cut is due to the fact that if the events present a slope the average pulse does not represent well such pulse and thus the truncated fit does not converge correctly.
    \item \textbf{MaxMinInWindow}<$1.6\cdot$\textbf{MaxBaseline}: the reason for this cut is to eliminate spurious events in which the total range of the pulse is much bigger than the trigger sample like a small triggering event immediately followed by a much more ample one. This actually presents quite a sizeable population of events with fitted amplitudes below $1000$~mV. This cut in particular enables the use of events with piled up pulses of similar amplitude since the average pulse only presents one peak and thus will be fitted to only one of the events in the acquisition window. The multiplicative constant was found via trial and error.
    \item \textbf{RightLeftBaseline}$<200$~mV : this cut serves to isolate only those pulses on which the truncated fit performs best.
\end{itemize}

At this point the only task left is to actually perform the calibration by converting the fitted amplitude to a deposited energy using the known energies of the iron peaks. In order to do so a linear relation is assumed and the following formula is used: 
\begin{equation}
E = \frac{\Delta E}{\Delta A}\cdot A+E_0
\end{equation}

where $\Delta E$ is the difference in energy between the $^{55}$Fe peaks, $\Delta A$ is the difference in amplitude of the gaussian centers and $E_0$ is the energy of the lower energy x-ray iron peak. This formula converts the fitted amplitude $A$ of an event to the corresponding energy $E$. The linearity assumption shows its validity in the comparison of the gaussian center ratio with the expected iron peak energy ratio, in fact these two numbers only coincide if a linear relation subsists between the two units.\\

\begin{figure}[H]
\centering
\includegraphics[scale=0.55]{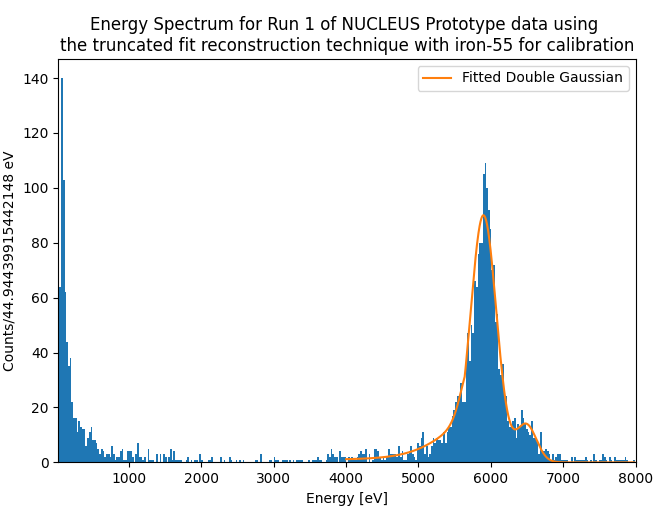}
\caption{Energy spectrum for the Run 1 NUCLEUS prototype data.}
\label{fig:energy_spectrum}
\end{figure}

The amplitude spectrum converted in energy spectrum via the calibration is presented in Figure \ref{fig:energy_spectrum} and is strikingly similar to the one presented in \cite{RUN1} that is calculated from the same starting data. The reason why the study of the energy spectrum from RUN 1 of the NUCLEUS prototype data was performed once again and the results from \cite{RUN1} where not simply used is because the aim was not to actually quantify the spectrum but to build an automatic data analysis protocol to be used in the DIANA analysis framework. In fact, in \cite{RUN1} the analysis was performed using detector stability and data cuts specific for RUN 1 as well as an almost event by event analysis to build the average pulse. Moreover, the procedure followed in \cite{RUN1} encountered the problem of not having enough pulses to build a seemingly noiseless template pulse and thus after building the average pulse it was fitted to the theoretical model shown in section \ref{chap:pulse_shape}. On the other hand, the analysis procedure described here is model independent and does not require any major theoretical inputs. Last but not least, this study served to make \textbf{DIANA} compatible with the data and analysis of the NUCLEUS experiment since until now the collaboration did not have a main analysis framework nor one as tested and versatile.

\section{Linearity response region detailed event characterization with optimum filter}\label{sec:optimum_filter}
To conclude the analysis and reach the same results as \cite{RUN1} what is left is to run the \textbf{optimum filter} in the linear response region. In fact, while the truncated fit method is necessary to extend the sensors range and to perform the energy calibration, to have a precise reading of the linear region the information from the noise acquisition must be used as well. This is exactly the role of the optimum filter that takes as inputs the noise power spectrum and the average pulse and looks for the maximum correlation of these with the acquired pulse. A more detailed description of the inner workings of the optimum filter can be found in \cite{optimum_filter} and is beyond the scope of this dissertation since the optimum filter utility was already present in \textbf{DIANA} and has already been extensively tested.\\

\begin{figure}[H]
\centering
\includegraphics[scale=0.45]{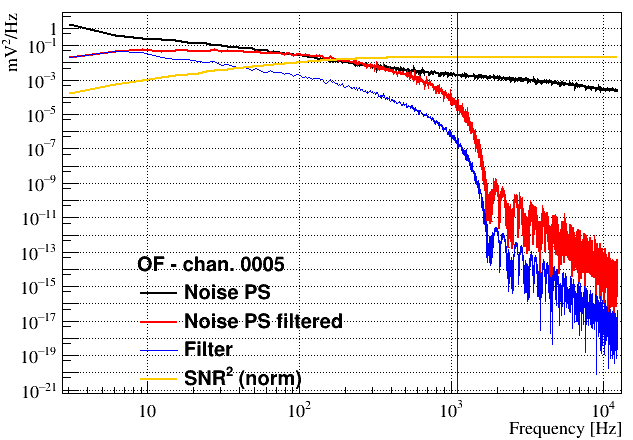}
\caption{Noise power spectrum for the RUN 1 acquisition.}
\label{fig:noise_power_spectrum}
\end{figure}

To build the noise power spectrum the "average\_noise\_power\_spectrum" sequence was used in \textbf{DIANA}. The sequence works only on noise acquisitions and filters them to find the optimal ones to build the average power spectrum. The first characteristic to be imposed is that the peak finding algorithm described in section \ref{sec:peak_identifier} must find no peaks at all, to reinforce this condition the cut $5~mV<$\textbf{MaxMinInWindowRMS}$<15$~mV is used. Similarly to before the requirement of a negative baseline is imposed. Finally, in order not to have slopes in the noise acquisitions and to not have extremely noisy and spurious samples the cut \textbf{BaselineRMS}$\leq 10$~mV and $-0.005\frac{\text{mV}}{\text{s}}<$\textbf{BaselineSlope} $<0.005\frac{\text{mV}}{\text{s}}$ are used. This cuts select a sample of 205 events out of the 1276 total noise acquisitions, the calculated noise power spectrum is presented in Figure \ref{fig:noise_power_spectrum}.\\

\begin{figure}[H]
\centering
\includegraphics[scale=0.42]{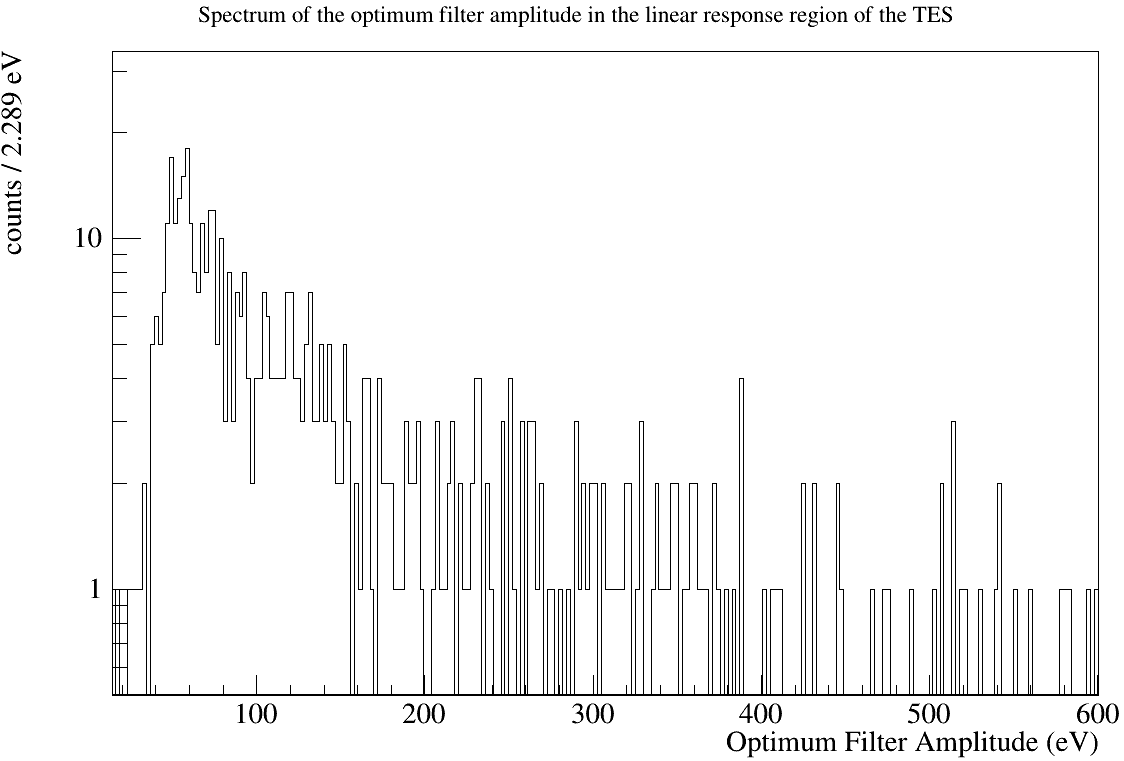}
\caption{Energy spectrum for the Run 1 NUCLEUS prototype data in the linear response region of the TES generated with the optimum filter.}
\label{fig:optimum_filter}
\end{figure}

After deriving the average noise power spectrum and the average pulse the optimum filter was ran on the data using the "Amplitude" DIANA sequence. The result is presented in Figure \ref{fig:optimum_filter} after undergoing the same energy conversion described for the truncated fit technique. The fact that the same energy conversion can be used is due to the fact that upon inspection the amplitudes calculated with the optimum filter and the truncated fit respectively give practically the same result up to a deviation of $\mathcal{O}$(1\%) which is definitely not the dominating uncertainty at this level of the analysis.

The resemblance of this spectrum with the inset in Fig. 2 of \cite{RUN1} is further proof of the good results obtained by the newly built analysis protocol described in detail in this chapter and the fact that error induced by the energy conversion is definitely not dominant.

\chapter{Sensitivity Studies}
\label{chap:6}

\lettrine[lines=2, findent=3pt, nindent=0pt]{A}{}n essential part of every experiment is the study of the expected statistical significance of a potential discovery. In this chapter the various elements of a sensitivity study for the NUCLEUS experiment will be described. Some preliminary results will also be presented using a simplistic toy model.\\

In the NUCLEUS experiment two different tools are used to perform sensitivity studies, \textit{PyCEnNS}, a  python based toolkit initially developed by J. Rothe for studies presented in the preliminary phases of NUCLEUS and heavily updated during the work presented in this dissertation, and \textit{FORECAST}, a C++ and ROOT based framework that is particularly oriented at the description of the nuclear reactor. Both tools have been used in parallel with different calculation algorithms in order to validate each other's results. The work presented in this chapter refers mostly to the \textit{PyCEnNS} framework unless specified otherwise.
\section{Significance and Likelihood Ratio Test}
The basic idea of a statistical test is to compare one or more measurements done during an experiment with the expected distributions calculated from theoretical and phenomenological considerations.\\

In the framework of the NUCLEUS experiment, as in most experiments, the hypothesis test needs to differentiate between two nested hypothesis: the \textit{null hypothesis}, which parametrizes the case in which only background phenomena are present, and the \textit{free hypothesis}, that considers both background and CE$\nu$NS to be present. The contributions due to CE$\nu$NS will often be label in this chapter as \textit{signal} contributions in order to keep the discussion as general as possible since most of the considerations derived here are common to other scenarios as well.\\

The most important definition in this field of study is the \textbf{likelihood function} $\mathcal{L}(D,\theta)$ which is the probability of observing the dataset $D$ given the values of the model parameters $\theta$. Two different sets of the $\theta$ parameters $\theta_{free}, \theta_{null}$ specify respectively the \textit{free} and \textit{null} models. In the case of the NUCLEUS experiment, as it is in most particle physics experiments, the idea is to consider the total event count and the energy spectrum to estimate what is the maximum probability of observing them with either the \textit{free} or the \textit{null} hypothesis. Once these values of the likelihood have been found their ratio is considered in order to evaluate with what certainty the observed signal is due to the desired physics.\\

The study of the statistical significance of a discovery can be performed following two different approaches that present different advantages and disadvantages. The first and most analytical approach is to use \textbf{unbinned statistics}, which is mathematically univocally defined and does not include any binning bias in the calculation. On the other hand, to produce a sensitivity estimate of the experiment several spectral measurements must be simulated which requires great computation time and can prove to be a limiting factor. The other approach is to use the \textbf{binned statistics} which is similar to the unbinned approach but compares the histograms of the spectra with the two hypothesis models. This second method has more or less the same computational demands as the first (depending on the binning used) if used in its basic formulation but allows the use of the \textit{Asimov} approach. The \textit{Asimov} method greatly cuts on computation time because it does not require to simulate a great number of experiments but instead uses the mean values for all the quantities that are randomically distributed; for example the \textit{Asimov energy spectrum} of an observation consists in a histogram whose bins do not contain an integer number but a real number corresponding to the expected number of counts that one would have averaging infinite experiments. The \textit{Asimov} approach requires only one computation cycle but is still subject to the same binning bias as the basic binned study.

\subsection{Unbinned Statistics}
As mentioned, one of the two approaches possible for the evaluation of the statistical significance is to use \textbf{unbinned statistics}. The probability of observing a certain spectrum is the product of the probability to observe the given number of total events (given by a poissonian distribution) and the probabilities of observing the individual events at their respective energies. This translates to a likelihood function:
\begin{equation}
\mathcal{L}(D,\theta) = \frac{\mu^n}{n!}e^{-\mu}\cdot \prod_{i=1,...,n}p(T_i)
\end{equation}
where $n$ is the number of total observed events each one with energy $T_i$, while $p(T)$ is the event rate $R(T)$ as a function of the energy normalized to unity and $\mu$ is the integral of $R(T)$, i.e. the expectation value of the number of observed events. \\

Since the quantity that estimates the statistical significance is function only of the ratio of the likelihoods corresponding to the \textit{null} and \textit{free} hypotheses all the multiplicative factors between the two models, such as the rate normalization or the number of events, get cancelled and can thus be taken away from the likelihood calculation which in turn becomes:

\begin{equation}\label{eq:unbin_like}
-\log(\mathcal{L}(\beta,\sigma))=\sum_{t\in isotopes}\left[\mu_{t}(\beta,\sigma)-\sum_{i=1,...,n_t}\log(R_{t}(T_i|\beta,\sigma))\right]
\end{equation}

where all the quantities presented are function of $\beta$ and $\sigma$ that are respectively the background and signal hypothesis relative contributions to the model. In the previous equation a sum over all the recoil isotopes is present in order to account for different target materials. The total expected number of events corresponds to:
$$\mu_t(\beta,\sigma)=\mu^{signal}_t(\sigma) + \mu^{background}_t(\beta)$$
where $\mu^{signal}_t(\sigma)$ and $\mu^{background}_t(\beta)$ are defined in section \ref{sec:unbin_spectrum} and are the expected number of counts of respectively \textit{signal} and \textit{background}.\\

To compare the likelihood of the different hypotheses one first defines the functions:
$$f_{null}(\beta)=-\log\left(\mathcal{L}(\beta,0)\right)\qquad f_{free}(\beta,\sigma) = -\log\left(\mathcal{L}(\beta,\sigma)\right)$$
which parametrize the background, or \textit{null}, and signal plus background, or \textit{free}, hypotheses. These two functions must then be minimized in $\beta$ and $\sigma$ to find the maximum observation probability with the two models:

$$\min\left[f_{null}(\beta)\right] = f_{null}(\beta_0) \qquad \min\left[f_{free}(\beta,\sigma)\right]=f_{free}(\beta_1,\sigma_1)$$

and the statistical significance, i.e. the number of standard deviations that separates the \textit{free} hypothesis from the \textit{background} one, of a discovery is given by the square root of the test statistics, and following Wilk's theorem it corresponds to:

\begin{equation}
Z = \sqrt{2\left(f_{null}(\beta_{0}) - f_{free}(\beta_{1},\sigma_{1})\right)}
\end{equation}

and whenever the difference is found to be negative the value of the statistical significance is set to 0. For more details on the statistical background of this brief discussion see \cite{johannes} and \cite{lista}.

\subsubsection{Recoil Spectra Simulation}\label{sec:unbin_spectrum}
In the above discussion an essential part of the calculation is the experimental recoil rate $R_t(T)$. When conducting a survey on the discovery potential of an experiment one must simulate a great number of experimental recoil rates in order to obtain the average statistical significance that the experiment will observe. Performing this great deal of simulations is one of the most computationally demanding steps in the whole procedure and must be performed with care.\\

In this section the calculations to derive the expected, or theoretical, recoil rate spectra $R^{signal}_{theo,t}(T)$ (see section \ref{sec:expected_spectra}) and $R^{background}_{theo,t}(T)$ are assumed to have been already performed and are described in the following sections (the subscript $t$ indicating the recoil target will be dropped for the rest of this section since the procedure is identical for all of them). \\

\begin{figure}[H]
\centering
\includegraphics[scale=0.5]{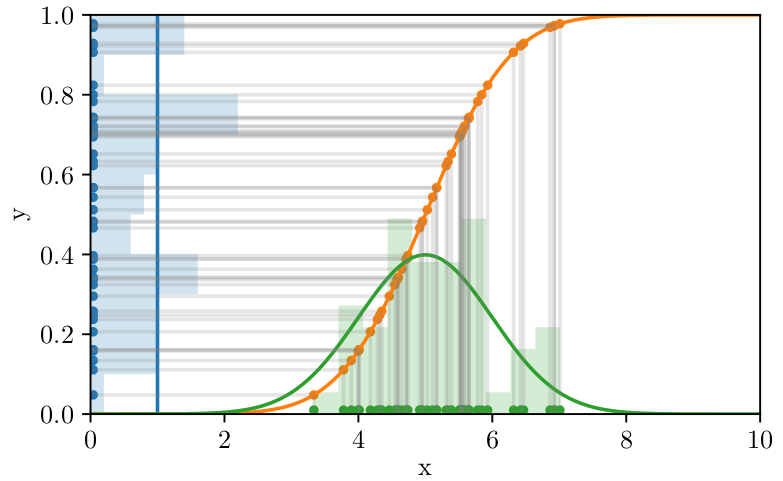}
\caption{Illustration of the inverse transform sampling procedure. The green solid line is the probability density function (pdf), in yellow the cumulative of the pdf. In blue the uniformly distributed samples that are projected on the cumulative and the green points are extracted following the pdf. Figure from \cite{johannes}}
\label{fig:inverse_transform}
\end{figure}

To randomly extract a distribution of events that follows the theoretical interaction rate the \textit{inverse transform sampling} technique is used to cut down on computation time. The procedure consists of extracting with a flat distribution a number of samples between 0 and 1, these values are then used to calculate the inverse of the cumulative function. The values extracted in this way follow the desired distribution, in this case the one given by theoretical recoil spectrum. The inverse transform sampling process is illustrated in Figure \ref{fig:inverse_transform}.

The first step to perform the inverse transform sampling of the spectrum is to calculate the cumulative function by integrating the recoil rate and normalizing the result:
$$F^{signal}(T)=\frac{\int_{0}^{T}R^{signal}(T')dT'}{\int_{0}^{E_{max}}R^{signal}(T')dT'}\qquad F^{background}(T) = \frac{\int_{0}^{T}R^{background}(T')dT'}{\int_{0}^{E_{max}}R^{background}(T')dT'}$$

The second step is to uniformly extract a set of samples to project on the cumulative distribution. The number of samples to be extracted samples is as well randomly distributed and is extracted from a poissonian distribution with average $\mu$ corresponding to the expected total number of counts in the energy detection region. Such energy range goes from $E_T$, a lower energy threshold (i.e. the lowest energy detectable, set conservatively to $20$~eV), to $E_{max}$ (which is just a maximum threshold arbitrarily decided from experimental considerations and in this dissertation is set to $1$~keV). The expected number of counts for signal and background is respectively: 

$$\mu^{signal}(\sigma) =\sigma \cdot\epsilon \int_{E_T}^{E_{max}}R^{signal}(T)dT\qquad\mu^{background}(\beta) = \beta \cdot\epsilon \int_{E_T}^{E_{max}}R^{background}(T)dT$$
where $\epsilon$ is the detector exposure in $\left[\text{kg}~\text{days}\right]$ and $\beta,\sigma$ are the hypothesis strength parameters mentioned before.\\

To take into account the lower energy threshold the uniformly distributed samples are not extracted between 0 and 1 but between $u_T$ and 1, where $u_T$ for the background and the desired signal is:

$$u_T^{signal} = \frac{\int_{0}^{E_T}R^{signal}(T)dT}{\int_{0}^{E_{max}}R^{signal}(T)dT}\qquad u_T^{background} = \frac{\int_{0}^{E_T}R^{background}(T)dT}{\int_{0}^{E_{max}}R^{background}(T)dT}$$

As shown here, the procedure of sample extraction of the recoil energy is the same for \textit{signal} and \textit{background}. To generate the simulated recoil spectrum the two sets of extracted recoil energy samples are then shuffled together loosing every memory of their origin (\textit{signal} or \textit{background}).\\

\subsection{Binned Statistics}
The main difference between the \textit{binned statistics} and the \textit{unbinned} one is the used of a binned recoil spectrum. The use of a binned spectrum  changes the loglikelihood expression to (with all the common factors already simplified):
\begin{equation}\label{eq:binned_like}
-\log(\mathcal{L}(\beta,\sigma))=\sum_{t\in isotopes}\left[\mu_t(\beta,\sigma)-\sum_{i,...,n_t}n_i\log(\mu_i(\beta,\sigma))\right]
\end{equation}
where $\mu_t$ has the same meaning of the unbinned case, $\mu_i$ is the expected event number in bin $i$ and $n_i$ is the measured, or simulated, event rate in bin $i$; the likelihood presented here assumes that the number of events present in each bin follows a poissonian distribution.\\

The rest of the steps for evaluating the statistical significance are exactly the same to the ones presented in the unbinned, i.e. find $\beta_0$ and $\beta_1,\sigma_1$ by minimizing $f_{null}$ and $f_{free}$ and then calculate $Z$ using the same formula:

\begin{equation}
Z = \sqrt{2\left(f_{null}(\beta_{0}) - f_{free}(\beta_{1},\sigma_{1})\right)}
\end{equation}

As before, for more detail look at \cite{johannes} and \cite{lista}, since a more detail description of this procedure is beyond the scope of this dissertation and has already been thoroughly presented in various books and articles.
\subsection{Binned Recoil Spectra Simulation}\label{sec:exp_binnig}
The simulation of the binned recoil spectrum is computationally different from the unbinned case and is worth to be described.\\

\paragraph{Binning:} The first step in such a computation is to decide the binning strategy. The choice of the bin size is a bargain between the keeping the computational time short, thus choosing a larger binning, and keeping the spectral characteristics, which requires the use of a smaller bin size. Since most of the spectral information due to CE$\nu$NS is in the low energy part of the recoil spectrum, an exponentially increasing bin size is chosen for this work. The edges of the bins are then given from:

\begin{equation}
bin_i = E_0 + w_0 \left(\frac{1-q^i}{1-q}\right)
\end{equation}
where $E_0$ is the minimum energy present in the spectrum, $w_0$ is the initial width and is set to 20\% of $E_0$ and $q$ is the argument of the exponential that increases the bin size and is evaluated by imposing:
$$\frac{E_{max}-E_{0}}{w_0} = \frac{1-q^{n_{bins}}}{1-q}$$
where $n_{bins}$ is the total number of bins and is priorly chosen by the user. This last condition to calculate $q$ is equivalent to requiring that the total number of bins spans exactly the specified energy range. \\

To explicitly see the exponential increase of the binning it is helpful to compute the difference between two neighboring ones:
$$\Delta bin_i = bin_{i+1}-bin_i = \cancel{E_0 }+ w_0 \left(\frac{1-q^{i+1}}{1-q}\right) - \cancel{E_0} - w_0 \left(\frac{1-q^{i+1}}{1-q}\right) =$$
$$=\frac{w_0}{1-q}\left(1-q^{i+1}-1+q^i\right) = \frac{w_0q^i}{\cancel{1-q}}(\cancel{1-q}) = w_0q^i$$

\paragraph{Spectrum Simulation:} Once the binning has been chosen for the recoil spectra of \textit{background} and \textit{signal}, experimental datasets must be simulated. As for the unbinned case, here it will be assume that a prior calculation to extract the theoretical event rate spectrum has been performed for both the CE$\nu$NS and background models. \\

The first step in the spectrum generation is to extract the total number of expected events, the procedure is exactly the same as the one presented in the unbinned case, i.e.:

$$\mu^{signal}(\sigma) =\sigma \cdot\epsilon \int_{E_T}^{E_{max}}R^{signal}(T)dT\qquad\mu^{background}(\beta) = \beta \cdot\epsilon \int_{E_T}^{E_{max}}R^{background}(T)dT$$

where $\epsilon$ is the detector exposure in $\left[kg~days\right]$ and $\beta,\sigma$ are the hypothesis strength parameters.\\

The second step consists in choosing a random number of events for each of the chosen bins. In order to do this the theoretical recoil rate is locally integrated in the range spanned by each bin in order to extract the expected number of counts for each bin. This expected number of counts for each bin is then used to extract, with a poissonian distribution, a random number of events present in each bin. This step is performed for both the \textit{background} and the \textit{signal} models alike.\\

Since the binning of the two models coincides, the two generated histograms are then summed in order to simulate an experimental event spectrum. At this point, all the ingredients needed for the likelihood calculations are present and the statistical significance can be extracted. Like the case of the unbinned procedure, several different spectra must be simulated in order to correctly quantify the expected discovery potential of the experiment.

\subsubsection{Asimov Spectrum}
The generation of the binned recoil spectrum is still a computationally demanding task, and the fact that it needs to be performed a great number of times in order to extract the right values puts some serious limitations to the amount of calculations that can be performed. Fortunately this can be overcome by the use of the \textit{Asimov spectrum}. \\

In the second step of the simulation of the binned spectra it is necessary to calculate the expected number of counts for each bin to be later used for random sampling purposes. On the other hand, if one averages an infinite amount of randomly sampled spectra the obtained results would be the starting spectrum with the bin-heights corresponding to the expectation values. This "expected" recoil spectrum goes under the name of \textbf{Asimov Spectrum}. Using this spectrum then allows to severely cut down on computation time since only one deterministic calculation needs to be performed instead of having the need to average several random generations. The validity of using such a method is discussed in \cite{lista} and \cite{Cowan_2011}.

\section{Estimation of expected signal and background interaction rates at VNS}\label{sec:expected_spectra}
\subsection{CE$\nu$NS cross-section parametrization}\label{sec:barranco}
The cross-section calculations described in section \ref{sec:theo} are useful for the understanding of the theoretical framework of CE$\nu$NS but do not prove to be the best formulation for sensitivity studies. The other differential cross-section parametrization that was taken into consideration for the purpose of this study is the one presented by J. Barranco et al. in \cite{barranco}:
\begin{equation}
\frac{d\sigma}{dT}=\frac{G_F^2 \cdot M}{2\pi}\left\{(G_V+G_A)^2+(G_V-G_A)^2\left(1-\frac{T}{E_\nu}\right)^2 -(G_V^2 -G_A^2)\frac{MT}{E_\nu^2} \right\}
\end{equation}
where $M$ and $T$ are respectively the mass and the kinetic energy of the recoiling nucleus, the $G_V$ and $G_A$ couplings are:
$$G_V = \left[(g^p_V+2\epsilon^u_V+\epsilon^d_V)\cdot Z + (g^n_V+\epsilon^u_V+2\epsilon^d_V)\cdot N\right]\cdot F^V_{nuc}(Q^2)$$
$$G_A=\left[(g^p_A+2\epsilon^u_A+\epsilon^d_A)(Z_+-Z_-)+(g^n_A+\epsilon^u_A+2*\epsilon^d_A)*(N_+-N_-)\right]\cdot F^A_{nuc}(Q^2)$$
where $Z$ and $N$ represent the number of protons and neutrons, while $Z_\pm$ and $N_\pm$ are respectively the number of protons and neutrons with spin up and down. The vector and axial nuclear form factors $F^V_{nuc},F^A_{nuc}$ can be assumed to be of order of unity for low energy transfers, thus satisfying the coherency condition described in section \ref{sec:xsec}. The $g^p_A$, $g^p_V$, $g^n_A$, $g^n_V$ parameters represent the vector and axial standard model couplings of protons and neutrons (with some non-standard model corrections that for the sake of this dissertation are set to zero). Moreover, the $\epsilon$ parameters describe non-standard neutrino interactions (NSI) like flavour changing contributions and non-universal terms.\\

This parametrization is then clearly much more suitable for further sensitivity studies beyond the standard model and has is the reason why it has been chosen by the NUCLEUS sensitivity studies team as the standard formulation of the cross-section. Since the sensitivity study tools are still being developed the results presented in this dissertation will only take into account standard model terms.\\

Another reason to choose this type of parametrization is that the spin of the recoiling nucleus is taken into account intrinsically by the two terms $(Z_+-Z_-)$ and $(N_+-N_-)$ which require nuclear spin statistics studies to be evaluated. For the sake of this study the two spin dependent terms where evaluated as $0$ if the number protons (neutrons) in the recoiling species is even and as $1$ if it is odd, since in first approximation one can assume that in a cryostat the nuclei are in their minimum energy state.\\

The values used in this study for the mentioned parameters are presented in Table \ref{tab:param_sens} below.\\

\begin{table}[H]
\centering
\caption{Numerical values of the cross-section parameters used in this dissertation.}
\begin{tabular}{cc}
\hline
Parameter               & Value  \\
\hline
\hline
$g_V^p$ & 0.03001\\
$g_V^n$ & -0.5116\\
$g_A^p$ & 0.4955\\
$g_A^n$ & -0.8224\\
$\epsilon_V^u$ & 0.0\\
$\epsilon_V^d$ & 0.0\\
$\epsilon_A^u$ & 0.0\\
$\epsilon_A^d$ & 0.0\\
$F^V_{nuc}$ & 1.0\\
$F^A_{nuc}$ & 1.0\\
\hline
\end{tabular}
\label{tab:param_sens}
\end{table}
 
\subsection{Neutrino Flux at experimental site}\label{sec:nu_flux}
The first element necessary for the sensitivity study is the expected neutrino flux at the measuring site (VNS). The flux at the reactor core is simulated using \textit{FORECAST}. The calculation takes into account the fission neutrino spectra of various radioactive isotopes present in typical nuclear reactor fuels. It also takes into account neutron capture phenomena that normally happen in reactor cores. The combination of the neutrino spectra coming from the various fuel components is performed using the BESTIOLE code from CEA, which combines multiple theoretical calculations, and differs from experimental data of about 10\%. Since the sensitivity studies for NUCLEUS are still being developed, in order not to over-complicate calculations, the time evolution of the fuel components are not considered but are instead set to their mean values for the rest of this dissertation. Moreover, an 80\% loading factor of the nuclear reactor is considered. More details can be found in \cite{Alex_thesis}.\\

\begin{figure}[H]
\centering
\includegraphics[scale=0.5]{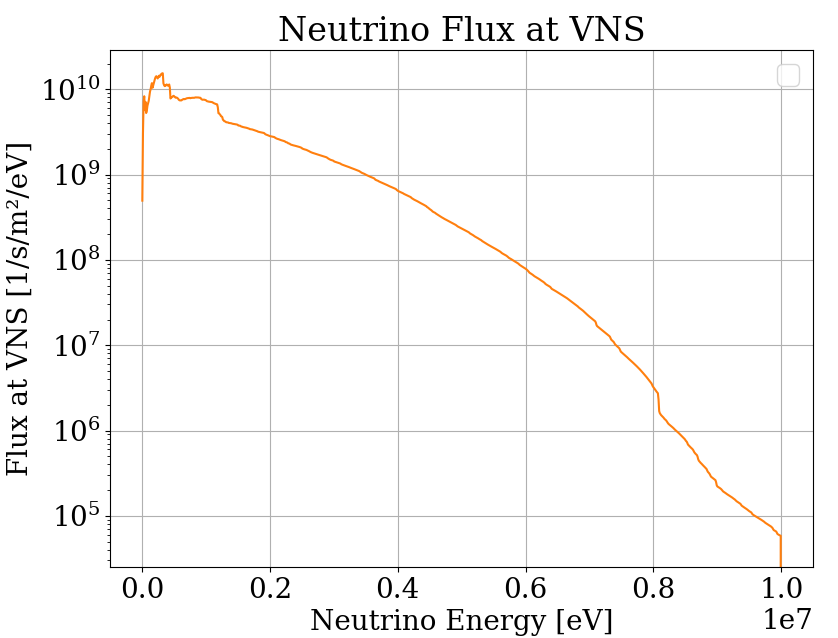}
\caption{Neutrino spectrum at detector site (VNS) considering an 80\% loading factor and no time evolution of the fuel components.}
\label{fig:VNS_flux}
\end{figure}

Once the neutrino flux at the reactor core is evaluated it needs to be scaled in order to account for the VNS distance from both reactor cores present at Chooz. To scale the flux an effective distance of $58.8$~m is used, this is the distance from a virtual reactor core whose flux is the combination of the two reactors present at the power plant. The resulting flux is presented in Figure \ref{fig:VNS_flux}, as mentioned in section \ref{sec:exp_binnig}, it can be seen that the most complex part of the neutrino spectrum is present in the low energy regions.

\subsection{Expected Interaction Rate at VNS}
The basis of all the statistical calculations presented in this chapter is the calculation of the expected interaction rate at the VNS. To do so the differential cross-section described in \ref{sec:barranco} is combined with the neutrino flux described in section \ref{sec:nu_flux} and integrated over the incoming neutrino energies in order to calculate the interaction rate:
\begin{equation}\label{eq:rate}
R_{t}(T) = \frac{N_A \cdot 10^3}{A_t}\int_{E_{min}^\nu(T)}^{E^\nu_{max}} \frac{d\sigma_t}{dT}(E_\nu) \cdot \phi_\nu(E_\nu)dE_\nu \qquad \left[\frac{1}{\text{eV~kg~day}}\right]
\end{equation}
where the term multiplying the integral converts the atomic mass number  $A_t$ of the nuclear recoiling species into the nuclear weight ($N_A$ is the Avogadro number), as before the subscript $t$ indicates a specific recoiling material. In the previous equation particular care must be taken when discussing the integration interval, in particular the lower limit. For a given recoil energy $T$ there is a minimum neutrino energy under which the neutrino does not have enough energy to induce the desired nuclear recoil, the kinematics relations give (see \cite{astro}):
$$E^\nu_{min}(T) = \frac{T}{2}\cdot \left[1+\sqrt{1+\frac{2M}{T}}\right]$$

When an absorber material is specified for the sensitivity calculations, one must take into account the fact that in the target material there are multiple isotopes of the same nuclear species. The way the different isotopes are taken into account in both the \textit{FORECAST} and the \textit{PyCEnNS} frameworks is to calculate the recoil rate $R_{t,t'}$ for each isotope $t'$ of the chosen target material $t$, and then average the single isotopic rates weighing them with the respective isotopic fractions $f_{t'}$ and masses $M_{t'}$:
\begin{equation}
R_{t}(T) = \frac{\sum_{t'} f_{t'}\cdot M_{t'}\cdot R_{t,t'}(T)}{\sum_{t'} f_{t'}\cdot M_{t'}}
\end{equation}

where $R_{t,t'}(T)$ is defined exactly the same as equation \ref{eq:rate} but all the quantites related to the target material are calculated specifically for the isotope $t'$.

\section{Updating the PyCEnNS sensitivity calculation framework}
One of the most demanding tasks during the \textit{PyCEnNS} upgrade was to make the two different sensitivity study frameworks agree on the interaction rate calculation. This task involved upgrading multiple parts of the \textit{PyCEnNS} toolkit that not only involved computational accuracy and efficiency but also restructuring the code in order to have a more user friendly interface for collaboration-wide use.\\

As it can been seen from the discussions above the simulation of an experimental recoil spectrum can be computationally very demanding, it requires two steps of random number extraction and a function inversion. Moreover the requirements on computational resources and time is increased by the fact that several experiments must be simulated in order to accurately calculate the expected statistical significance. \\

\subsubsection{Reaching numerical agreement between PyCEnNS and FORECAST}
As mentioned, several improvements were made to \textit{PyCEnNS} in order to have the best possible agreement with the \textit{FORECAST} predictions and to cut on computation time and resources. The various updates are described in this section along with their effects on the output results.

\paragraph{Numerical Integration algorithm} The first upgrade made was the substitution of the integration method, both the old and new integration functions where chosen from the ones present in the \textit{Scipy} python package due to its extensive testing and maintenance. The old method used for this calculation was the \textit{quadrature} function  \cite{quad} which is derived from the \textit{QUADPACK} Fortan 77 numerical integration library. This method is extremely complex, it uses adaptive integration steps and other computational algorithms to deal with functions that are not easily integrated. On the other hand, the functions that need to be numerically integrated for CE$\nu$NS predictions are not extremely complex and the \textit{quadrature} method proves to be too computationally demanding without reaching satisfying results.\\

\begin{figure}[H]
\centering
\includegraphics[scale=0.5]{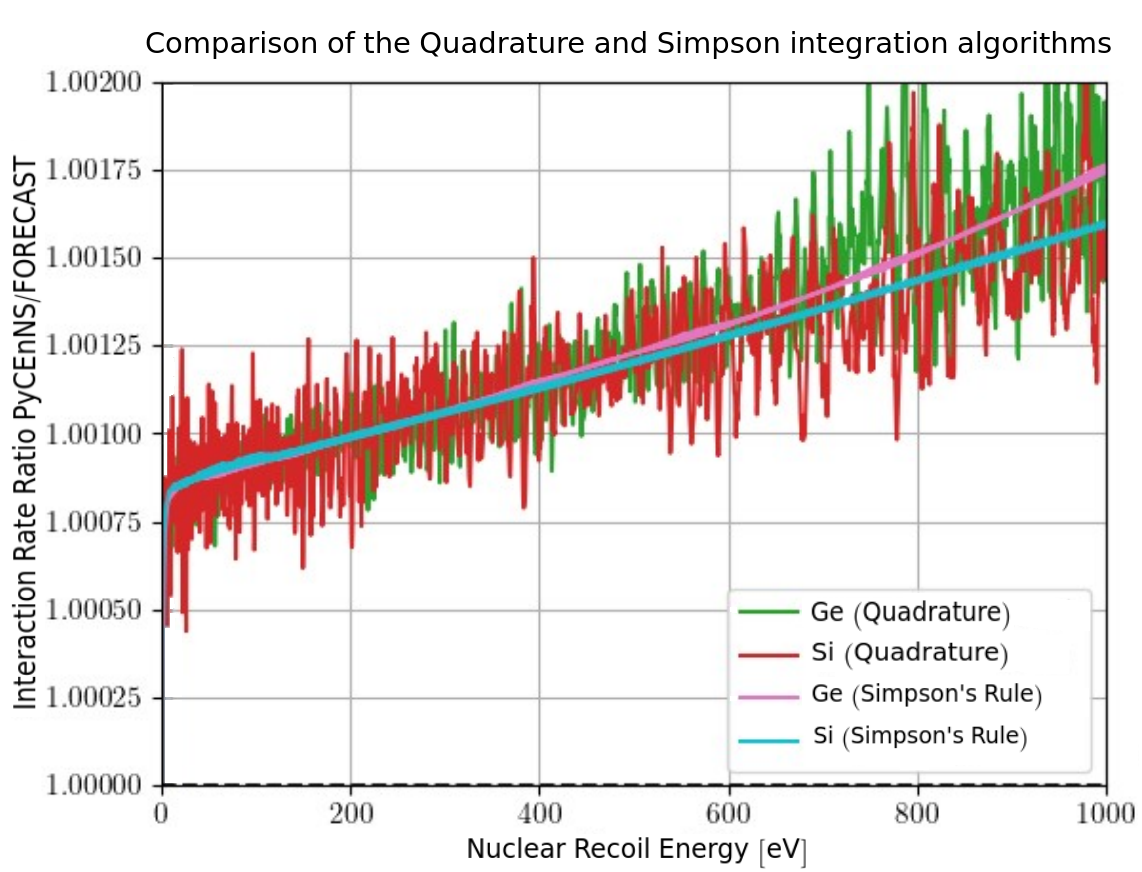}
\caption{Comparison between two different integration methods in the \textit{PyCEnNS} toolkit, \textit{quadrature} (green and red) and \textit{Simpson's rule} (pink and light blue). The results are presented for two different target materials, germanium and silicon, and are compared to \textit{FORECAST}'s results }
\label{fig:quad}
\end{figure}

For this reason the \textit{Simpson's rule} for integration \cite{simp} now substitutes the \textit{quadrature} method throughout the whole framework. This method is computationally much less demanding since it consists only in a weighted sum of the samples and is based on polynomial approximation of the integration function in small sub-ranges of the whole integration region. Obviously, this integration method is not particularly suitable for the integration of extremely complex functions but works extraordinarily well with the functions present in the sensitivity studies. In Figure \ref{fig:quad} a comparison of the two integration methods is presented, in this picture the ratio between the interaction rates calculated from both \textit{PyCEnNS} and \textit{FORECAST} is shown for the two integration methods and for two different target materials, germanium and silicon. From this figure it is clear that \textit{Simpson's rule} presents much more stable results with respect to the \textit{quadrature} method. Moreover, being the \textit{Simpson's rule} a sum of samples it allows to use a matricial implementation of the calculations following the typical Python approach using the \textit{Numpy} library. Using this vector algebra approach made the \textit{PyCEnNS} code much quicker, with approximately an $80~\%$ reduction in computation time. Moreover, since the low energy region of the neutrino flux is the most structured from a spectral point of view, the integration is performed with two different integration steps, a finer integration is used in the low energy regime ($E_\nu\leq 20$~eV) while a 10 times larger integration step is used for the rest of the spectrum since the neutrino flux presents a much smoother behaviour.

\paragraph{Updating cross-section parametrization: }From Figure \ref{fig:quad} it is also visible a certain degree of disagreement between the \textit{PyCEnNS} and the \textit{FORECAST} results, in particular a $\sim 0.1\%$ offset and a rising slope in their ratio. Ideally one would expect that the two softwares would give exactly the same results, giving a constant ratio of 1. \\

The constant offset of about 0.1\% is due to the different precision in the output data and the physical constants used and has been quickly fixed. On the other hand the rising slope was due to a more fundamental issue caused by approximations in the cross-section physics. In fact, most CE$\nu$NS papers will present the differential cross-section of this process under two assumptions: the recoiling nucleus is not relativistic and the recoil energy is negligible with respect to the neutrino energy. While the first assumption can be considered true to a very high degree of precision (the maximum recoil energy is $\mathcal{O}(\text{keV})$ while the nuclear masses are of order of the tens of GeV) the second assumption is true only up to a precision of 0.1\% (the neutrino energies are typically in the MeV region while the recoil energies are of the keV order, which gives a factor 0.1\% between the two energy scales). The cross-section with only the non-relativistic nuclei approximation is:
\begin{equation}
\frac{d\sigma}{dT}\propto \left(1-\frac{T}{E_\nu} - \frac{T\cdot M}{2E_\nu^2}\right)
\end{equation}
and has an additional factor proportional to $\frac{T}{E_\nu}$ when compared to the normally presented differential cross-section:
\begin{equation}
\frac{d\sigma}{dT}\propto \left(1- \frac{T\cdot M}{2E_\nu^2}\right)
\end{equation}
this additional factor is already contained in the expression for the cross-section given in section \ref{sec:barranco} but is worth to be mentioned since the next generation of CE$\nu$NS experiments are devoted to high precision measurements of this process.\\

\begin{figure}[H]
\centering
\includegraphics[scale=0.5]{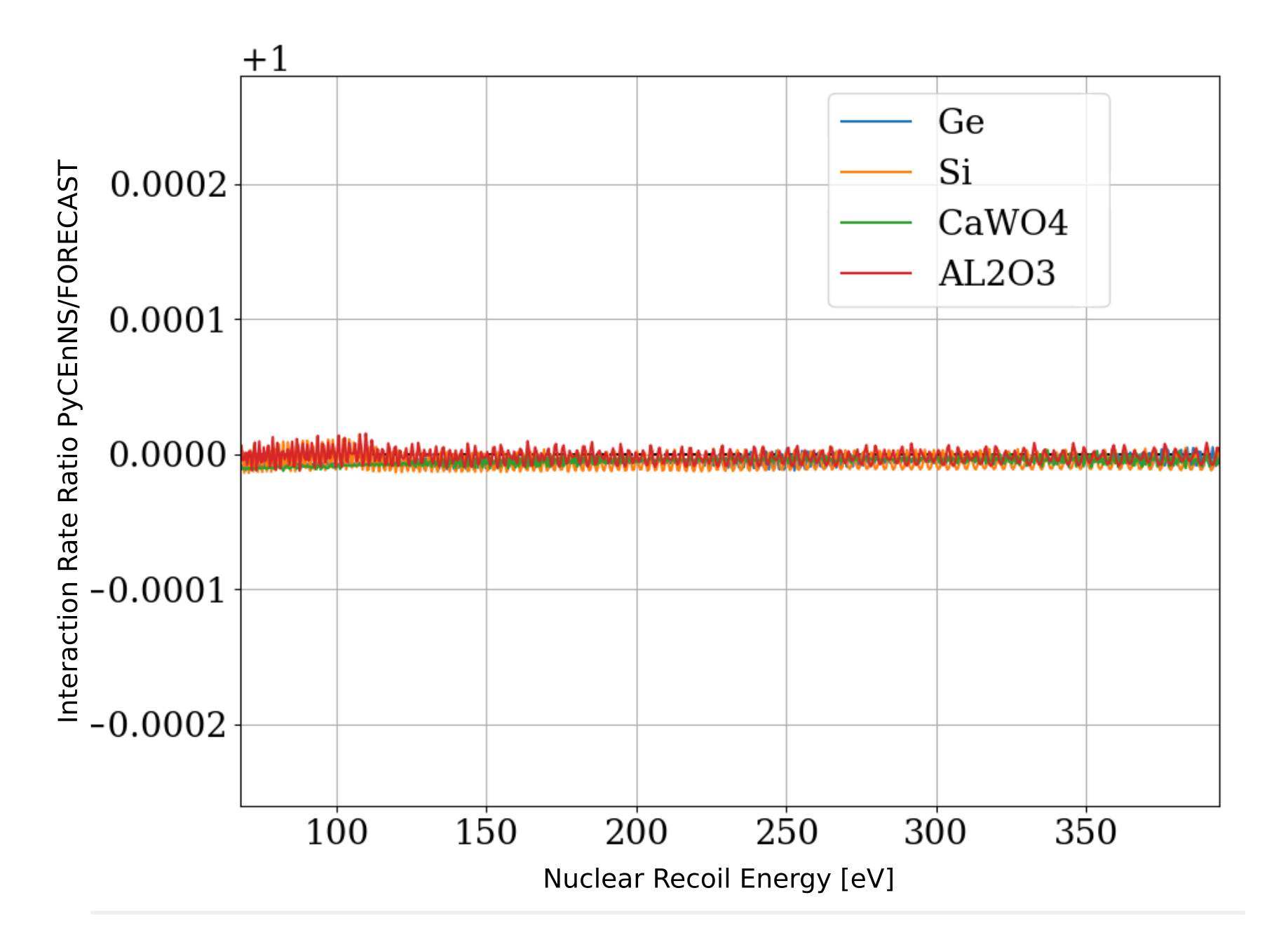}
\caption{Interaction rate at VNS ratio (\textit{PyCEnNS} over \textit{FORECAST).} }
\label{fig:osc}
\end{figure}

\paragraph{Different Approaches between PyCEnNS and FORECAST: } Once an agreement on the physical constants, the cross-section parametrization and the neutrino flux between the two toolkits was reached, another problem needed to be address when the results of the interaction rates were compared. From Figure \ref{fig:osc} it is noticeable that when the resulting interaction rates from \textit{FORECAST} and \textit{PyCEnNS} are compared a sort of modulation is present that can be only due to a difference in the computation algorithms used. Besides the modulation, a disagreement in the low energy regime is noticeable as well.
After thorough investigation a difference in the calculation approach was noticed and it was determined to be the source of these two last discrepancies. In fact, while \textit{FORECAST} uses a more complex way of calculating the recoil spectra using and inverting a transfer matrix, \textit{PyCEnNS} instead uses a much more simple and direct approach that allows for much faster computation times. The two different approaches are translated in the computed results in two opposite ways: the \textit{FORECAST} approach tends to conserve the integral of the recoil rate, which means that when a sampling is chosen the value of the extracted points is adjusted in order to take into account for this choice, so that the integral of the interaction rate is sampling independent. The approach used by the \textit{PyCEnNS} framework is somewhat the opposite and consists in calculating the interaction rate at given nuclear recoil energies without considering the sampling. In this way the integral of the rate is not conserved between different samplings but the analytical shape remains consistent. This feature of the \textit{PyCEnNS} toolkit can be then used to speed up computation time and use less resources. In fact, since the analytical shape is preserved one can use a smaller amounts of samples when performing the interaction rate calculation, since it is a very computationally demanding task, and then interpolate in between the extracted samples whenever an accurate estimation of rate's integral is needed.\\

\begin{figure}[H]
\centering
\includegraphics[scale=0.5]{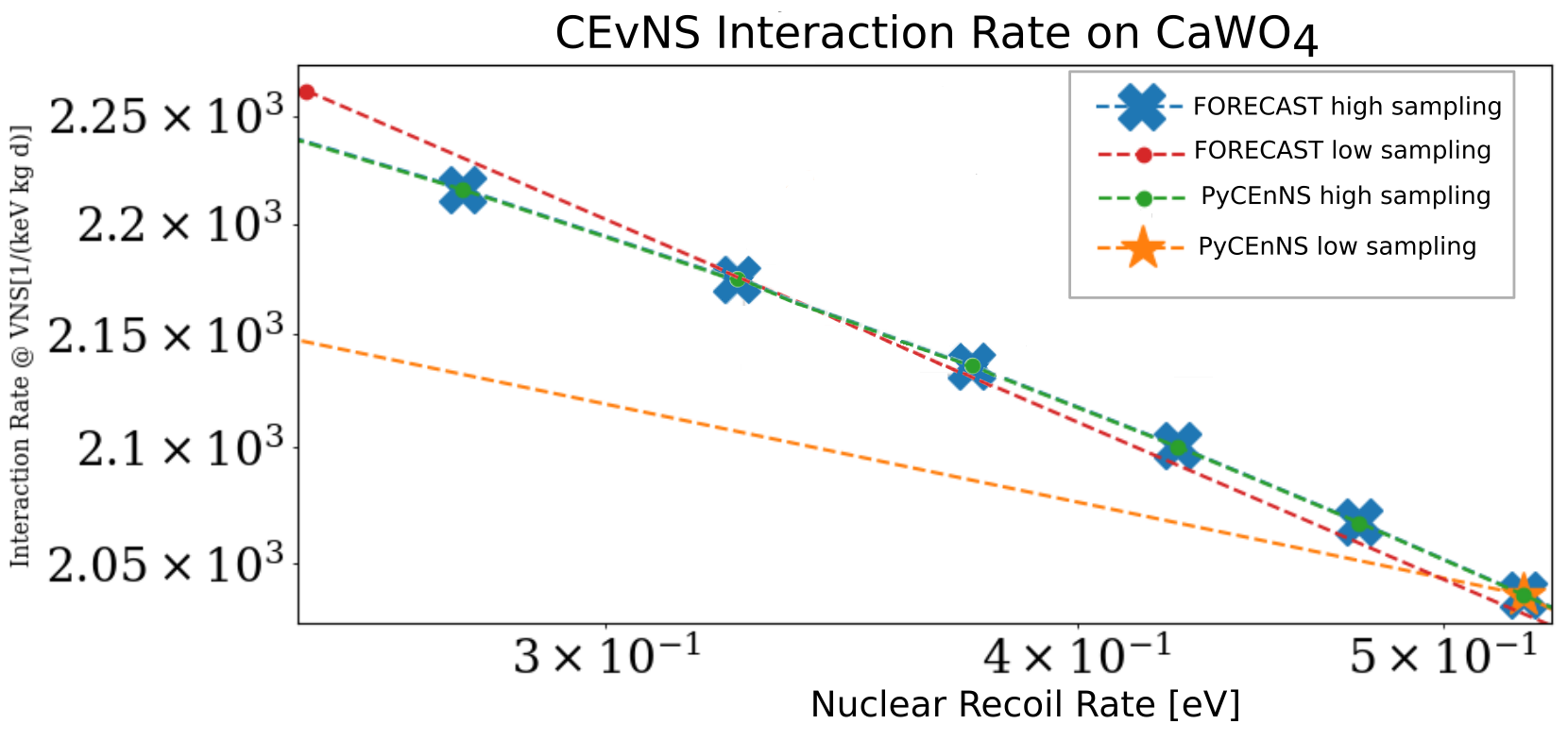}
\caption{Close-up look of the interaction rate predictions between \textit{PyCEnNS} and \textit{FORECAST} for two different sampling levels.}
\label{fig:analytical}
\end{figure}

The difference in behaviour between the two frameworks is shown in Figure \ref{fig:analytical}, it can be clearly seen that two different samplings yield the same results for the \textit{PyCEnNS} toolkit whether the \textit{FORECAST} results differ. In particular, the rate with the more sparse sampling has slightly increased values in order to compensate for the less amount of points.

After this behaviour was understood, it has been decided to keep the two distinct approaches in order to have a better validation of the results. \\

\begin{figure}[H]
\centering
\includegraphics[scale=0.47]{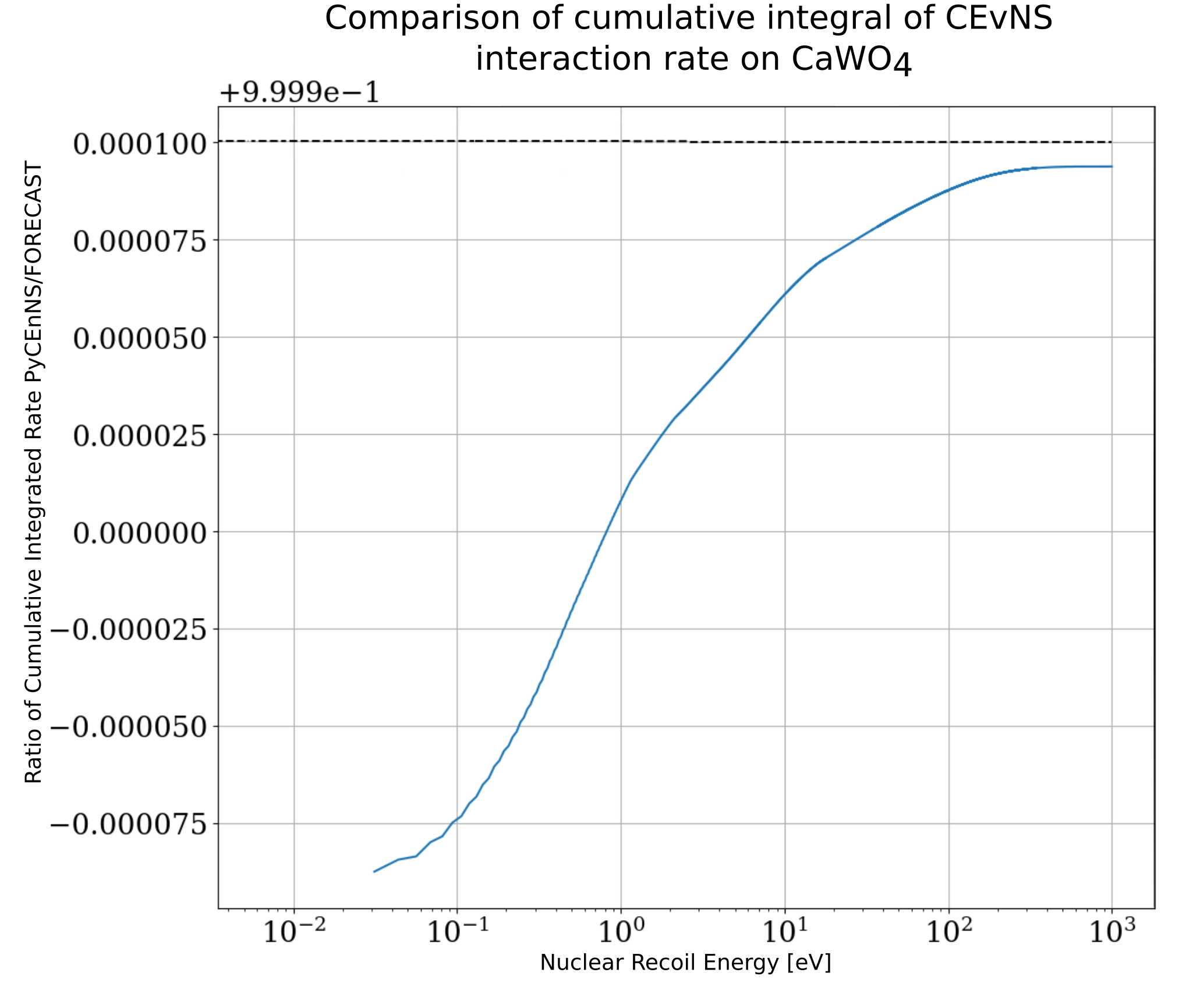}
\caption{ Comparison of cumulative integral of CE$\nu$NS interaction rate on CaWO$_4$ at VNS (blue). The dashed black line is constant at 1 and would be the expected result if the two sensitivity toolkits gave exactly the same results.}
\label{fig:best_sens}
\end{figure}

At this point a new type of comparison is in order, in fact, comparing the cumulative integral of the interaction rate and using the interpolation technique allows to fully understand the agreement between the two codes. In this comparison (Figure \ref{fig:best_sens}), an agreement under 0.005\%  was reached in the energy range above the foreseen lower threshold for NUCLEUS ($T\geq 10$~eV), which is a major improvement when compared with the initial 0.2\% disagreement. This high level of compatibility in the results also proves the power of the interpolation technique to further speed up the \textit{PyCEnNS} framework and save less amount of data virtually without loosing any amount of precision.

\subsubsection{PyCEnNS Experience Update}
Besides the improvements done in order to have the best possible agreement between the two sensitivity toolkits, \textit{PyCEnNS} underwent a serious makeover and redesign in order to be usable from all the members of the NUCLEUS collaboration and in general to make the overall usage experience more enjoyable.

\paragraph{User Interface Update: } The first set of upgrades concerns the interface of the code. One of the most important features that has been added to \textit{PyCEnNS} is the automatic code documentation produced with the \textit{Sphinx} program. All the functions throughout the code have been heavily commented and described compatibly with the documentation program that produces a user manual in both pdf and HTML formats.\\

Another important upgrade was the use of a logging program and loading bars for the communication with the user, not only these make the outputs of the code easily understandable, their use drastically decreased the computation time of the code since it has to make much less I/O requests to the operating system and all the requests are made through specified Python packages that have been thoroughly optimized (like \textit{TQDM} and \textit{COLOREDLOGS}). In order to improve the multi-user experience as well all the outputs of the code are forced to be stored in specified paths that are different for each section of the code. Moreover, all the settings used to generate such outputs are thoroughly logged and saved along with the generated data in order to easy reproduce the results in future calculations. \\

Some restyling of the code structure has also been done keeping in mind that in the near future several additional elements will be added in order to have more accurate calculations. The first upgrade concerns the positioning of all the constants used in the code in a dedicated module that can easily be imported, moreover, the source used for the constants' values have all been referenced along with a brief description and the units. Finally, one last feature added to \textit{PyCEnNS} is the dynamical choice of reactor neutrino flux and background models, now these can be chosen from the user at execution time instead of having to be manually changed in the source code as it was in the beginning. This last feature is particularly important since in the near future better background models and detector descriptions will be available and must be used in future sensitivity calculations.\\

All these upgrades might not look as important as the previous one that tackled more fundamental issues but are actually the backbone of the \textit{PyCEnNS} framework. These newly added features evolved the first version of this toolkit, developed for \cite{johannes}, that was computationally inefficient and hard to use, into a sensitivity studies framework that can rival, and in many aspects surpass, with the more professionally developed \textit{FORECAST} and can now be used by any member of the collaboration without any prior knowledge of the inner-workings of the code.

\paragraph{Computation Time and Resources Improvements: } Besides the upgrades on the user interface, several more upgrades have been done to make the \textit{PyCEnNS} code faster and less computationally demanding. The most important upgrade done on this front was rewriting all the code in order to function with vectorial computing allowing to perform the calculations on a vast number of samples simultaneously. This upgrade made extensive use of the \textit{Numpy} and \textit{Pandas} libraries and allowed for a drop in computation time between 50\% and 80\%, in particular during the extraction of the expected interaction rate. As mentioned above, the implementation of the vectorial calculations was accompanied with the use of the \textit{Simpons} numerical integration rule which also contributed to severely speed up the execution of the code.\\

Another upgrade made to speed up \textit{PyCEnNS} is the use of \textit{multiprocessing} and multicore calculations. This strategy is particularly useful when dealing with a great number of independent computations and is thus mostly suitable for the likelihood ratio test where several independent experiments are simulated and then go through the statistical machinery in order to extract the expected statistical significance of the discovery. Implementing the multiprocessing computations has been an arduous task since it is in conflict with another strategy used to make the code more performing, the use of \textit{lambda} functions. These functions are a peculiarity of the Python language and allow the user to define temporary functions. In \textit{PyCEnNS} the \textit{lambda} functions are used to rapidly access precomputed data by interpolating it and storing it into a function or by combining different models (chosen at runtime) in a single function that must then be integrated allowing for an extremely versatile modularity of the code.\\

Finally, the last newly implemented improvement is the fact that all the calculations described in this chapter are performed on batches of samples in order not to use up the whole computer memory, making \textit{PyCEnNS} able to deal with an arbitrary number of samples. This was done following machine learning techniques used to deal with great amounts of data that need to be loaded for the training of neural networks algorithms. The size of the batches was found through a trail and error procedure in order to maximize the benefits of vectorial and multiprocessing computations, since these ideally work best when all the data is loaded in the memory.\\

All these upgrades allow \textit{PyCEnNS} to perform sensitivity calculations with a great amount of samples in a computation time that ranges from $15~min$ to the couple of hours depending on the sampling, while \textit{FORECAST} would take almost half a day for the same calculations and has shown limitations in the amount of data it can handle. On the other hand, one must not forget that the C++/ROOT based framework is still extremely necessary to the collaboration since it implements accurate calculations for the extraction of the reactor neutrinos spectrum and is fundamental for all the cross-checks that can be easily used to validate the results.

\section{Toy Model Results}
\begin{figure}[H]
\centering
\includegraphics[scale=0.45]{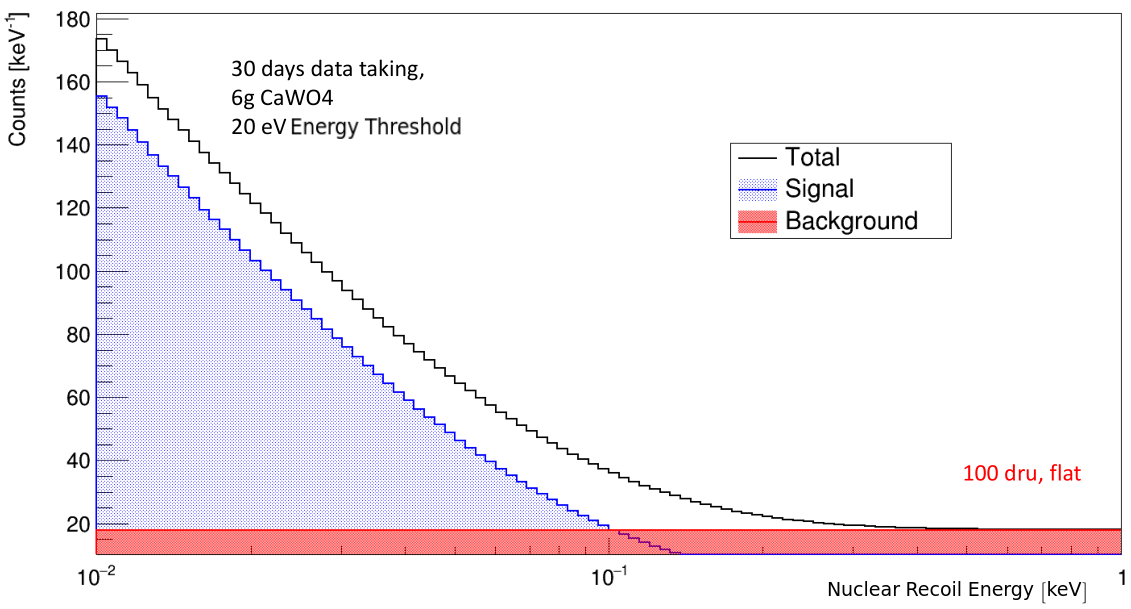}
\caption{Asimov spectra for the considered toy model, the CE$\nu$NS (blue) and background (red) contributions to the total spectrum (black) are presented. Spectrum generated from FORECAST and presents linear binning (from \cite{Alex}).}
\label{fig:toy_model}
\end{figure}

\begin{figure}[H]
\centering
\includegraphics[scale=0.48]{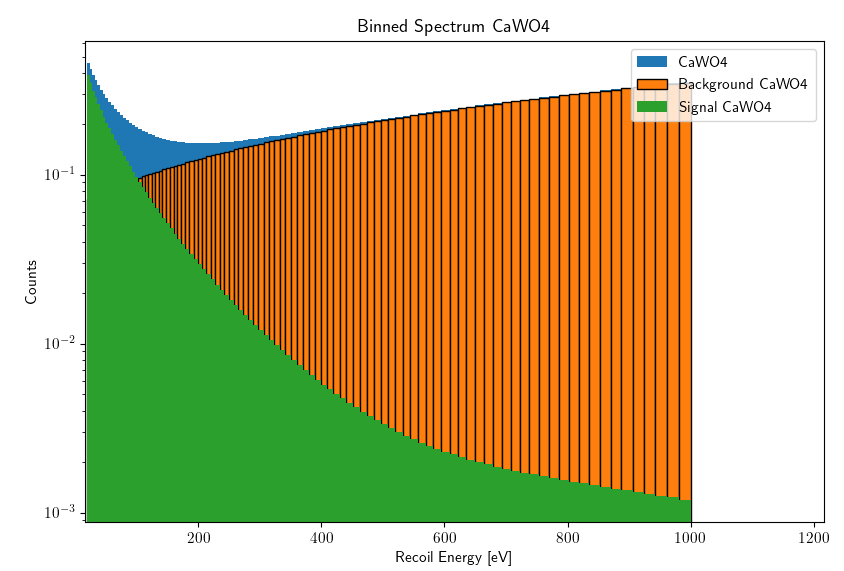}
\caption{Asimov spectra for the considered toy model presented with exponential binning. The binning is explicitly shown only for the background spectrum but is equivalent on all the presented spectra.}
\label{fig:toy_model2}
\end{figure}

In order to check the correct behaviour of the newly implemented and revised statistical significance study in both the \textit{PyCEnNS} and \textit{FORECAST} frameworks a toy model scenario was decided. The chosen experimental scenario consists of a $100$~dru flat background($1~\text{dru} = \frac{count}{\text{keV~kg~day}}$), a 30 day data taking period, a $6$~g CaWO$_4$ absorber for the detector and a lower nuclear recoil energy threshold of $20$~eV. The model scenario is summarized in Figures \ref{fig:toy_model} and \ref{fig:toy_model2}, where the latter presents the same scenario and spectra but with the exponential binning discussed previously. \\

All three types of calculations (using binned, unbinned and Asimov spectra) where performed in order to study eventual differences in the results. While the binned and Asimov calculations where conducted with both \textit{PyCEnNS} and \textit{FORECAST}, the unbinned procedure is specific of the \textit{PyCEnNS} framework and does not have any counterpart but can still be compared with the binned results. The studies that require random extractions (so the normal binned and unbinned procedures) where conducted using 100000 simulated experiments in order to have a big enough number of samples to calculate the statistical significance.\\

\begin{figure}[H]
\centering
\includegraphics[scale=0.67]{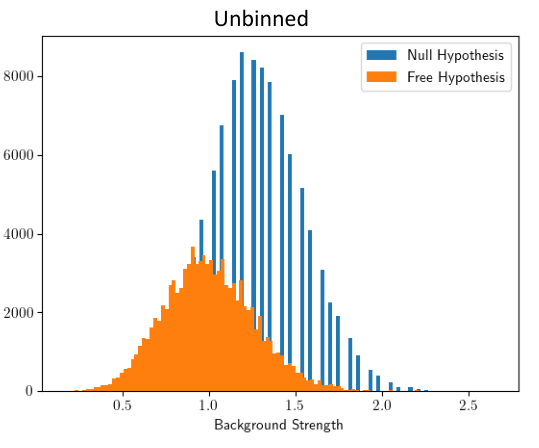}
\caption{Distribution of the $\beta_0$ parameter for the \textit{null} (blue) and \textit{free} (yellow) hypothesis. The results are only for the unbinned calculation.}
\label{fig:bcknd_str}
\end{figure}

The quantity that gives a first comparison between all the stages of the calculation is the strength of the background model (i.e. the $\beta_0$ parameter). The convergence value of this parameter is shown in Figure \ref{fig:bcknd_str}, it can be immediately seen that the average value of this parameter is greater for the background only hypothesis when compared to the \textit{free} one.  This is an expected behaviour since it needs to account for all those events that are due to CE$\nu$NS . Another noticeable characteristic is the fact that for the background only hypothesis the value of $\beta_0$ has a very discrete spectrum, this is most likely due to the fact that a flat background hypothesis is too simplistic to explain the excess of events due to neutrino-nuclear scattering and thus the minimization converges always in the same set of points that are local minima common for most generated spectra. This convergence behaviour is completely absent for the \textit{free} hypothesis which points to the correct functioning of the code. Moreover, another indication of the correct working of the code is the fact that no visible difference between the binned and unbinned hypothesis is present, since these are very distinct calculations and could yield very different results in case of errors.

\begin{figure}[H]
\centering
\includegraphics[scale=0.5]{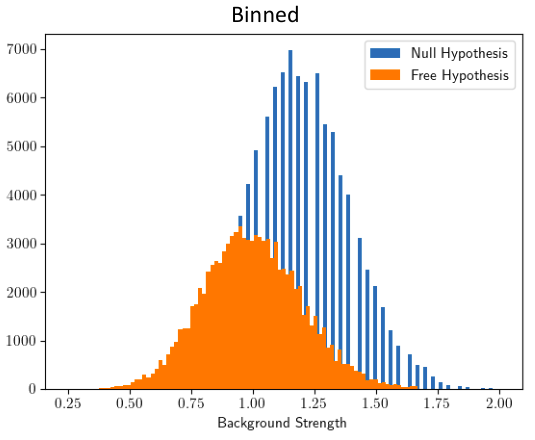}
\caption{Distribution of the $\beta_0$ parameter for the \textit{null} (blue) and \textit{free} (yellow) hypothesis. Results only for the binned calculation.}
\label{fig:bcknd_str2}
\end{figure}

The results of the toy model studies are presented in Table \ref{tab:toy_model}, no real difference between the results from \textit{PyCEnNS} and \textit{FORECAST} is visible. A slight increase in the statistical significance is present in both codes when using the Asimov technique, but this is to be expected since the generated spectrum fits perfectly with the \textit{free} hypothesis.

\begin{table}[H]
\centering
\caption{Toy model results for 30 day exposure with $6$~g of CaWO$_4$ and $100$~dru flat background.}
\begin{tabular}{cccccc}
\hline
               & $\beta_0$ & $\beta_1$ & $\sigma_1$ & $Z$ & $Z$ (Asimov)  \\
\hline
\hline
\makecell{\textbf{FORECAST}\\(binned)} & $1.29\pm0.27$ & $1.00\pm 0.25$ & $1.04\pm 0.51$& $2.21 \pm 1.19$ & 2.31\\
\makecell{\textbf{PyCEnNS}\\(binned)} & $1.28\pm0.27$& $1.00\pm 0.26$ & $1.00\pm 0.59$& $2.21 \pm 1.15$ & 2.28\\
\makecell{\textbf{PyCEnNS}\\(unbinned)} & $1.28\pm 0.27$& $1.00\pm 0.26$ & $1.00\pm 0.60$& $2.20 \pm 1.16$ & --\\
\hline
\end{tabular}
\label{tab:toy_model}
\end{table}

\begin{figure}[H]
\centering
\includegraphics[scale=0.5]{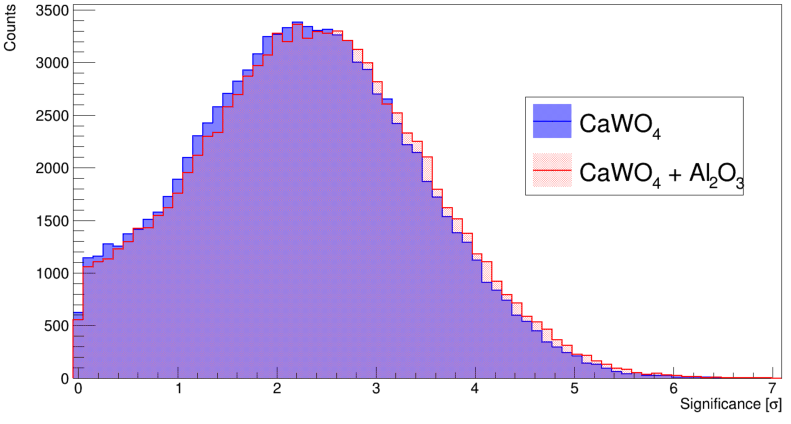}
\caption{Distribution of statistical significance with only calcium tungstate (blue) and with the addition of sapphire (red). Figure from \cite{Alex}.}
\label{fig:saffire}
\end{figure}

A further study on this toy model scenario has been done by adding $4$~g of Al$_2$O$_3$ to the calcium tungstate detector, with the same energy threshold as before. From Figure \ref{fig:saffire}, it can been seen that the use of additional sapphire slightly increases the statistical significance but the 30 day measuring period is not sufficient to drastically improve the detection. The reason why the addition of sapphire increases the discovery power of the experiment is because the background is mostly independent on the detector material but CE$\nu$NS instead scales with the square of the number of neutrons in the target, thus sapphire being lighter than calcium tungstate primarily probes the background rather than neutrino-nucleus scattering events.

\chapter{Conclusions}
\label{chap:7}
\lettrine[lines=2, findent=3pt, nindent=0pt]{T}{}he NUCLEUS experiment aims to use gram-scale cryogenic detectors to measure antineutrinos emitted by commercial PWR reactors via the coherent elastic neutrino nucleus scattering (CE$\nu$NS) with an unprecedented low threshold on the nuclear recoil energy of $10$~eV. For the first phase of the experiment, denoted as NUCLEUS-10g, a 10~g cryogenic detector made with a combination of absorber cubes of CaWO$_4$ and Al$_2$O$_3$ each coupled with a transition-edge sensor will be deployed at the Chooz nuclear power plant.\\

The work described in this thesis focuses on the development of several new tools to be used in the NUCLEUS experiment in order to overcome several problems that arise when using ultra-low threshold cryogenic detectors. The first activity being described in this dissertation focuses on the development of the setup of an optical calibration system. The calibration of low threshold detectors is not an easy task since most of the radioactive sources saturate the detector response. With the optical calibration setup developed in this work the detector response can be accurately characterized by delivering a small amount of energy on each pixel of the detector. The calibration procedure described in this work is based on the absolute calibration  which exploits photon statistics to measure the responsivity of the detector under test without requiring the knowledge of the amount of photons, i.e. the amount of energy, delivered during the calibration.\\

The calibration electronics is comprised of a UV LED Driver that generates light pulses when triggered, for example, with a signal generator. The output of the UV LED driver is connected to the input channel of the Fiber Optical Switch which allows to connect one incoming optical channel to multiple outgoing ones. A dedicated computer software was developed in order to fully exploit the features of these three devices when they are used together. This software allows to perform several characterizations of cryogenic array detectors such as cross-talk studies and single pixel calibration. The main use of this electronics and software is to fully automatize the absolute calibration of an array of cryogenic detectors in order to be performed swiftly and often throughout the data acquisition. The electronics setup and piloting software have been tested using kinetic inductance detectors developed in the BULLKID R$\&$D experiment and have shown extremely good performance both regarding the easy usage and the available functionalities. This calibration setup, both hardware and software, is already integrated in the experimental setup of the BULLKID experiment and will soon be merged in NUCLEUS during the upcoming detector assembly in the end of 2021.\\

Using the absolute calibration the first prototype of the \textit{mirror wafer} was characterized. This device is used to split in two the light pulses coming from an optical fiber and allows to use half the amount of optical channels needed if each pixel of the detector array needed a dedicated fiber. Using a lower amount of fibers not only allows to use simpler electronics but mainly serves to better preserve the thermal isolation of the lowest temperature plate of the cryostat. The test performed on the \textit{mirror wafer}, besides showing quite a big imbalance in the light splitting, is considered a success since the device worked without requiring any particular care. The unbalanced light division is probably due to a misalignment of the optical fiber which is the main aspect that needs to be perfect in further tests.\\

The second activity performed in this thesis work is the development of an analysis software and protocol for the NUCLEUS experiment. At the current stage the NUCLEUS experiment does not have an official analysis software this is why during the making of this thesis the analysis framework DIANA has been proposed. DIANA has been initially developed for the CUORE experiment at LNGS, and in order to be used by the NUCLEUS collaboration several new upgrades have been made both to make it compatible with the data and to make the analysis procedures more accurate. Moreover, a new automatic analysis protocol has been developed in order to replace the old analysis procedure which required a great deal of user interaction. In particular, the new protocol efficiently identifies events in the linear response region of the detector and uses them to build the average pulse which corresponds to an almost noise free description of the detector response. Once the average pulse has been built, it is used to accurately reconstruct events that saturate the dynamical range of the cryodetector in order to measure their amplitude. This reconstructing algorithm was newly developed and tested on the data acquired with a prototype NUCLEUS run. The results achieved with this new protocol are more than satisfactory since it efficiently replicates the results in \cite{RUN1} that have been obtained with an almost event by event model dependent analysis of the data.\\

Lastly, a python based software for the sensitivity studies on the discovery potential of the NUCLEUS experiment has been developed following the PhD work of J. Rothe. The work mainly focused on the upgrade of PyCEnNS, the python toolkit for the sensitivity studies, in order to achieve optimal agreement with the FORECAST software. The agreement between the two softwares was reach below the 0.005\% level on the CE$\nu$NS interaction rate calculation for the NUCLEUS-10g detector. Moreover, PyCEnNS was intensively upgraded in order to cover one of the failings of the FORECAST toolkit which concerns the computation time and the amount of data it can handle. The computation time of the PyCEnNS toolkit has been reduced of about 80\% with respect to its initial stage and can handle virtually an infinite amount of data without saturating the available resources using machine learning techniques on data handling (obviously using more data influences the computation time). The two softwares were also tested on a toy model scenario for the CE$\nu$NS detection using a 100~dru flat background and obtained strikingly similar results even when implementing drastically different calculations. With respect to the FORECAST software, PyCEnNS also implements studies using unbinned statistics that, while being more computationally demanding, do not introduce any bias due to the binning and can be used to further validate the results.\\

The whole work presented in this thesis serves as a guide for these newly developed tools that need to be soon integrated in the NUCLEUS experiment considering the upcoming NUCLEUS-10g phase (end of 2021 - beginning of 2022). Moreover, since the calibration setup and the analysis protocol are not necessarily specific for the NUCLEUS experiment they can be exploited by the whole low energy particle physics community. In particular, since the expected WIMP signature is extremely similar to the CE$ \nu $NS one, these tools could be exploited in future dark matter experiments involving cryogenic detectors.

\printbibliography
[
heading=bibintoc,
title={Bibliography}
]
\chapter*{Acknowledgements}
\label{chap:8}
It is always a not so easy task to write the acknowledgments, thinking back to these last couple of peculiar years of university, a great deal of people that have helped me comes to mind. For sure, a particular thanks must be dedicated to Prof. Vignati, Dr. Tomei and Dr. Casali who have helped and guided me through the developments of this thesis and much more. Special thanks are also in order for Prof. Strauss, J. Rother and A. Wex and, in general, the whole of the NUCLEUS collaboration which have been a wonderful team of researchers to work along.\\

Giovanni (Isopi), Andrea (Falzetti), Marco (Giammei) and Daniele (Delicato) require a special mention, since they have been my support and study group through practically all the phases of my master's degree. I also need to thank greatly Leo (Pancheri), Paolo (Forti) and Daniele (Colombi) for being my moral support and for keeping me grounded throughout these pandemic years. In general, I have to thank profoundly all of my friends which have been wonderfully supportive, in more ways than one, throughout all of my university years.\\

An extremely felt thank you must also be issued for Ines which has been wonderfully helpful, supportive, and is also particularly patient with all of my weird side projects and ideas. The same, obviously, can also be said about my parents and brother, that always support and help me with my decisions.\\

If you are reading this and feel left out... well you know me... I am bad with timing and as usual I am writing the acknowledgments at the last moment possible. That is why, I want to deeply thank everyone who throughout these years has helped me both to achieve my goals and to relax after a long day of studying.
\end{document}